\documentclass{jpp}
\usepackage{graphicx}
\usepackage[export]{adjustbox}
\usepackage{comment, color}

\usepackage[utf8]{inputenc}
\usepackage[T1]{fontenc}

\newcommand{\changed}[1]{{#1}}
\newcommand{\vect}[1]{\mbox{\boldmath $#1$}}
\newcommand{\Gsign}{ s_G }
\newcommand{\psisign}{ s_{\psi} }

\newcommand{\iotaN}{\iota_N}
\newcommand{\etabar}{\bar{\eta}}

\usepackage{amsmath}
\usepackage{amssymb}

\shorttitle{Constructing stellarators with quasisymmetry to high order}
\shortauthor{M. Landreman and W. Sengupta}

\title{Constructing stellarators with quasisymmetry to high order}

\author{Matt Landreman\aff{1}
\corresp{\email{mattland@umd.edu}}
\and
 Wrick Sengupta\aff{2}}

\affiliation{\aff{1}Institute for Research in Electronics and Applied Physics, University of Maryland, College Park MD 20742, USA
\aff{2}Courant Institute of Mathematical Sciences, New York University, New York NY 10012, USA}

\begin{document}

\maketitle

\begin{abstract}
A method is given to rapidly compute quasisymmetric stellarator magnetic fields for plasma confinement,
without the need to call a three-dimensional magnetohydrodynamic equilibrium code
 inside an optimization iteration.
The method is based on direct solution of the equations of magnetohydrodynamic
equilibrium and quasisymmetry using Garren and Boozer's expansion about the magnetic axis (Phys Fluids B 3, 2805 (1991)), and it is several orders of magnitude faster than
the conventional optimization approach.
The work here extends the method of Landreman, Sengupta and Plunk (J Plasma Phys 85, 905850103 (2019)),
which was limited to flux surfaces with  elliptical cross-section,
to higher order in the aspect ratio expansion. As a result, configurations can be generated with strong shaping that achieve quasisymmetry to high accuracy.
Using this construction, we give the first numerical demonstrations of Garren and Boozer's ideal scaling of quasisymmetry-breaking with the cube of inverse aspect ratio.
We also demonstrate a strongly nonaxisymmetric
configuration (vacuum $\iota > 0.4$) in which symmetry-breaking mode amplitudes throughout a finite volume are $< 2\times 10^{-7}$, the smallest ever reported.
To generate boundary shapes of finite-minor-radius configurations, a careful
analysis is given of the effect of substituting a finite minor radius into the near-axis expansion.
The approach here can provide analytic insight into the space of possible
quasisymmetric stellarator configurations, and it can be used to generate good
initial conditions for conventional stellarator optimization.

\end{abstract}

\section{Introduction}

Quasisymmetry is a type of continuous symmetry in the strength 
of a toroidal magnetic field $B=|\vect{B}|$ that does
not require continuous symmetry of the magnetic field vector $\vect{B}$ \citep{Boozer83,NuhrenbergZille,Boozer1995,HelanderReview}. As a consequence of the conservation laws associated with quasisymmetry or full axisymmetry of $\vect{B}$, both symmetries enable confinement of charged particles and plasma. However, confinement with axisymmetric $\vect{B}$
requires a large electric current in the confinement region that is prone to instabilities and hard to sustain, while quasisymmetric confinement does not. Hence, non-axisymmetric toroidal magnetic fields (``stellarators'') with quasisymmetry offer the promise of stable and efficient confinement of high-temperature plasma for fusion energy.
Quasisymmetric stellarators would also enable magnetic confinement of plasmas with density that is too low to support substantial electric current, such as  electron-positron plasmas for basic physics studies \citep{Pedersen}.

Several quasisymmetric magnetic field configurations have been found numerically,
mostly by using optimization over the space of boundary magnetic surface shapes to minimize
symmetry-breaking Fourier modes of $B$ \citep{NuhrenbergZille,HSX,Garabedian,NCSX,KuBoozerQHS,ESTELL,Henneberg}. While this optimization approach is effective,
it does not provide much insight into the size and character of the solution space, and
it requires good initial guesses for the numerical iteration. Hence, one can never be sure that all the interesting regions of parameter space have been found, for perhaps a different initial guess would yield a new solution. The optimization approach requires significant computation time, and so it is expensive to generate parameterized families of solutions.

An alternative to the optimization approach is to construct quasisymmetric configurations directly using an analytic expansion, in the smallness of either the departure from  axisymmetry of $\vect{B}$ \citep{PlunkHelander} or in the distance from the magnetic axis (equivalent to an expansion in large aspect ratio). Near-axis expansions have been explored by several authors \citep{Mercier,SolovevShafranov,LortzNuhrenberg}, with the particular case of quasisymmetry examined by \citet{GB1,GB2}. 
The near-axis expansion, though it is an approximation, is always accurate in the core of any stellarator, even stellarators for which the aspect ratio of the outermost surface is not large.
In a recent series of papers \citep{PaperI, PaperII, PaperIII}, the near-axis expansion was developed into practical procedures for constructing  fields with quasisymmetry, or the more general condition of omnigenity. It was also shown that close to the axis, quasisymmetric configurations obtained by conventional optimization closely match configurations generated by the construction \citep{fitToGarrenBoozer}. The configurations presented to date from this 
near-axis construction
have been quasisymmetric to first order in $r/\mathcal{R}$, where $r$ is the typical distance from the axis and $\mathcal{R}$ denotes a scale length of the magnetic axis (such as a typical radius of curvature).
At this order, due to regularity conditions at the magnetic axis, the magnetic surfaces must have an elliptical cross-section. It was found that the space of quasisymmetric configurations
to this order can be parameterized by the shape of the axis together with three other  numbers.

In the present paper, we extend the construction to next order in $r/\mathcal{R}$. 
The equations describing quasisymmetry to $O((r/\mathcal{R})^2)$ were derived in the appendix of \citet{GB2}, but no solutions were presented before now.
At this order, several important effects appear for the first time, including triangularity and Shafranov shift. By extending the construction to $O((r/\mathcal{R})^2)$, more complicated and realistic stellarator shapes will be generated, and quasisymmetry will be achieved to higher accuracy.
The extension of the model to $O((r/\mathcal{R})^2)$ only slightly increases the computational cost of solving the equations, which remains on the level of milliseconds on one CPU. This time is far faster than a traditional 3D equilibrium calculation, which typically requires on the order of 10 seconds or more.

In Garren and Boozer's original work, it was argued quasisymmetry can be achieved (in the absence of axisymmetry) to $O((r/\mathcal{R})^2)$ but not to $O((r/\mathcal{R})^3)$, so departures from quasisymmetry should scale as the cube of the inverse aspect ratio. However, despite various numerical calculations of quasisymmetric configurations using optimization since 1991, 
there does not appear to have been a numerical demonstration of this predicted scaling. Using the construction here we are able to numerically demonstrate this predicted ideal scaling for the first time (figure \ref{fig:GarrenBoozerScaling}). An implication of this scaling is that quasisymmetry can be achieved to arbitrary precision, in the following sense. Given any desired small level of symmetry-breaking Fourier modes in $B$, and given any desired axis shape (constrained only by the requirement that its curvature cannot vanish), there is some aspect ratio above which quasisymmetry can be achieved to the desired precision.
To emphasize this point, we will present examples of nonaxisymmetric configurations in which quasisymmetry is realized to unprecedented precision, with the symmetry-breaking mode amplitudes orders of magnitude smaller than in previously reported configurations.

A primary application of the work here is to generate input data for stellarator equilibrium codes such as VMEC  \citep{VMEC1983} or optimization codes such as STELLOPT \citep{StelloptSpong,StelloptReiman} or ROSE \citep{ROSE}. For these applications, the input we must generate is the shape (or initial shape) of a boundary magnetic surface. In the $O(r/\mathcal{R})$ construction \citep{PaperII,PaperIII}, it was possible to plug a finite value of $r$ into the near-axis expansion to obtain the boundary surface. In turns out that at $O((r/\mathcal{R})^2)$, this substitution requires some care. We will show that in fact, a part of the $O((r/\mathcal{R})^3)$ shape must be retained in order to generate a boundary surface
that is consistent with the desired field strength to $O((r/\mathcal{R})^2)$. Once this step is taken, we will construct boundary surfaces, then use the VMEC code to generate 3D equilibria inside the boundaries, and show that quasisymmetry-breaking modes
of $B$ in these equilibria are small and scale with the aspect ratio as expected.

In the following section, notation will be introduced and the near-axis expansion will be outlined. Section \ref{sec:finite_r} describes the analysis of generating a finite-aspect-ratio boundary
surface from the near-axis expansion, and the need for including some $O((r/\mathcal{R})^3)$ terms.
The numerical method for solving the equations is detailed in section \ref{sec:numerical},
and several examples of constructed quasi-axisymmetric and quasi-helically symmetric configurations are presented in section \ref{sec:results}. We discuss the results and conclude in section \ref{sec:conclusions}. Several detailed analytic calculations can be found in the appendices. Appendix \ref{sec:2nd_order_field_strength} gives the equations for $O((r/\mathcal{R})^2)$ quasisymmetry, derived using a new method that reduces the algebra required. Appendix \ref{sec:finite_r_appendix} gives a detailed proof of results presented in section \ref{sec:finite_r}. Finally, one method for converting the constructed boundary shapes to cylindrical coordinates is presented in appendix \ref{sec:transformation}.


\section{Near-axis expansion}
\label{sec:expansion}

Our goal is to relate the three-dimensional shapes of flux surfaces to the magnetic field strength in Boozer coordinates $(\theta,\varphi)$. In this section, we introduce the main features of the expansion, and many of the explicit expressions are given in appendix \ref{sec:2nd_order_field_strength}. While the expansion here is equivalent to the one in \cite{GB1,GB2}, our approach in appendix \ref{sec:2nd_order_field_strength} provides a streamlined method to derive the equations at each order. Throughout the analysis, we assume that good nested flux surfaces exist in the region of interest near the axis.

In Boozer coordinates, the magnetic field has the forms
\begin{align}
\label{eq:BoozerCoords}
\vect{B} = &\nabla\psi \times\nabla\theta + \iota \nabla\varphi \times\nabla\psi, \\
 = &\beta \nabla\psi + I \nabla\theta + G \nabla\varphi, \nonumber
\end{align}
where $2\pi\psi$ is the toroidal flux, $\iota(\psi)$ is the rotational transform, $\theta$ and $\varphi$ are the poloidal and toroidal Boozer angles, and $I$ and $G$ are constant on $\psi$ surfaces. To consider quasi-helical symmetry later in the analysis, it is convenient to introduce a helical angle
$\vartheta = \theta - N \varphi$ where $N$ is a constant integer. Then
\begin{align}
\vect{B} = &\nabla\psi \times\nabla\vartheta + \iota_N \nabla\varphi \times\nabla\psi,
\label{eq:straight_field_lines_h}
\\
 =& \beta \nabla\psi + I \nabla\vartheta + (G+NI) \nabla\varphi,
\label{eq:Boozer_h}
\end{align}
where $\iota_N = \iota - N$.

The position vector $\vect{r}$ at a general point in a neighborhood of the axis can be described by
\begin{align}
\label{eq:positionVector}
\vect{r}(r,\vartheta,\varphi) = \vect{r}_0(\varphi)
+X(r,\vartheta,\varphi) \vect{n}(\varphi)
+Y(r,\vartheta,\varphi) \vect{b}(\varphi)
+Z(r,\vartheta,\varphi) \vect{t}(\varphi),
\end{align}
where $r$ is the flux surface label defined by $2 \pi \psi = \pi r^2 \bar{B}$ with $\bar{B}$ a constant reference field strength, and $\vect{r}_0(\varphi)$ is the position vector along the magnetic axis. Here, the orthonormal vectors $(\vect{t},\vect{n},\vect{b})$ give the Frenet-Serret frame of the magnetic axis. These vectors satisfy
\begin{align}
\frac{d\varphi}{d\ell}
\frac{d\vect{r}_0}{d\varphi} = \vect{t}, 
\hspace{0.3in}
\frac{d\varphi}{d\ell}
\frac{d\vect{t}}{d\varphi} = \kappa \vect{n}, 
\hspace{0.3in}
\frac{d\varphi}{d\ell}
\frac{d\vect{n}}{d\varphi} = -\kappa \vect{t} + \tau \vect{b}, 
\hspace{0.3in}
\frac{d\varphi}{d\ell}
\frac{d\vect{b}}{d\varphi} = -\tau \vect{n}, 
\label{eq:Frenet}
\end{align}
and $\vect{t}\times\vect{n}=\vect{b}$,
where $\ell$ is the arclength along the axis, $\kappa(\varphi)$ is the axis curvature, and $\tau(\varphi)$ is the axis torsion. (Garren and Boozer use the opposite sign convention for torsion.) Using the dual relations,
\begin{align}
\label{eq:dual}
\nabla \varphi = \frac{1}{\sqrt{g}} \frac{\partial\vect{r}}{\partial r}\times\frac{\partial\vect{r}}{\partial\vartheta}\hspace{0.2in}
\text{\& cyclic permutations,}
\end{align}
where $\sqrt{g}=(\partial\vect{r}/\partial r)\cdot(\partial\vect{r}/\partial\vartheta) \times
(\partial\vect{r}/\partial\varphi)$ is the Jacobian, then
(\ref{eq:straight_field_lines_h})-(\ref{eq:Boozer_h}) can be expressed in terms
of the $(\vect{t},\vect{n},\vect{b})$ vectors and derivatives of $(X,Y,Z)$.
Equating (\ref{eq:straight_field_lines_h})-(\ref{eq:Boozer_h}) then gives three scalar equations,
(\ref{eq:YZ})-(\ref{eq:XY}).
The field strength can be expressed in terms of derivatives of $(X,Y,Z)$ using the square of either (\ref{eq:straight_field_lines_h}) or (\ref{eq:Boozer_h}). The former turns out to be more useful, and is given in (\ref{eq:modB}).

These equations are supplemented by the equilibrium condition $[\nabla \times \text{(\ref{eq:Boozer_h})}] \times \text{(\ref{eq:straight_field_lines_h})} = \mu_0 \nabla p$, where $p(r)$ is the pressure. The average of this condition over $\vartheta$ and $\varphi$ gives
\begin{align}
\frac{dG}{dr}+\iota\frac{dI}{dr}=-\frac{\mu_0}{(2\pi)^2} (G+\iota I) \frac{dp}{dr}  \int_0^{2\pi}d\vartheta \int_0^{2\pi}d\varphi \frac{1}{B^2},
\label{eq:MHD_avg}
\end{align}
while the $\vartheta$ and $\varphi$ dependence  of the equilibrium condition implies
\begin{align}
\frac{\partial\beta}{\partial\varphi}+\iotaN\frac{\partial\beta}{\partial\vartheta}
=\frac{\mu_0}{r \bar{B}} \frac{dp}{dr} (G+\iota I) \left[
 \frac{1}{B^2}- \frac{1}{(2\pi)^2} \int_0^{2\pi}d\vartheta \int_0^{2\pi}d\varphi \frac{1}{B^2} \right].
\label{eq:beta}
\end{align}

The near-axis expansion is then introduced by writing
\begin{align}
X(r,\vartheta,\varphi)
= r X_1(\vartheta,\varphi) + r^2 X_2(\vartheta,\varphi) + r^3 X_3(\vartheta,\varphi) + \ldots,
\label{eq:radial_expansion}
\end{align}
with analogous expressions for $Y$ and $Z$. 
Other than $r$, all scale lengths in the system are ordered as $\mathcal{R}$, so (\ref{eq:radial_expansion})
represents an expansion in $r/\mathcal{R}$.
The field strength is expanded similarly but with an $O((r/\mathcal{R})^0)$ term:
\begin{align}
\label{eq:radial_expansion_B}
B(r,\vartheta,\varphi)
= B_0(\varphi) + r B_1(\vartheta,\varphi) + r^2 B_2(\vartheta,\varphi)+ r^3 B_3(\vartheta,\varphi) + \ldots,
\end{align}
and $\beta(r,\vartheta,\varphi)$ is expanded in the same way. 
The profile functions
$G(r)$, $I(r)$, $p(r)$, and $\iota_N(r)$ are analytic functions of $\psi$, so their expansions contain only even powers of $r$:
\begin{align}
G(r) = G_0 + r^2 G_2 + r^4 G_4 + \ldots.
\end{align}
Since $I(r)$ is proportional to the toroidal current inside the surface $r$, then $I_0 = 0$.
From analyticity considerations near the axis (see appendix A of \cite{PaperI}), the expansion coefficients have the form
\begin{align}
\label{eq:poloidal_expansions}
X_1(\vartheta,\varphi) = &X_{1s}(\varphi) \sin(\vartheta) + X_{1c}(\varphi) \cos(\vartheta), \\
X_2(\vartheta,\varphi) = &X_{20}(\varphi) + X_{2s}(\varphi) \sin(2\vartheta) + X_{2c}(\varphi) \cos(2\vartheta), \nonumber \\
X_3(\vartheta,\varphi) = &X_{3s3}(\varphi) \sin(3\vartheta) + X_{3s1}(\varphi) \sin(\vartheta) + X_{3c3}(\varphi) \cos(3\vartheta) + X_{3c1}(\varphi) \cos(\vartheta).
 \nonumber
\end{align}
The expansions of $Y$, $Z$, $B$, and $\beta$ have the same form.
The expansions (\ref{eq:radial_expansion})-(\ref{eq:poloidal_expansions})
are then substituted into (\ref{eq:YZ})-(\ref{eq:XY}),  (\ref{eq:modB}), and 
(\ref{eq:MHD_avg})-(\ref{eq:beta}), and terms at each order in $r/\mathcal{R}$ are collected.
We can thereby relate the surface shape coefficients $(X,Y,Z)$ to the field strength through
a desired order in $r/\mathcal{R}$. Explicit results through $O((r/\mathcal{R})^2)$ are given in
(\ref{eq:G0})-(\ref{eq:beta1}).

Quasisymmetry is achieved through order $O((r/\mathcal{R})^j)$ when $\partial B_k/\partial\varphi=0$ for $k \le j$.
At $O((r/\mathcal{R})^1)$, the analysis in appendix \ref{sec:2nd_order_field_strength} shows quasisymmetric fields are described by
\begin{align}
\label{eq:position_vector_r1}
\vect{r}(r,\vartheta,\varphi) = \vect{r}_0(\varphi)
+ \frac{r \etabar}{\kappa(\varphi)}\cos\vartheta \vect{n}(\varphi)
+ \frac{r s_\psi s_G \kappa(\varphi)}{\etabar} \left[ \sin\vartheta + \sigma(\varphi) \cos\vartheta \right] \vect{b}(\varphi)
+O(r^2/\mathcal{R}),
\end{align}
where $s_\psi = \mathrm{sign}(\psi)$, $s_G = \mathrm{sign}(G_0)$, $\etabar$ is a constant, and $\sigma(\varphi)$ is a solution of
\begin{align}
\label{eq:sigma_quasisymmetry}
\frac{d\sigma}{d\varphi} + (\iota_0 - N) \left[ \frac{\bar{\eta}^4}{\kappa^4} + 1 + \sigma^2 \right]
-\frac{2 G_0 \bar{\eta}^2}{B_0 \kappa^2} \left[ \frac{I_2}{B_0} - s_\psi \tau \right] = 0.
\end{align}
Here, the constant $I_2$ is the leading term in the coefficient $I(r)$
of (\ref{eq:BoozerCoords}), and is proportional to the on-axis toroidal current density, which is typically zero. 
The constant $\bar\eta = B_{1c} / B_0$ (introduced by \citet{GB1}) reflects the magnitude by which $B$ varies on surfaces:
\begin{align}
B \approx B_0 \left[ 1 + r \bar{\eta} \cos\vartheta + O((r/\mathcal{R})^2)\right].
\end{align}
The surface shapes corresponding to (\ref{eq:position_vector_r1}) are ellipses
in the plane perpendicular to the axis (the $\vect{n}$-$\vect{b}$ plane). The ellipses
are centered on the magnetic axis, so there is no Shafranov shift to this order.

It can be proved \citep{PaperII} that quasisymmetric fields to $O(r/\mathcal{R})$ can be parameterized by the shape of the axis - which can be any curve for which $\kappa$ never vanishes - along with three real numbers: $I_2$, $\bar\eta$, and $\sigma(0)$. 
Varying $\bar\eta$ has the effect of varying the average elongation.
The parameter $\sigma(0)$ is the value of $\sigma(\varphi)$ at $\varphi=0$, and it reflects the angle by which the major and minor axes of the elliptical flux surfaces are oriented with respect to $\vect{n}$ at $\varphi=0$. Stellarators typically posses `stellarator symmetry' (unrelated to quasisymmetry) which implies $\sigma(0)=0$. The pressure profile
does not appear in the model to this order.

Proceeding to $O((r/\mathcal{R})^2)$, nine new functions of $\varphi$ arise in the surface shapes:
$X_{20}$, $X_{2s}$, $X_{2c}$,
$Y_{20}$, $Y_{2s}$, $Y_{2c}$,
$Z_{20}$, $Z_{2s}$, and $Z_{2c}$. 
The corresponding surface shapes can now posses triangularity, and $X_{20}$ and $Y_{20}$
enable the center of the surfaces at any given $(r,\varphi)$ to be shifted in the $\vect{n}$-$\vect{b}$ plane with respect
to the axis (Shafranov shift.)
These nine functions are constrained by 10 new $\varphi$-dependent equations:
(\ref{eq:Z20})-(\ref{eq:Z2c}), 
(\ref{eq:2nd_order_tangent_sin})-(\ref{eq:X2c}), and (\ref{eq:3rd_order_1})-(\ref{eq:3rd_order_2}) (with $X_{1s}=\beta_0=\beta_{1c}=0$ \changed{as explained in \ref{sec:symmetry_reduction}}). Therefore, as noted by \citet{GB1,GB2}, there is a net loss of one $\varphi$-dependent degree of freedom, and so most axis shapes are not consistent with quasisymmetry
through this order, in contrast to the $O(r/\mathcal{R})$ case. 

Also, four new scalar parameters appear at $O((r/\mathcal{R})^2)$ that are independent of $\varphi$. One is $p_2$, providing the first information about the pressure profile. Also appearing are the numbers $B_{20}$, $B_{2s}$, and $B_{2c}$,
describing the variation of $B$ with $\vartheta$.
The global magnetic shear does not yet enter the system of equations; it first appears at $O((r/\mathcal{R})^3)$.


\section{Generating a finite-minor-radius boundary}
\label{sec:finite_r}

Our goal is ultimately to construct the shape of a boundary magnetic surface such that the magnetic field
in the interior is quasisymmetric to a good approximation. 
The interest in constructing boundary surfaces comes from the fact that magnetohydrodynamic (MHD) equilibrium codes such as VMEC
naturally take the shape of a boundary magnetic surface as an input. Conventional stellarator optimization
codes such as STELLOPT and ROSE are built upon VMEC, so if we can construct a boundary surface for a quasisymmetric
configuration, then we can construct a good initial condition for optimization.

Given a solution of the equations of the near-axis expansion through some order, how can a boundary
surface be generated? A natural approach is to plug a small but finite value $a$ of the expansion parameter $r$ into the series for
$(X,Y,Z)$, which yields a toroidal surface with average minor radius $a$. This approach was applied successfully to the $O((r/\mathcal{R})^1)$ equations in \cite{PaperII,PaperIII}.
However, this approach of setting $r \to a$ turns out to require modification when applied to the $O((r/\mathcal{R})^2)$ equations. 
It can be seen that some care is required when setting $r$ to a finite value, as follows.
When the expansion is truncated at a finite order in $r/\mathcal{R}$, a finite value of $r$ chosen, and MHD equilibrium computed within the resulting surface, a configuration results that has a slightly different axis shape and $(X,Y,Z)$ coefficients than the ones
assumed in the original expansion. The difference turns out to be unimportant for the $O((r/\mathcal{R})^1)$ construction but critical for the $O((r/\mathcal{R})^2)$ construction. The fact that plugging in a finite value of $r$ to obtain a boundary results in a slight change to the equilibrium can be seen in figures
3, 8, and 10 of \cite{PaperII}. These figures, showing the $B$ spectrum when a finite $r$ is substituted into the $O((r/\mathcal{R})^1)$ Garren-Boozer expansion and the equilibrium is computed inside the resulting boundary, show that there is a small but finite mirror mode on the magnetic axis, even though the on-axis mirror mode amplitude is precisely zero in the original expansion.

In this section, we introduce a systematic method to examine the effect of setting the expansion parameter $r$ equal to a finite number $a$. Technical details of the calculation are given in appendix \ref{sec:finite_r_appendix}.
Here and in the appendix, we consider a more general problem of trying to construct a configuration with any desired field strength
$B(r,\vartheta,\varphi)$, which may or may not be quasisymmetric; therefore the analysis also applies to more general optimizations such
as omnigenity.
The basic approach is to introduce a second Garren-Boozer-type expansion, denoted with tildes, that describes the configuration
which results from computing an MHD equilibrium inside the boundary constructed from the original Garren-Boozer expansion. 
In contrast to the original expansion, the ``tilde'' expansion is not truncated, since it describes MHD equilibrium in a finite volume.
The axis shapes and $(X,Y,Z)$ coefficients for the tilde and non-tilde expansions are similar but not identical. Their differences diminish as $a \to 0$. The non-tilde expansion represents a single idealized configuration we would like to achieve, whereas the tilde expansion represents a family of different `real' configurations, (real in the sense that they are what is computed by solving for MHD equilibrium without a near-axis expansion), parameterized by the finite value $a$ used to construct their boundaries.
From the fact that the (truncated) non-tilde expansion and (non-truncated) tilde expansions \changed{describe the same surface} at $r=a$, and exploiting an expansion in $r / \mathcal{R} \ll 1$ with the ordering $a \sim r$, we can derive the magnetic field strength that results for the constructed configurations.
We will thereby rigorously show that substituting $r \to a$ in an $O(r/\mathcal{R})$ Garren-Boozer solution yields a configuration that has the desired $B$ to $O(r/\mathcal{R})$, but substituting $r \to a$ in an $O((r/\mathcal{R})^2)$ Garren-Boozer solution yields a configuration that only
has the desired $B$ through $O(r/\mathcal{R})$, not $O((r/\mathcal{R})^2)$. However, the achieved $B$ can be made to match the desired one at $O((r/\mathcal{R})^2)$ by a small modification of the construction,
in which $X_3$ and $Y_3$ terms are included.

Beginning the formal analysis, we consider a family of equilibria parameterized by $a$, in which the position vector is
\begin{align}
\vect{r} = \tilde{\vect{r}}_0(a,\tilde\varphi)
+\tilde{X}(a,r,\tilde\vartheta,\tilde\varphi) \tilde{\vect{n}}(a,\tilde\varphi)
+\tilde{Y}(a,r,\tilde\vartheta,\tilde\varphi) \tilde{\vect{b}}(a,\tilde\varphi)
+\tilde{Z}(a,r,\tilde\vartheta,\tilde\varphi) \tilde{\vect{t}}(a,\tilde\varphi),
\end{align}
analogous to (\ref{eq:positionVector}). The Boozer angles $(\tilde\vartheta,\tilde\varphi)$ for the real configuration generally differ somewhat from the angles $(\vartheta,\varphi)$ of the ideal configuration, with the differences denoted by single-valued functions $t$ and $f$:
\begin{align}
\tilde\vartheta(a,\vartheta,\varphi) = \vartheta + t(a,\vartheta,\varphi),
\hspace{0.5in}
\tilde\varphi(a,\vartheta,\varphi) = \varphi + f(a,\vartheta,\varphi).
\end{align}
Analogous to (\ref{eq:radial_expansion}), we have
\begin{align}
\tilde{X}(a,r,\tilde\vartheta,\tilde\varphi) = \sum_{j=1}^{\infty} r^j \tilde{X}_j(a,\tilde\vartheta,\tilde\varphi),
\end{align}
with similar expansions for $\tilde{Y}$ and $\tilde{Z}$,
and analogous to (\ref{eq:radial_expansion}), the field strength in the real configurations is
\begin{align}
\tilde{B}(a,r,\tilde\vartheta,\tilde\varphi) = \sum_{j=0}^{\infty} r^j \tilde{B}_j(a,\tilde\vartheta,\tilde\varphi),
\end{align}
with a similar expansion for $\tilde{\beta}$.
All quantities in the tilde configurations are assumed to have $a$-dependence in the form of a Taylor series, with coefficients denoted by superscripts in parentheses:
\begin{align}
\tilde{\vect{r}}_0(a,\tilde\varphi) = \sum_{k=0}^{\infty} a^k \tilde{\vect{r}}_0^{(k)}(\tilde\varphi),
\end{align}
with analogous expansions for $\tilde{\vect{n}}$, $\tilde{\vect{b}}$, and $\tilde{\vect{t}}$, and
\begin{align}
\tilde{X}_j(a,\tilde\vartheta,\tilde\varphi) = \sum_{k=0}^{\infty} a^k \tilde{X}_j^{(k)}(\tilde\vartheta,\tilde\varphi),
\end{align}
with analogous expansions for $\tilde{Y}_j$, $\tilde{Z}_j$, $\tilde{B}_j$, and $\tilde{\beta}_j$. We similarly assume
\begin{align}
t(a,\vartheta,\varphi) = \sum_{k=0}^{\infty} a^k t^{(k)}(\vartheta,\varphi),
&\hspace{0.5in}
f(a,\vartheta,\varphi) = \sum_{k=0}^{\infty} a^k f^{(k)}(\vartheta,\varphi).
\end{align}
To reiterate, subscripts refer to an expansion in distance from the axis in a fixed configuration, whereas superscripts in parentheses indicate a distinct expansion in the finite value of minor radius substituted into the original near-axis expansion.

The boundary of a tilde configuration, i.e. its $r=a$ surface, by definition is the surface obtained by setting $r=a$ in the non-tilde expansion. The  equation representing this fact is
\begin{align}
\label{eq:positionVector_tilde}
& \vect{r}_0(\varphi)
+X(a,\vartheta,\varphi) \vect{n}(\varphi)
+Y(a,\vartheta,\varphi) \vect{b}(\varphi)
+Z(a,\vartheta,\varphi) \vect{t}(\varphi)
\\
&= 
\tilde{\vect{r}}_0(a,\tilde\varphi)
+\tilde{X}(a,a,\tilde\vartheta,\tilde\varphi) \tilde{\vect{n}}(a,\tilde\varphi)
+\tilde{Y}(a,a,\tilde\vartheta,\tilde\varphi) \tilde{\vect{b}}(a,\tilde\varphi)
+\tilde{Z}(a,a,\tilde\vartheta,\tilde\varphi) \tilde{\vect{t}}(a,\tilde\varphi),
\nonumber
\end{align}
and it  plays a central role in the analysis.

To the order of interest, the field strength in the real configurations is
\begin{align}
\label{eq:constructedB}
\tilde{B}(a,r,\tilde\vartheta,\tilde\varphi)
= &\tilde{B}_0^{(0)}(\tilde\varphi)
+r \tilde{B}_1^{(0)}(\tilde\vartheta,\tilde\varphi)
+a \tilde{B}_0^{(1)}(\tilde\varphi) \\
&
+r^2 \tilde{B}_2^{(0)}(\tilde\vartheta,\tilde\varphi)
+r a \tilde{B}_1^{(1)}(\tilde\vartheta,\tilde\varphi)
+a^2 \tilde{B}_0^{(2)}(\tilde\varphi)
+O((r/\mathcal{R})^3). \nonumber
\end{align}
The quantities in this expression are computed in terms of the non-tilde expansion in appendix \ref{sec:finite_r_appendix}, by systematically examining (\ref{eq:positionVector_tilde})  at each order in $a/\mathcal{R} \sim r/\mathcal{R}$. 
There, assuming only that the construction is carried out through $(X_1,Y_1,Z_1)$ or higher, it is found that 
$\tilde{B}_0^{(0)}(\varphi) = B_0(\varphi)$,
$\tilde{B}_1^{(0)}(\vartheta,\varphi) = B_1(\vartheta,\varphi)$,
and
$\tilde{B}_0^{(1)}(\varphi) = 0$.
Hence, if a finite value $a$ is substituted into $r$ in a $O((r/\mathcal{R})^1)$ Garren-Boozer solution to construct a boundary surface, the real configuration inside this boundary will have the desired magnetic field strength in Boozer coordinates through $O((r/\mathcal{R})^1)$.
This finding is consistent with the results in \cite{PaperII}.
However, the results at next order are more complicated. 
Assuming now that the construction is carried out through $(X_2,Y_2,Z_2)$ or higher,
it is found in appendix \ref{sec:finite_r_appendix} that
$\tilde{B}_2^{(0)}(\vartheta,\varphi) = B_2(\vartheta,\varphi)$,
$\tilde{B}_1^{(1)}(\vartheta,\varphi) = 0$,
and
\begin{align}
\label{eq:B02}
\tilde{B}_0^{(2)}(\varphi) = \hat{B} B_0 - f^{(2)}B'_0 
\end{align}
where
\begin{align}
\label{eq:BHat}
\hat{B}(\varphi) =& \left(X_{3s1} Y_{1c} - X_{3c1}Y_{1s} + X_{1s} Y_{3c1} - X_{1c} Y_{3s1} - Q \right) \frac{s_G B_0}{\bar{B}},
\\
\label{eq:Q}
Q(\varphi) =& \frac{(G_2 + I_2 N)\bar{B} \ell'}{2G_0^2} 
  +2(X_{2c} Y_{2s} - X_{2s} Y_{2c}) 
 + \frac{\bar{B}}{2G_0} \left( \ell'  X_{20} \kappa - Z'_{20}\right) 
 \\
& +\frac{I_2}{4G_0} \left(  -\ell'   \tau V_1 + Y_{1c} X'_{1c} - X_{1c}Y'_{1c} +Y_{1s} X'_{1s} -X_{1s}Y'_{1s}\right)
 \nonumber \\
& +\frac{\beta_0 \bar{B}}{4G_0} \left(   X_{1s} Y'_{1c} + Y_{1c} X'_{1s} - X_{1c} Y'_{1s} - Y_{1s} X'_{1c}\right), \nonumber
\end{align}
primes denote $d/d\varphi$, and 
\begin{align}
f^{(2)}(\varphi) = 
\left( \int_0^{\varphi} d\bar\varphi \, \hat{B}(\bar\varphi) \right)
+\left( \frac{1}{2} - \frac{\varphi}{2\pi}\right) \left( \int_0^{2\pi} d\bar\varphi \, \hat{B}(\bar\varphi) \right)
-\frac{1}{2\pi} \int_0^{2\pi} d\bar{\bar\varphi}  \int_0^{\bar{\bar\varphi}} d\bar\varphi \, \hat{B}(\bar\varphi) .
\label{eq:p2}
\end{align}
Thus, if the Garren-Boozer solution is carried out through $(X_2,Y_2,Z_2)$ but then truncated so $(X_3,Y_3,Z_3)$ are all set to zero when constructing the boundary, (\ref{eq:BHat}) will be nonzero. The $a^2 \tilde{B}_0^{(2)}$ term in (\ref{eq:constructedB}) will then be nonzero
and will cause a  difference between the desired and achieved field strength that is comparable to the desired $r^2 B_2$ term. Generally (\ref{eq:B02}) will depend on
$\varphi$ and so it will break quasisymmetry.

Fortunately, a workaround can be achieved that does not require a full solution
of the  Garren-Boozer equations for $(X_3,Y_3,Z_3)$. It can be preferable to avoid a full solution through $O((r/\mathcal{R})^3)$ 
 because the equations are extremely complicated, because
one then has to choose additional parameters for the construction,
and because the presence of squareness that grows with $r$ at this order can limit the minimum aspect ratio.
In the workaround, we take  $(X_3,Y_3)$ to be  
$(X_1,Y_1)$ scaled by some function $\lambda(\varphi)$:
\begin{align}
\label{eq:lambda_definition}
X_{3c1} = \lambda X_{1c},
\hspace{0.2in}
X_{3s1} = \lambda X_{1s},
\hspace{0.2in}
X_{3c3} = X_{3sc} =0,
\end{align}
with analogous expressions for $Y$, and $Z_3=0$.
 In other words, we introduce a small correction to the $O(r/\mathcal{R})$ elliptical flux surface shape. Setting (\ref{eq:BHat})$=0$, substituting (\ref{eq:lambda_definition}), and using (\ref{eq:flux_area}),
we find 
 \begin{align}
 \label{eq:lambda_solution}
 \lambda(\varphi) = - Q B_0 / (2 s_G \bar{B}).
 \end{align}
 Adding these $(X_3,Y_3)$ terms to the constructed boundary surface therefore results in $\tilde{B}_0^{(2)}=0$, so the real configurations have the same Boozer spectrum as the ideal target configuration through $O((r/\mathcal{R})^2)$.

Some physical intuition for this result can be given. The leading-order field strength $B_0$ is approximately the toroidal flux divided by the cross-sectional area of the flux surfaces. Indeed, this interpretation of (\ref{eq:flux_area}) is shown precisely in \cite{PaperI}. The area of the surfaces is primarily determined by the $\sin\vartheta$ and $\cos\vartheta$ modes of $X$ and $Y$, which generate ellipses, and not by $\sin 2\vartheta$, $\cos 2\vartheta$, and independent-of-$\vartheta$ modes which distort and shift the ellipses but do not expand or contract them. The $\sin\vartheta$ and $\cos\vartheta$ modes of $X$ and $Y$ that affect the cross-sectional area arise at orders $O(r/\mathcal{R})$ and $O((r/\mathcal{R})^3)$ but not at $O((r/\mathcal{R})^2)$, due to analyticity. Thus, if the Garren-Boozer solution is truncated by setting $X_3=Y_3=Z_3=0$, there is an $O((r/\mathcal{R})^2)$ error in the cross-sectional area of the surfaces, which (since $B \sim \mathrm{flux} / \mathrm{area}$) implies an $O((r/\mathcal{R})^2)$ error in  $B_0$. This error can vary toroidally, spoiling quasisymmetry or whatever other pattern of $B$ is desired. To solve the problem, we note (\ref{eq:XY3}) is a correction to (\ref{eq:flux_area}) (they both derive from (\ref{eq:XY})), relating the cross-sectional area and $B$ to flux. Thus, (\ref{eq:XY3}) indicates how much the surfaces should be expanded or contracted to give the correct $B_0$ through $O((r/\mathcal{R})^2)$.
Indeed, the same result (\ref{eq:lambda_solution}) can be obtained by
setting $2\pi\psi = \int d^2\vect{a} \cdot \vect{B}$ at $O((r/\mathcal{R})^2 r^2 B)$, where $\int d^2\vect{a}$ is an integral over a constant-$\varphi$ cross-section, using (\ref{eq:Boozer_h}) and (\ref{eq:lambda_definition}).

Note that by a modified choice of $\lambda$, $\tilde{B}_{0}^{(2)}$ can be made to cancel the toroidal dependence of $\tilde{B}_{20}^{(0)}$ at a nonzero value $r_0$ of $r$. In the case of quasisymmetry, such a choice has the effect of
introducing an $O((r/\mathcal{R})^2)$ mirror mode on the axis, with the mirror mode amplitude vanishing at radius $r_0$. There may be advantages in this approach, for as found in a recent numerical study \citep{Henneberg}, fast particle confinement in a quasi-axisymmetric configuration was best when the quasisymmetry was optimized off-axis rather than on-axis. 


\section{Numerical formulation}
\label{sec:numerical}

\subsection{Inputs and outputs}

We now describe a practical numerical implementation of the equations derived in the preceding sections and associated appendices.
The \changed{parameters of the algorithm} here are a superset of the inputs to the $O((r/\mathcal{R})^1)$ construction detailed in \cite{PaperII}.
The latter are the shape of the magnetic axis, which must have nonvanishing curvature everywhere, and the numbers $\bar{\eta}$, $I_2$, and $\sigma(0)$.
The parameter $\bar{\eta}$ effectively controls the elongation; $I_2$ indicates the on-axis toroidal current and is typically zero for stellarators; $\sigma(0)$ controls the angle of elongation at $\varphi=0$ and is zero for stellarator-symmetric configurations.
In the $O((r/\mathcal{R})^2)$ construction, three new constant input parameters are needed: $p_2$, $B_{2c}$, and $B_{2s}$. We will not take $B_{20}$ as an input parameter for reasons explained shortly. The parameter $p_2$ defines the pressure profile to this order. The parameters $B_{2c}$ and $B_{2s}$ set the desired $\cos 2\vartheta$ and $\sin 2\vartheta$ modes in the field strength. These two parameters have the effect of controlling the stellarator-symmetric and stellarator-asymmetric parts of the triangularity. For stellarator-symmetric configurations, $B_{2s}=0$.

The outputs of the calculation include the shapes of the magnetic surfaces and the
rotational transform on axis, $\iota_0$. As noted by  \citet{GB1,GB2} and discussed above, the number of scalar $\varphi$-dependent unknowns exceeds the number of $\varphi$-dependent equations by only one if quasisymmetry is imposed through $O((r/\mathcal{R})^2)$, whereas an axis shape represents two $\varphi$-dependent quantities (e.g. $\kappa$ and $\tau$, or $R(\phi)$ and $z(\phi)$, where $(R,\phi,z)$ are cylindrical coordinates.) Therefore, to achieve quasisymmetry through $O((r/\mathcal{R})^2)$, one needs to solve for part of the axis shape.
We proceed by temporarily relaxing the requirement that $B_{20}$ must be independent of $\varphi$. By reducing the number of equations by one in this way, any axis shape becomes allowed. One can still make $B_{2s}$ and $B_{2c}$ independent of $\varphi$, achieving quasisymmetry partially through $O((r/\mathcal{R})^2)$. Then $B_{20}(\varphi)$ is an output of the calculation, and it generally has some toroidal variation. We can then numerically optimize the input parameters (including not only the axis shape but also $\left\{\bar{\eta},\;\sigma(0),\; B_{2c},\;B_{2s}\right\}$) such that the toroidal variation of $B_{20}$ is minimal.
\changed{While we have not proved rigorously that solutions exist in which $B_{20}$ is exactly constant, experience with the numerical solutions described in the next section suggests $B_{20}'$ can be made arbitrarily small as the number of degrees of freedom in the axis shape is increased.}

As explained in detail in section 5.2 of \cite{PaperI}, quasi-axisymmetry versus quasi-helical symmetry is determined by the choice of axis shape. In particular, the integer $N$ is the number of times the axis normal vector rotates poloidally around the axis as the axis is traversed toroidally.


\subsection{Numerical solution of the equations}

Given the input parameters described in the previous subsection,
the $O(r/\mathcal{R})$ equations are solved as described in section 3 of \cite{PaperII}.
As a result, $X_1$ and $Y_1$ are computed on a uniform grid in the standard toroidal angle $\phi$ covering one field period with $N_\phi$ points, and $\iota_0$ is obtained.
Then $Z_2$ is computed from (\ref{eq:Z20})-(\ref{eq:Z2c}),
$X_{2s}$ is computed from (\ref{eq:X2s}), and 
$X_{2c}$ is computed from (\ref{eq:X2c}).

Next, a $(2N_\phi) \times (2N_\phi)$ linear system is solved.
The unknowns for this system are the values of $X_{20}$ and $Y_{20}$ on the $N_\phi$ grid points. The rows of the linear system represent (\ref{eq:3rd_order_1}) and 
 (\ref{eq:3rd_order_2}) imposed at the $N_\phi$ grid points.
In these equations, $d/d\varphi$ derivatives are discretized using the same
pseudospectral differentiation matrix described in \cite{PaperII}.
In this system, $Y_{2s}$ and $Y_{2c}$ are eliminated using 
(\ref{eq:2nd_order_tangent_sin})-(\ref{eq:2nd_order_tangent_cos}).
The dense linear system is solved with direct factorization (LAPACK).
With $X_{20}$ and $Y_{20}$ thereby determined, $Y_{2s}$ and $Y_{2c}$
are computed from 
(\ref{eq:2nd_order_tangent_sin})-(\ref{eq:2nd_order_tangent_cos}),
and then (\ref{eq:X20}) gives $B_{20}$.
Finally, (\ref{eq:lambda_definition})-(\ref{eq:lambda_solution}) give $X_3$ and $Y_3$.

Note that although the $O((r/\mathcal{R})^1)$ equations are nonlinear in the unknowns,
a unique solution always exists, as proved in \cite{PaperII}, so the solution by Newton's method is extremely robust. Furthermore, once the $O((r/\mathcal{R})^1)$ solution is determined, the equations of appendix \ref{sec:equations_through_r2} are linear in the higher-order unknowns, so the $O((r/\mathcal{R})^2)$ construction is equally robust.
At a typical resolution ($N_\phi \sim 30)$, solution of the equations for the $O((r/\mathcal{R})^2)$ construction takes $<$ 2 ms on one core of a modern laptop, many orders of magnitude lower computational cost than a general 3D equilibrium calculation used in each iteration of traditional stellarator optimization.


\subsection{Optimization of input parameters}
\label{sec:optimization}

For many sets of input parameters, the model results in configurations
that are not of practical interest because they are limited to 
extremely high aspect ratio. This limitation arises because for any solution
of the model equations, beyond a certain value of $r$, the constant-$r$ surfaces will begin to self-intersect.
If $X_2$, $Y_2$, $X_3$, or $Y_3$ are large, this critical $r$ will be small.
From another perspective, the near-axis expansion is only accurate at values of
$r$ sufficiently small that the terms of successive orders in $r$ in the expansion are decaying.
If $X_2$, $Y_2$, $X_3$, or $Y_3$ are large, the accuracy of the expansion is then limited to smaller values of $r$.

Therefore, for some of the examples below, we use optimization
- either numerical or by hand - to find solutions with small $X_2$, $Y_2$, $X_3$, or $Y_3$. We also minimize the toroidal derivatives  of these quantities, anticipating
that large derivatives would drive large symmetry-breaking terms at next order.
All the quantities targeted for minimization are squared, averaged over $\phi$, and combined in a weighted sum to form a single objective function.
For some examples, to improve quasisymmetry, we also include in the sum a term minimizing the toroidal variation of $B_{20}$.
While doing these optimizations, it may be necessary to penalize parameters
for which $\iota_0$ becomes too small.
Note that when optimization is applied to the $O((r/\mathcal{R})^2)$ quasisymmetry model,
the objective function can be evaluated in milliseconds, roughly four orders
of magnitude or more faster than the objective function evaluations
in conventional stellarator optimization.
Also, in principle analytic derivatives are available for the $O((r/\mathcal{R})^2)$ model,
although we will not exploit them here to accelerate optimization.
For the examples below, we use Matlab's derivative-free algorithm `fminsearch',
 a variant of the Nelder-Mead simplex algorithm by \citet{Lagarias}.


\subsection{Conversion to cylindrical coordinates}

A principal aim of the construction is to generate boundary shapes that
can be provided as input to an MHD equilibrium code such as VMEC.
VMEC requires as input the boundary surface shape defined by its cylindrical coordinates $(R,z)$ expressed as a double Fourier series in the toroidal angle $\phi$ and any poloidal angle.
As discussed in section 4 of \cite{PaperII}, there are several ways this boundary
description can be obtained from our representation (\ref{eq:positionVector}).
One approach is to derive the transformation between the two representations order by order in $a$. This approach was developed in section 4.1 of \cite{PaperI} to $O(r/\mathcal{R})$. In appendix \ref{sec:transformation}, the transformation is extended to $O((r/\mathcal{R})^3)$.
The advantage of this approach is that it results in explicit expressions for $R(\theta,\phi)$ and $z(\theta,\phi)$ that can be evaluated extremely rapidly.

A second approach to transforming from the Frenet representation to the representation required by VMEC is described in section 4.2 of \cite{PaperII}. In this approach, nonlinear root-finding is applied to (\ref{eq:positionVector}). At a grid of points in $\theta$ and $\phi$, one solves for the value of $\varphi$ such that the position vector has a standard toroidal angle $\phi$. The nonlinear root-finding requires additional computation time. This approach tends to result in slightly lower magnitude of symmetry-breaking, so we use it for the numerical examples that follow.


\section{Numerical results}
\label{sec:results}

Several examples of constructed quasi-axisymmetric and quasi-helically symmetric configurations will now be presented. 
The examples are all generated ``from scratch'', in that no fitting was done to 
previously optimized equilibria.
All the examples are scaled such that
the zero-frequency component of the axis major radius $R_{00}=(2\pi)^{-1}\int_0^{2\pi} d\phi\,R_0$ is 
1 meter, and the on-axis field strength $B_0$ is 1 Tesla.
In each VMEC calculation shown, the pressure profile specified was $p(r) = (1-r^2/a^2) p_2$ for the same constant $p_2$ used in the construction. Also, the current profile for VMEC calculations was specified as $\mathcal{I}(s)=2\pi s a^2 I_2/\mu_0$, where $\mathcal{I}(s)$ is the toroidal current inside the surface with normalized toroidal flux $s=(r/a)^2$, and $I_2$ is the constant used in the construction.

We will describe the configurations using two different measures of effective aspect ratio. The measure that is most convenient for the construction is $A=R_{00}/a$, where 
again $\pi a^2$ is the toroidal flux. We will also quote the effective aspect ratio $A_{vmec}$ used in the VMEC code, since this measure is often reported in the literature. Its definition is $A_{vmec} = (\mathtt{Aminor\_p})/(\mathtt{Rmajor\_p})$, where
the effective minor radius $\mathtt{Aminor\_p}$ is defined by $\pi (\mathtt{Aminor\_p})^2=\bar{S}$, with $\bar{S} = (2\pi)^{-1} \int_0^{2\pi}d\phi\, S(\phi)$ the toroidal average of the
area $S(\phi)$ of the outer surface's cross-section in the $R$-$z$ plane, and the effective major radius $\mathtt{Rmajor\_p}$ is defined by $[2 \pi (\mathtt{Rmajor\_p})][(\pi (\mathtt{Aminor\_p})^2]=V$ with $V$ the volume of the outer surface.

\subsection{Quasi-axisymmetry partially through $O((r/\mathcal{R})^2)$}
\label{sec:partiallyOptimized}

\begin{figure}
  \centering
\includegraphics[width=2.5in]{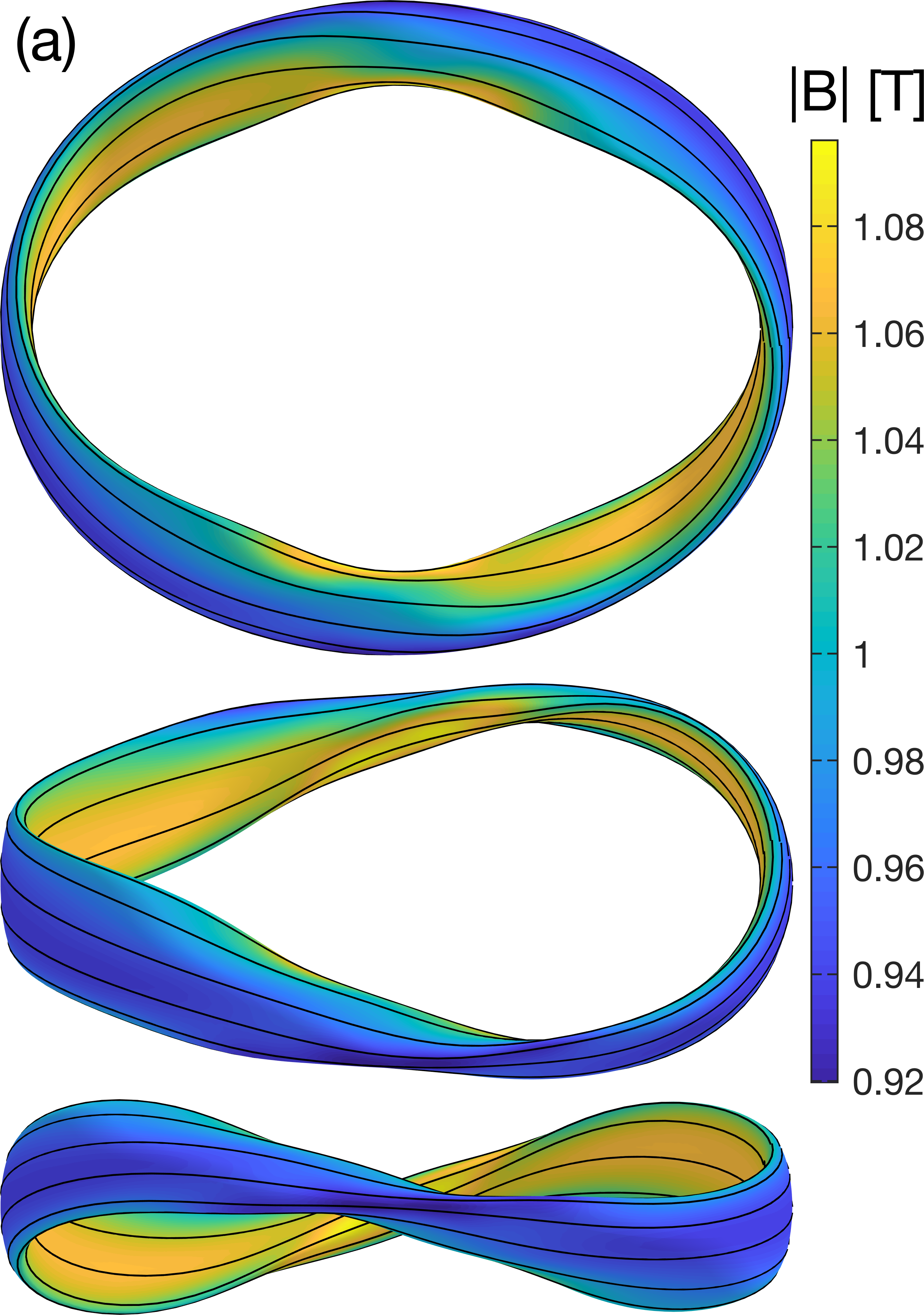}
\raisebox{1.3in}{\includegraphics[width=2.5in]{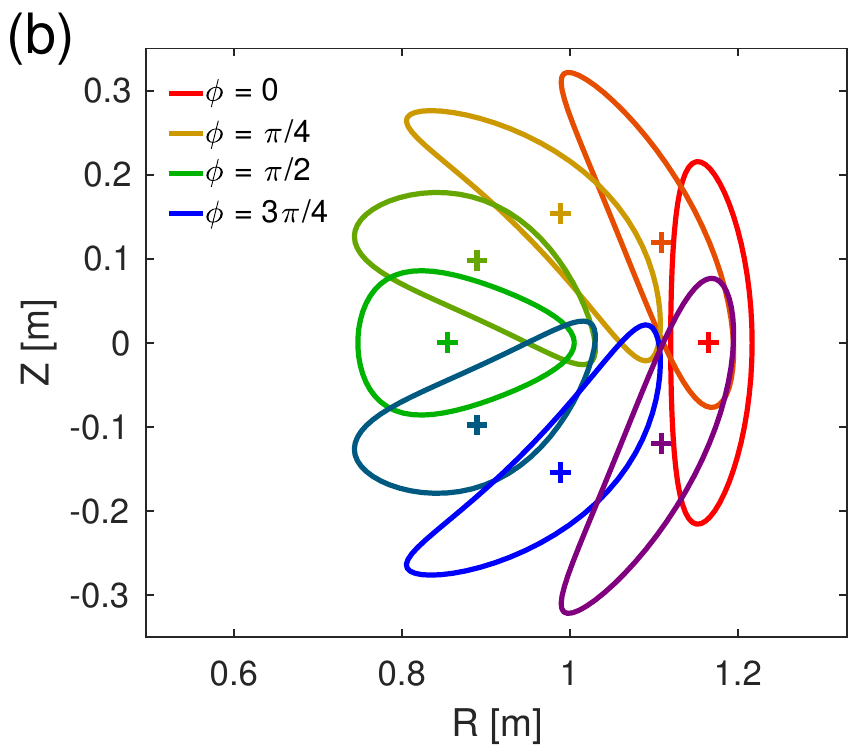}}
  \caption{
The partially-quasi-axisymmetric example of section \ref{sec:partiallyOptimized}, 
for aspect ratio $A=10$, $A_{vmec}=9.75$.
The 3D surface shape in (a), shown from three angles, and the cross-sections in (b), are generated by the construction. In (a), magnetic field lines are shown as black lines, and color indicates the field strength computed by VMEC.}
\label{fig:partialQA_example}
\end{figure}

\begin{figure}
  \centering
  \includegraphics[width=3in]{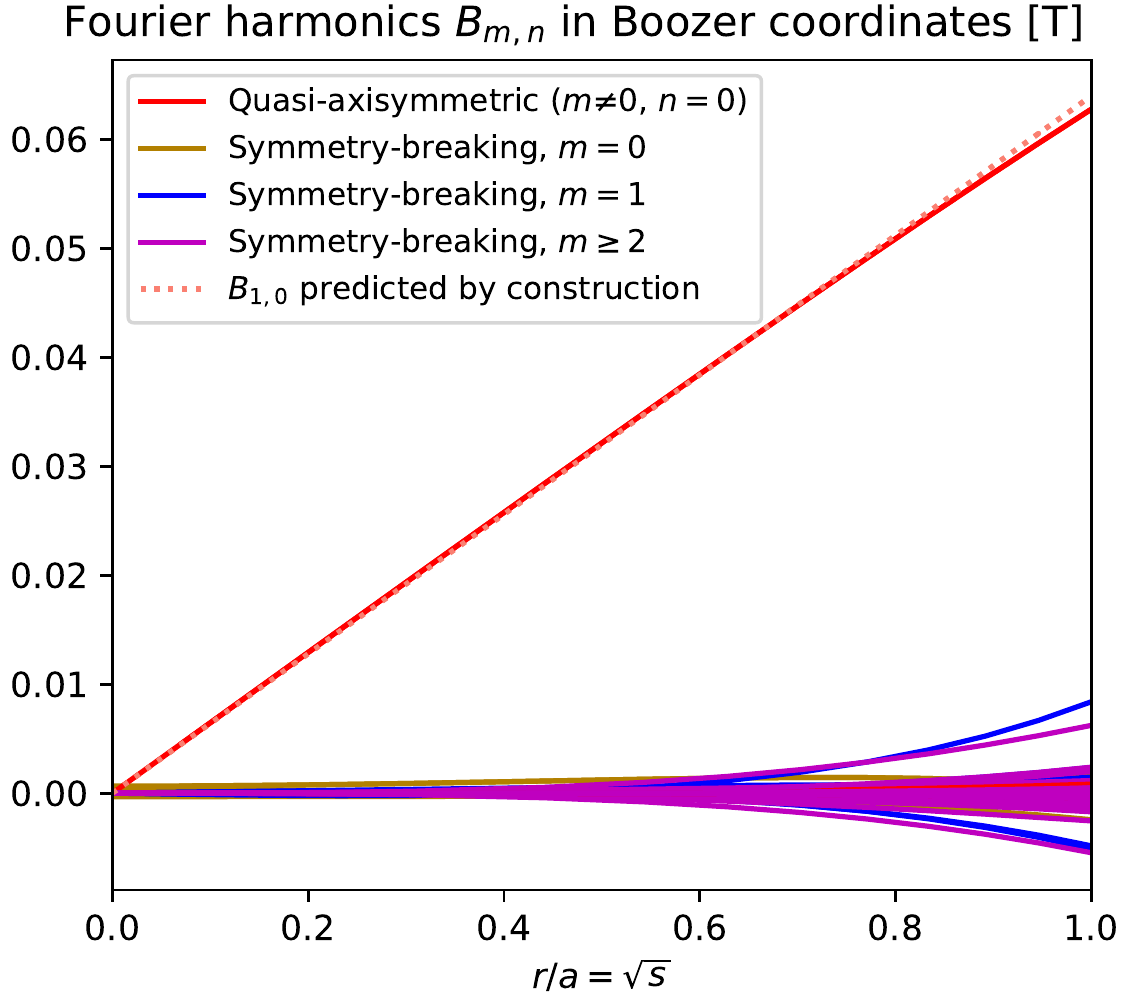}
  \caption{
The spectrum of $B$ for the partially-quasi-axisymmetric example of section \ref{sec:partiallyOptimized}, computed by running the VMEC and BOOZ\_XFORM codes inside the constructed boundary surface for aspect ratio $A=10$, $A_{vmec}=9.75$.}
\label{fig:partialQA_spectrum}
\end{figure}

The first set of input parameters we consider includes the axis shape
\begin{align}
R_0(\phi) \;\mathrm{[m]}=& 1 + 0.155 \cos(2\phi) + 0.0102 \cos(4\phi) , \\
z_0(\phi) \;\mathrm{[m]} = & \hspace{0.25in}0.154 \sin(2\phi) + 0.0111 \sin(4\phi) ,\nonumber
\end{align}
corresponding to two field periods.
We also choose $\etabar=0.640$ m$^{-1}$ and $B_{2c}=-0.00322$ T$/$m$^2$.
These values and axis shape were obtained by 
minimizing $X_2$, $Y_2$, $X_3$, and $Y_3$ as described in section \ref{sec:optimization}, subject to a lower bound $\iota_0 \ge 0.42$.
The parameters $\sigma(0)$ and $B_{2s}$ were set to zero so the configuration is stellarator-symmetric. We also choose $I_2=0$ and $p_2=0$ so the configuration is a vacuum field.
For this first configuration, no attempt was made to make $B_{20}$ independent of $\varphi$,
so the configuration is only partially quasisymmetric at $O((r/\mathcal{R})^2)$.
The resulting configuration has $\iota_0=0.420$ as desired, and
the boundary shape for $A=10$ is shown in figure \ref{fig:partialQA_example}.  VMEC is then run to compute the equilibrium inside the finite-thickness boundary without making any near-axis approximation, and then the BOOZ\_XFORM code \citep{Sanchez} is run to transform the VMEC result to Boozer coordinates. 
Figure \ref{fig:partialQA_spectrum} shows the resulting Fourier coefficients $B_{m,n}(r)$ defined by $B(r,\theta,\varphi) = \sum_{m,n} B_{m,n}(r)\cos(m\theta-n\varphi)$.
It can be seen that the dominant $B_{m,n}$ mode is the quasi-axisymmetric term $B_{1,0}$ as desired. The  magnitude of this mode predicted by the construction, $B_{1,0} = r \etabar B_0$, is displayed for comparison, and it is nearly identical to the VMEC result. 

\begin{figure}
  \centering
  \includegraphics[width=3in]{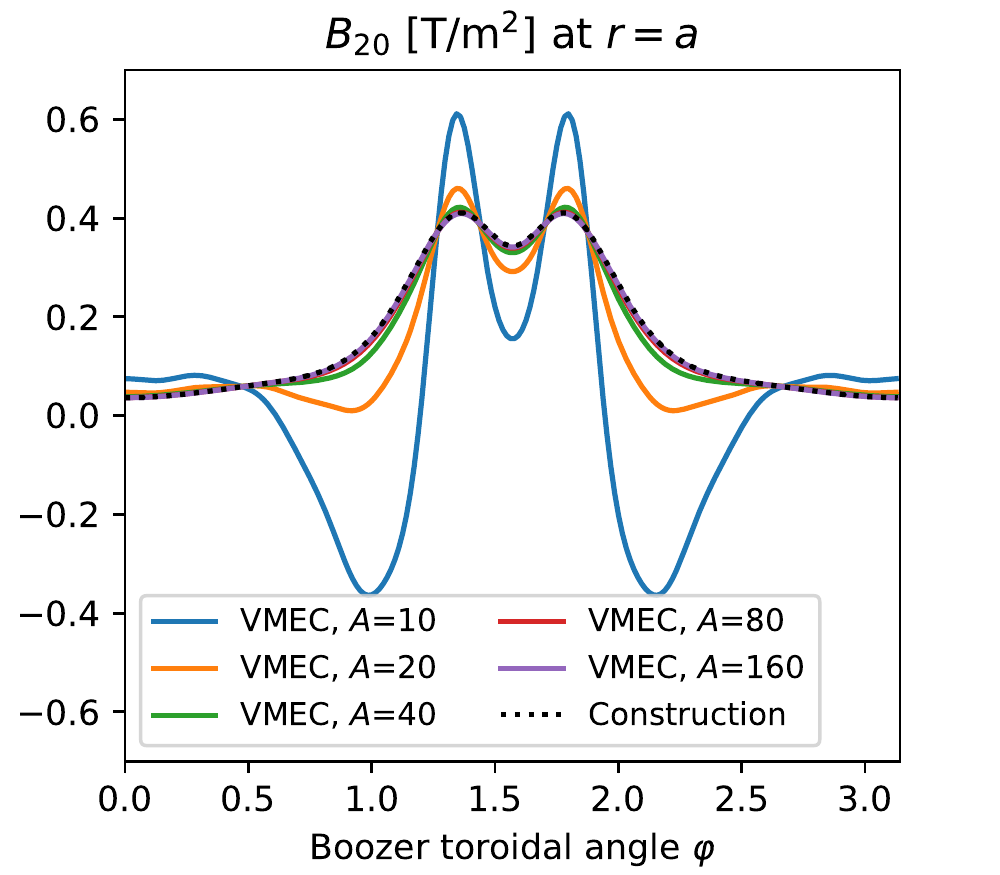}
  \caption{
As the aspect ratio $A$ increases, the $B_{20}(\varphi)$ component of the field strength of the numerical VMEC configurations converges to the function predicted by the Garren-Boozer construction. Data here are for the partially quasi-axisymmetric configuration of section \ref{sec:partiallyOptimized}.  }
\label{fig:partialQA_B20Convergence}
\end{figure}

Keeping all input parameters fixed except for the boundary aspect ratio, we then construct boundary surfaces at a sequence of increasing aspect ratios, and repeat the VMEC and BOOZ\_XFORM calculations for each case.
In figure \ref{fig:partialQA_B20Convergence}, the quantity $[B_{m=0}(\varphi,r=a) - B(\varphi,r=0)]/a^2$ is displayed
for this sequence of numerical calculations. According to the construction,
this quantity should be $B_{20}(\varphi)$. Indeed, it can be seen that 
the VMEC/BOOZ\_XFORM results converge to the Garren-Boozer prediction.
Thus, in the limit $A \gg 1$, the full 3D equilibrium calculations
achieve the expected field strength in Boozer coordinates.

\begin{figure}
  \centering
  \includegraphics[width=3in]{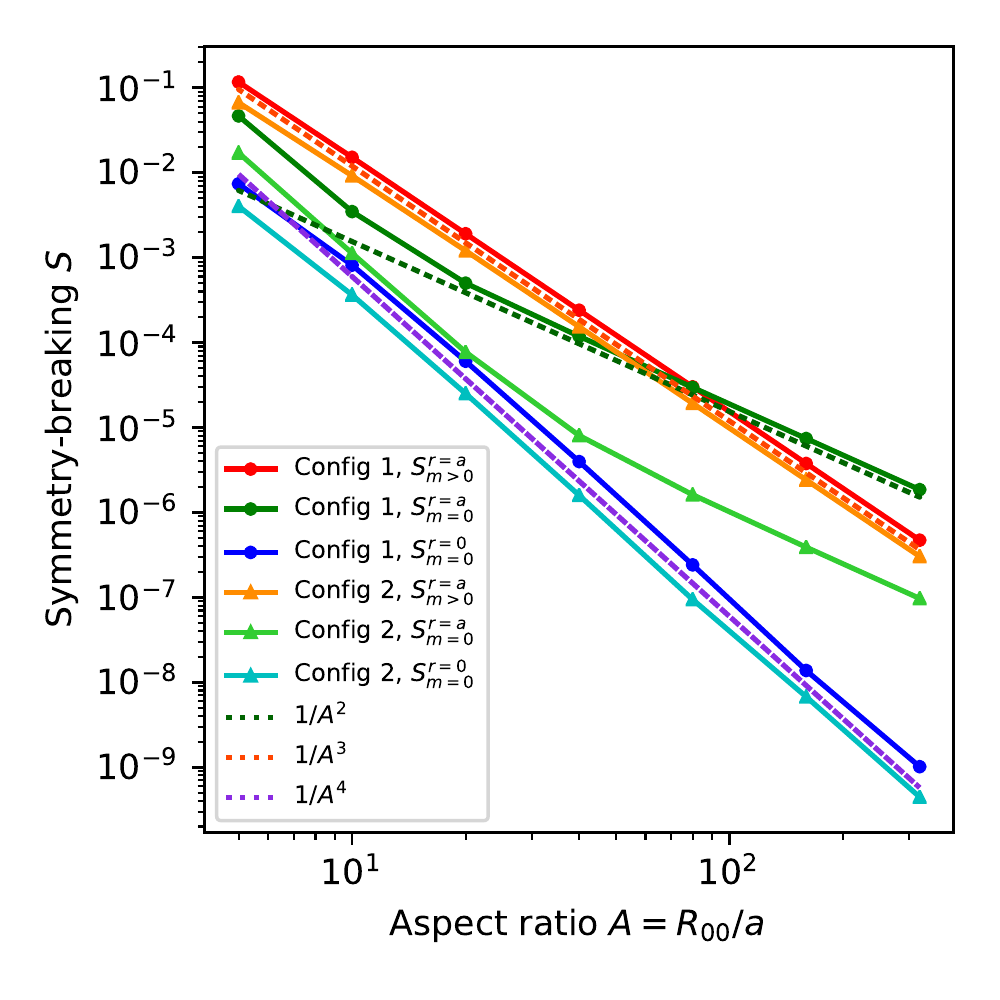}
  \caption{
 The measures of quasisymmetry-breaking (\ref{eq:S_def}), computed by running the VMEC and BOOZ\_XFORM codes inside the constructed boundary surfaces, scale as the expected power of aspect ratio. Data here are for the partially quasi-axisymmetric and optimized quasi-axisymmetric examples
 of sections \ref{sec:partiallyOptimized} (`Config 1') and \ref{sec:QA} (`Config 2').
   }
\label{fig:scaling}
\end{figure}

As another verification of the construction, figure \ref{fig:scaling} displays three symmetry-breaking measures
\begin{align}
\label{eq:S_def}
&S_{m>0}^{r=a} = \frac{1}{B_{0,0}}\sqrt{\sum_{m>0,n\ne Nm}B_{m,n}^2(r=a)},
\hspace{0.3in}
S_{m=0}^{r=a} = \frac{1}{B_{0,0}}\sqrt{\sum_{n \ne 0}B_{0,n}^2(r=a)},
\\
&\hspace{1in}
S_{m=0}^{r=0} = \frac{1}{B_{0,0}} \sqrt{\sum_{n \ne 0}B_{0,n}^2(r=0)},
\nonumber
\end{align}
computed from the VMEC and BOOZ\_XFORM results for the aspect ratio scan. 
(`Config 1' in the figure refers to the present section, while `Config 2' will be described in the next section, \ref{sec:QA}.) 
It can be seen that $S_{m>0}^{r=a}$ scales as $1/A^3$, consistent 
with the fact that the corresponding modes were constructed to be zero through $O((r/\mathcal{R})^2)$, so symmetry-breaking generally arises at next order. The on-axis mirror modes, measured by $S_{m=0}^{r=0}$, are found to have an even stronger scaling, $\propto 1/A^4$. This scaling arises because the $m=0$ modes were constructed to be zero through $O((r/\mathcal{R})^2)$,
and they are automatically zero at $O((r/\mathcal{R})^3)$ (only $m=1$ and $m=3$ modes exist at this order), so the first nonvanishing contribution occurs at $O((r/\mathcal{R})^4)$.
Finally, $S_{m=0}^{r=a}$ shows an asymptotic scaling $\propto 1/A^2$, associated with the fact that $B_{20}$ is not independent of $\varphi$.  Thus, all three symmetry-breaking measures scale as predicted by the construction.

\subsection{Quasi-axisymmetry fully through $O((r/\mathcal{R})^2)$}
\label{sec:QA}

\begin{figure}
  \centering
  \includegraphics[width=2.5in]{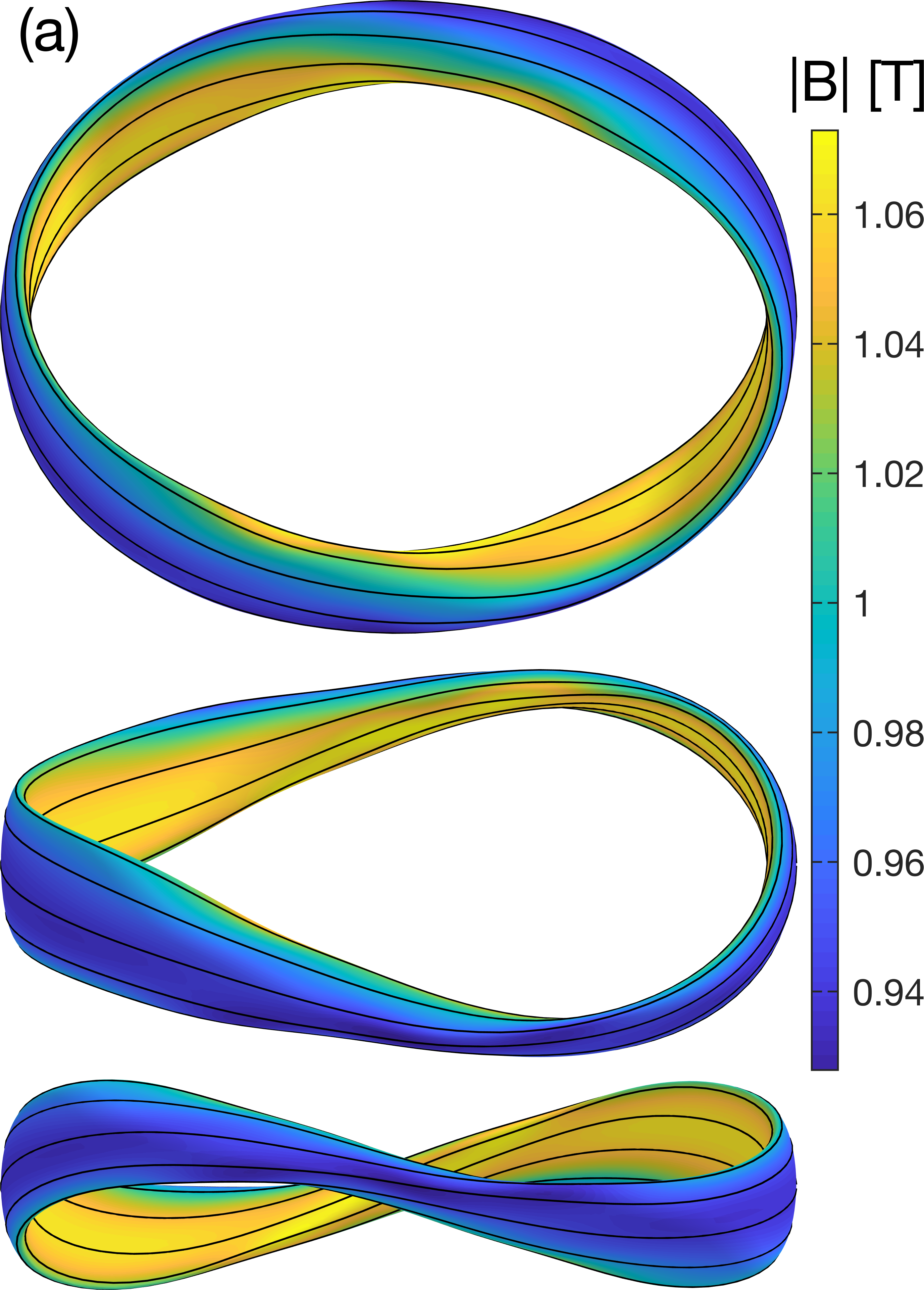}
\raisebox{1.3in}{  \includegraphics[width=2.5in]{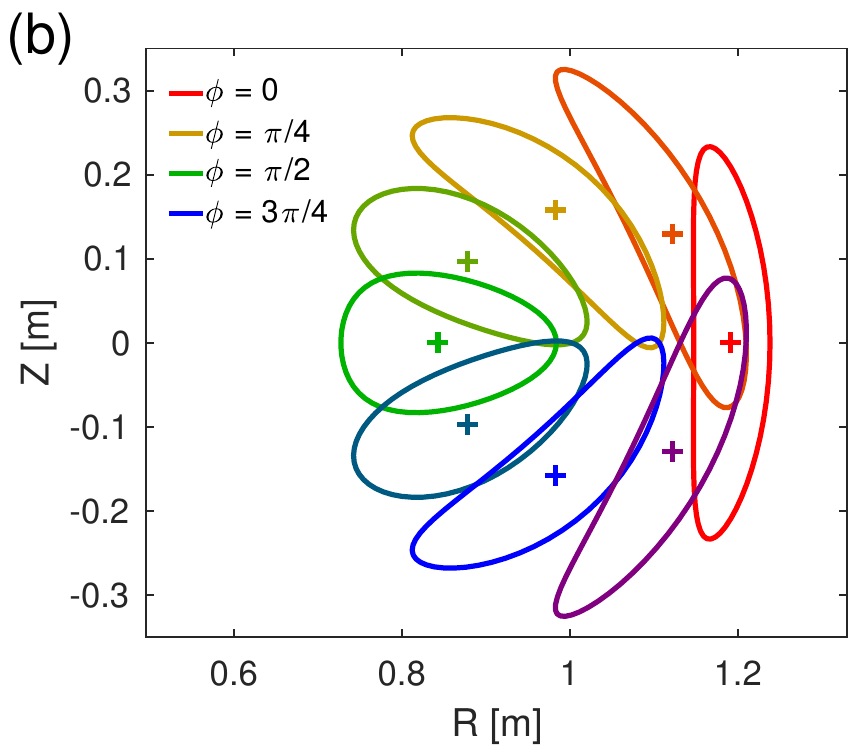}}
  \caption{
The quasi-axisymmetric example of section \ref{sec:QA}, 
for aspect ratio $A=10$, $A_{vmec}=9.71$.
The 3D surface shape in (a), shown from three angles, and the cross-sections in (b), are generated by the construction. In (a), magnetic field lines are shown as black lines, and color indicates the field strength computed by VMEC.}
\label{fig:QA_example}
\end{figure}

\begin{figure}
  \centering
  \includegraphics[width=3in]{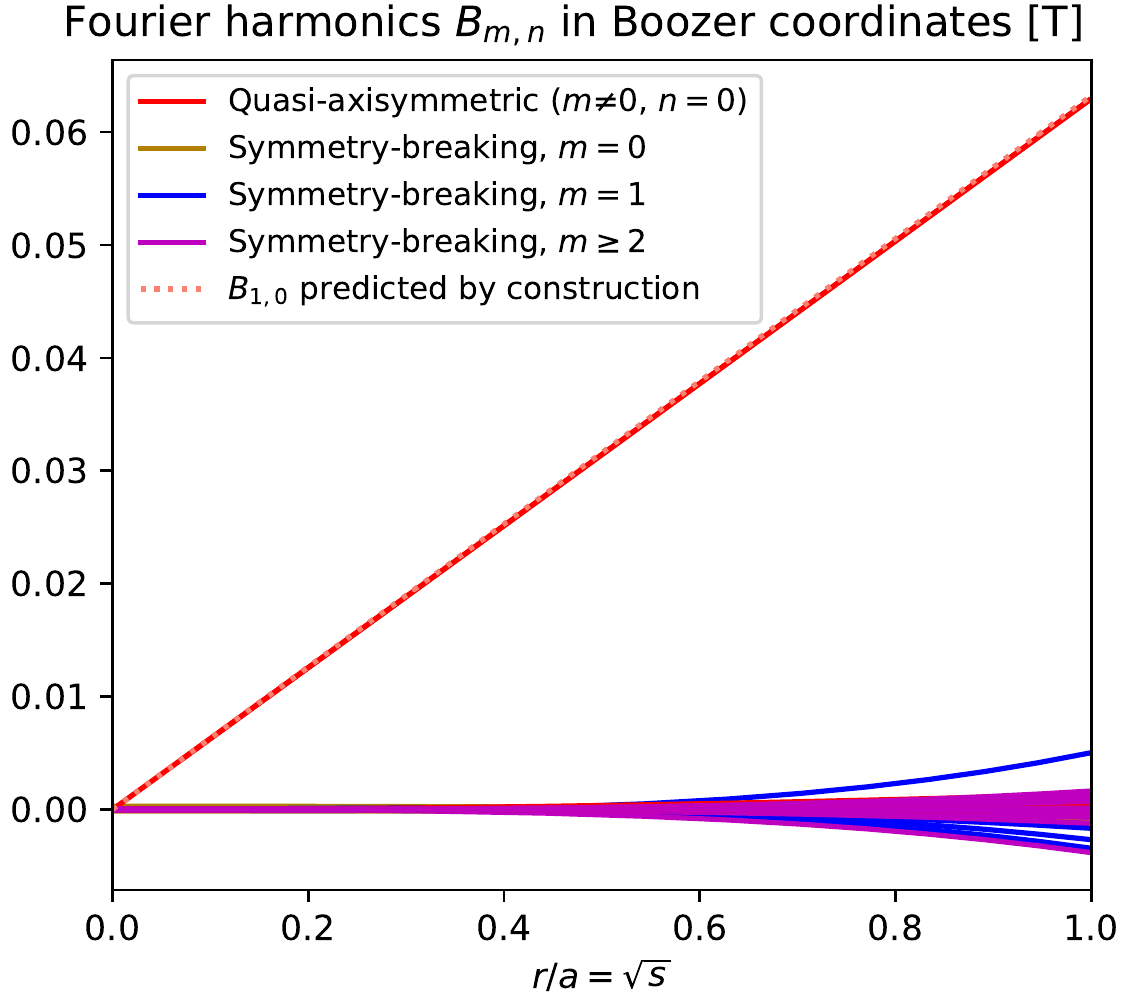}
  \caption{
The spectrum of $B$ for the quasi-axisymmetric example of section \ref{sec:QA}, computed by running the VMEC and BOOZ\_XFORM codes inside the constructed boundary surface for aspect ratio $A=10$, $A_{vmec}=9.71$}
\label{fig:QA_spectrum}
\end{figure}

\begin{figure}
  \centering
  \includegraphics[width=3in]{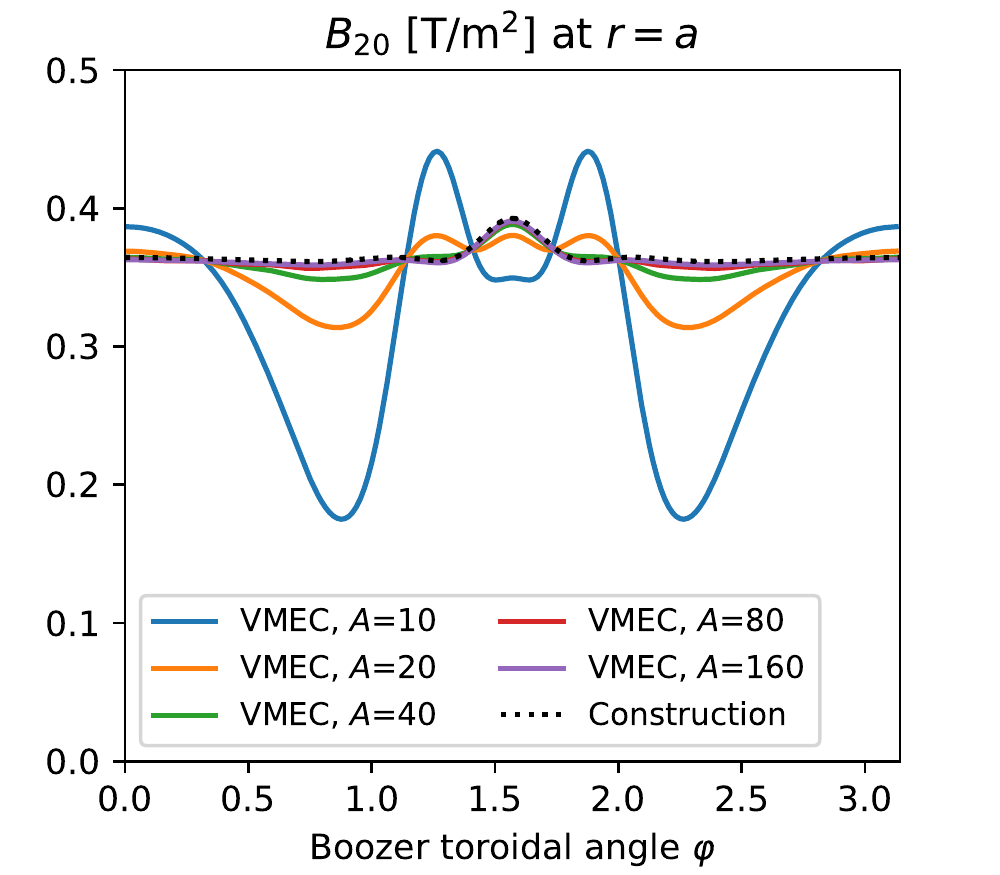}
  \caption{As the aspect ratio $A$ increases, the $B_{20}(\varphi)$ component of the field strength of the numerical VMEC configurations converges to the function predicted by the Garren-Boozer construction. Data here are for the quasi-axisymmetric configuration of section \ref{sec:QA}.
  }
\label{fig:QA_B20Convergence}
\end{figure}

We next consider a configuration that is similar to the one of section \ref{sec:partiallyOptimized}, but with
slightly different parameters such that $B_{20}$ has significantly reduced toroidal variation,
resulting in improved quasi-axisymmetry.
We again consider a two-field-period device, with axis shape
\begin{align}
R_0(\phi)  \;\mathrm{[m]}=& 1 + 0.173 \cos(2\phi) + 0.0168 \cos(4\phi) + 0.00101 \cos(6\phi), \\
z_0(\phi)  \;\mathrm{[m]}= & \hspace{0.25in}0.159 \sin(2\phi) + 0.0165 \sin(4\phi) + 0.000985 \sin(6\phi).\nonumber
\end{align}
The other nonzero input parameters are
$\etabar=0.632$ m$^{-1}$ and $B_{2c}=-0.158$ T$/$m$^2$.
These values and axis shape were obtained by the optimization
procedure of section \ref{sec:optimization},
again minimizing $X_2$, $Y_2$, $X_3$, and $Y_3$, but now also minimizing the toroidal variation of $B_{20}$.
The parameters $\sigma(0)$ and $B_{2s}$ were again set to zero so the configuration is stellarator-symmetric, and $I_2$ and $p_2$
were set to zero so the configuration is a vacuum field.
The resulting configuration has $\iota_0=0.424$.
The function $B_{20}(\varphi)$ for these parameters is shown as the black dotted curve in figure \ref{fig:QA_B20Convergence},
and it can be seen that the toroidal variation is greatly reduced compared to figure \ref{fig:partialQA_B20Convergence}.
The small remaining toroidal variation of $B_{20}$ could presumably be further reduced if additional Fourier modes were included in the axis shape.
The constructed boundary shape for $A=10$ is shown in figure \ref{fig:QA_example}, and it is only slightly different from the previous example (figure \ref{fig:partialQA_example}.)
Running VMEC and BOOZ\_XFORM inside this boundary results in the magnetic spectrum of figure \ref{fig:QA_spectrum}. Again, 
the desired mode $B_{1,0}$ dominates, and its magnitude is nearly identical to the prediction.
Figure \ref{fig:QA_B20Convergence} shows that
$[B_{m=0}(\varphi,r=a) - B(\varphi,r=0)]/a^2$ again
converges to the predicted function, $B_{20}(\varphi)$.

\begin{figure}
  \centering
  \includegraphics[width=3in]{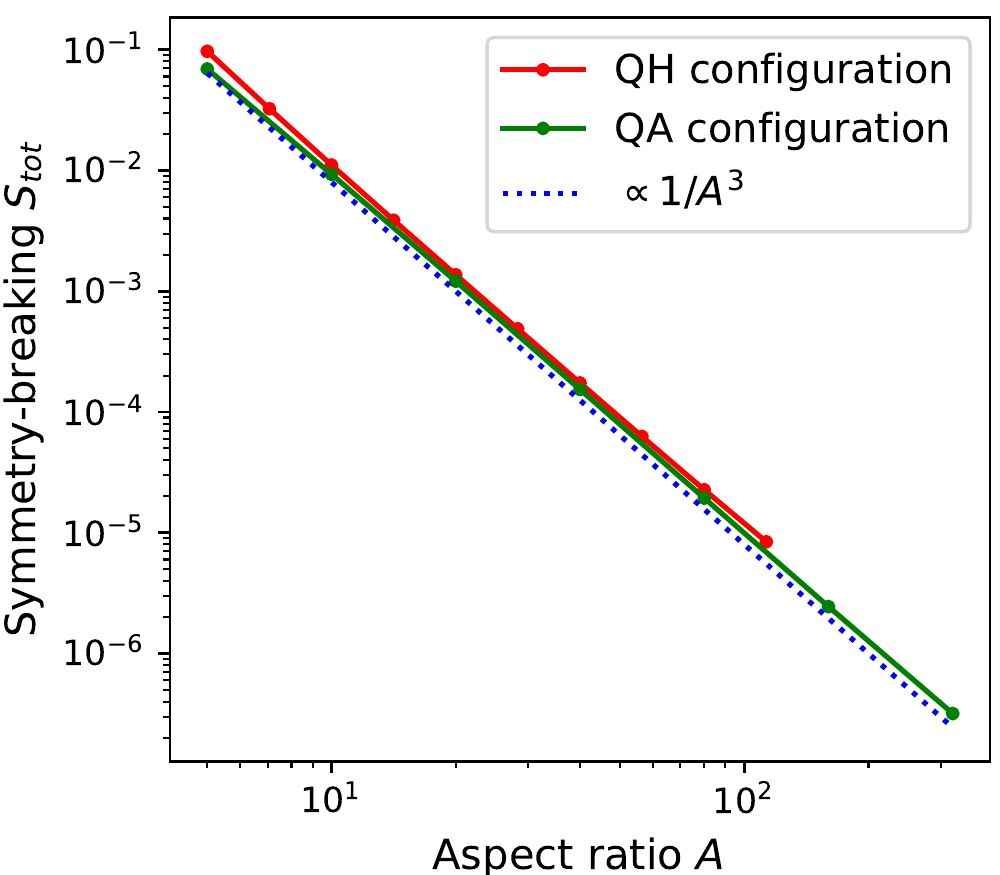}
  \caption{\changed{
A numerical demonstration of 
the prediction by \cite{GB2} that deviations
from quasisymmetry can be made to scale as $1/A^3$. Here,
the deviations are measured by (\ref{eq:Stot})
for the configurations of 
  sections \ref{sec:QA} and \ref{sec:QH}.}
  }
\label{fig:GarrenBoozerScaling}
\end{figure}

The three symmetry-breaking measures for this second configuration
are displayed in figure \ref{fig:scaling}, labeled as `Config 2'.
It can be seen that $S_{m>0}^{r=a}$ and $S_{m=0}^{r=0}$ are not much changed from the first configuration,
scaling as $1/A^3$ and $1/A^4$ as before. However, $S_{m=0}^{r=a}$
is significantly changed, still scaling as $1/A^2$ for sufficiently large $A$, but with the leading constant reduced by over an order of magnitude. This reduction reflects the reduced variation of $B_{20}$. For $A<40$, $S_{m=0}^{r=a}$ now scales as $1/A^4$ since it is dominated by the on-axis variation of $B$ measured by $S_{m=0}^{r=0}$. Thus, $B_{20}$ is constant enough that it is not the dominant source of symmetry-breaking for the entire range of aspect ratios shown, $A \in [5,320]$.
\changed{This point is apparent also in figure \ref{fig:GarrenBoozerScaling}, in which the quantity plotted
\begin{equation}
    S_{tot} = \frac{1}{B_{0,0}} \sqrt{\sum_{m, n \ne N m} B_{m,n}^2(r=a)}
    \label{eq:Stot}
\end{equation}
includes all quasisymmetry-breaking modes.
This figure more clearly shows the scaling
predicted by \cite{GB2} that the total deviation from quasisymmetry can be made to scale as $\propto 1/A^3$.}
The quasisymmetry of this quasi-axisymmetric configuration is sufficiently good that for $A \ge 60$, the deviation from quasisymmetry $S_{tot} B_0$ is smaller than the Earth's magnetic field of $\sim 0.5$ Gauss.
At the rightmost point ($A=320$),
$S_{tot}$ is $< 4\times 10^{-7}$, and the largest single symmetry-breaking Fourier mode has an amplitude $< 2\times 10^{-7}$ T.

\subsection{Tokamak-stellarator hybrid}
\label{sec:hybrid}

\begin{figure}
  \centering
  \includegraphics[width=2.5in]{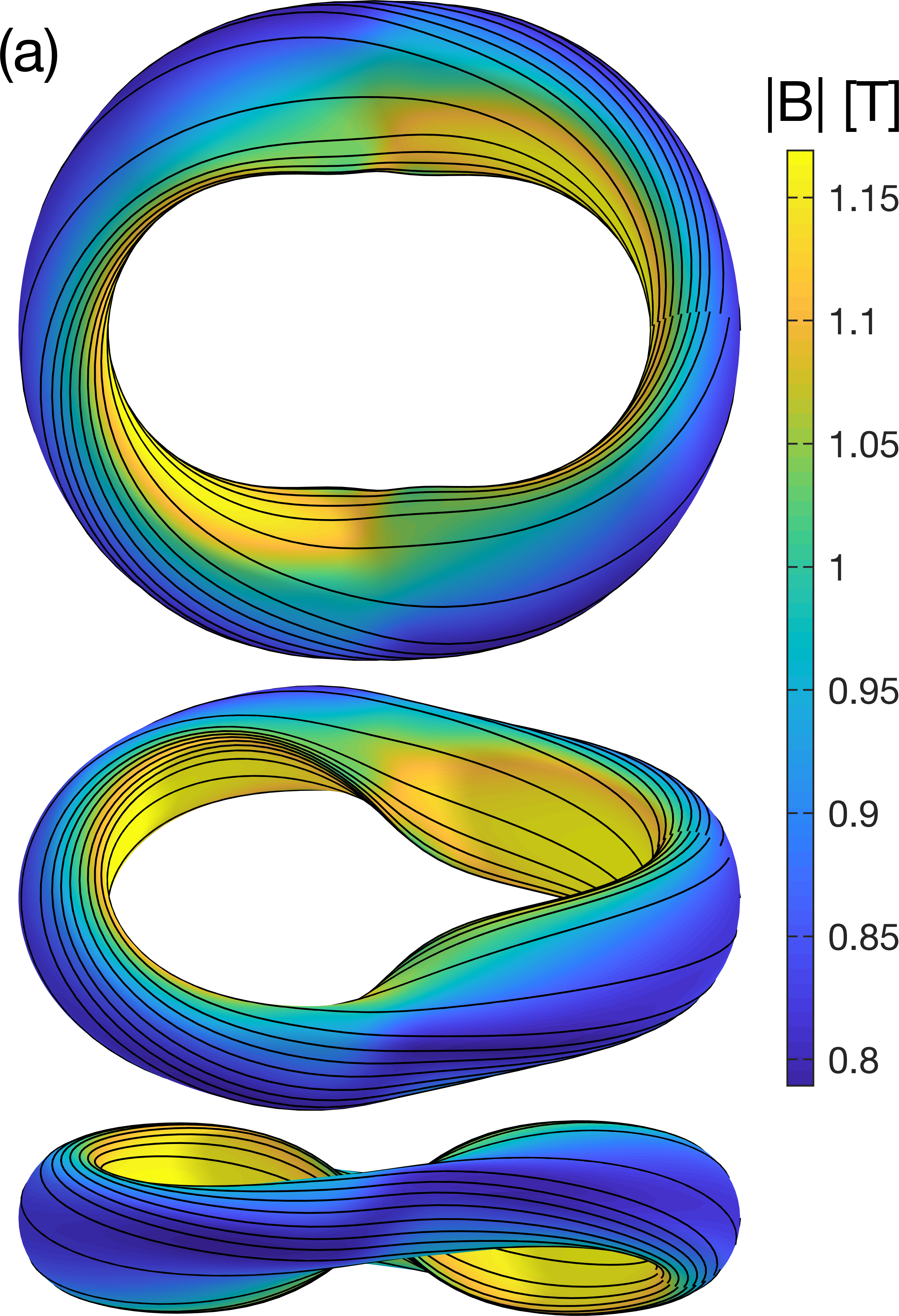}
\raisebox{1.3in}{  \includegraphics[width=2.5in]{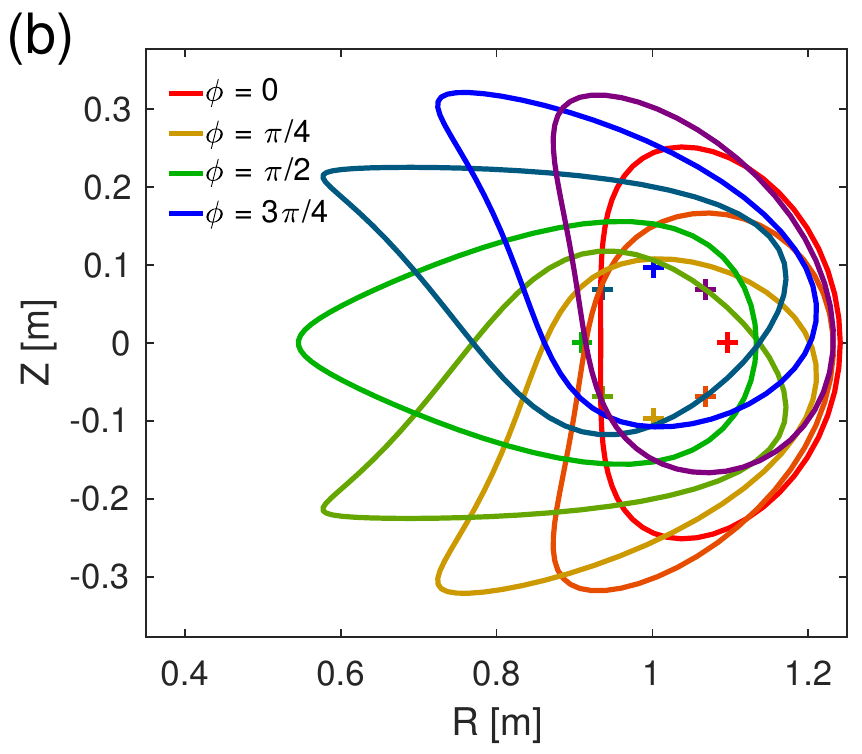}}
  \caption{
The tokamak-stellarator hybrid example of section \ref{sec:hybrid}, 
for aspect ratio $A=5$, $A_{vmec}=4.87$.
The 3D surface shape in (a), shown from three angles, and the cross-sections in (b), are generated by the construction. In (a), magnetic field lines are shown as black lines, and color indicates the field strength computed by VMEC.}
\label{fig:tokamakStellaratorHybrid_example}
\end{figure}
\begin{figure}
  \centering
    \includegraphics[width=3in]{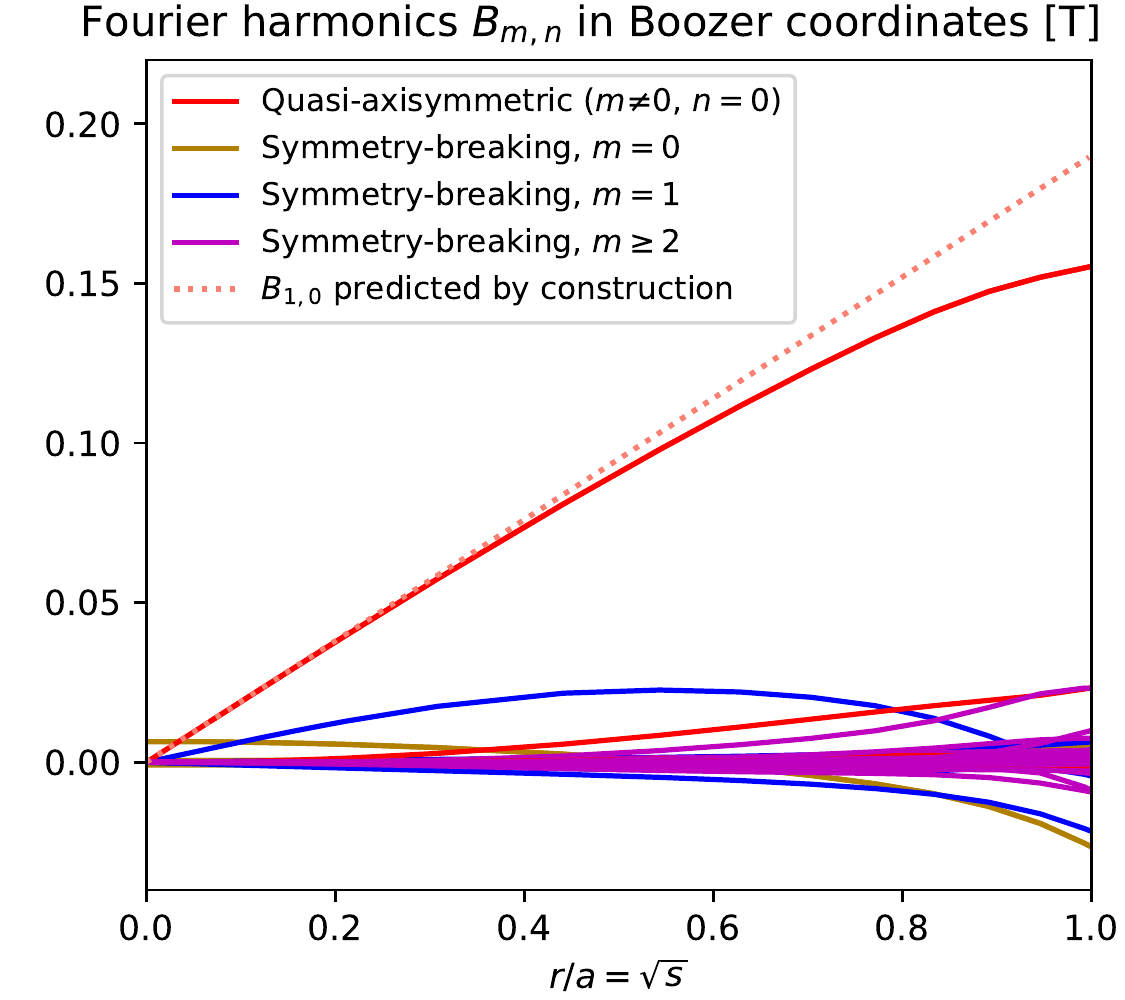}
  \caption{
The spectrum of $B$ for the tokamak-stellarator hybrid example of section \ref{sec:hybrid}, computed by running the VMEC and BOOZ\_XFORM codes inside the constructed boundary surface for aspect ratio $A=5$, $A_{vmec}=4.87$.}
\label{fig:hybrid_spectrum}
\end{figure}
\begin{figure}
  \centering
  \includegraphics[width=3in]{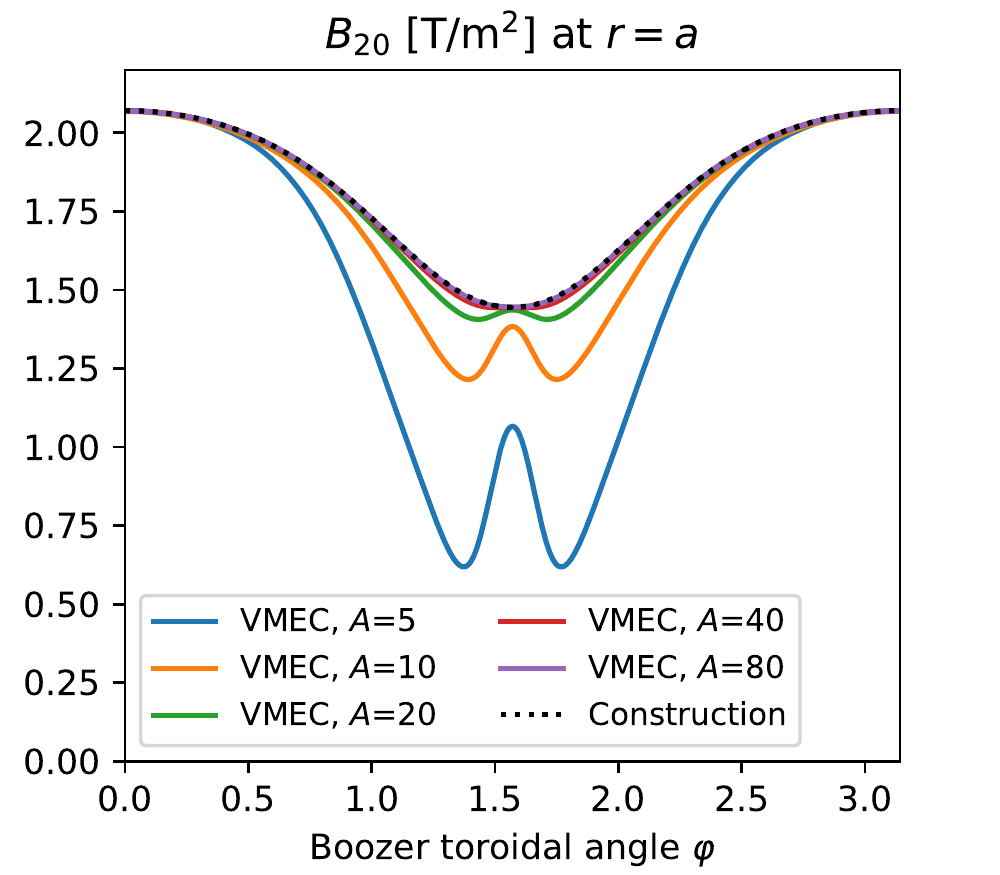}
    \caption{
As the aspect ratio $A$ increases, the $B_{20}(\varphi)$ component of the field strength of the numerical VMEC configurations converges to the function predicted by the Garren-Boozer construction. Data here are for the tokamak-stellarator hybrid configuration of section \ref{sec:hybrid}.  }
\label{fig:hybrid_B20Convergence}
\end{figure}
\begin{figure}
  \centering
  \includegraphics[width=3in]{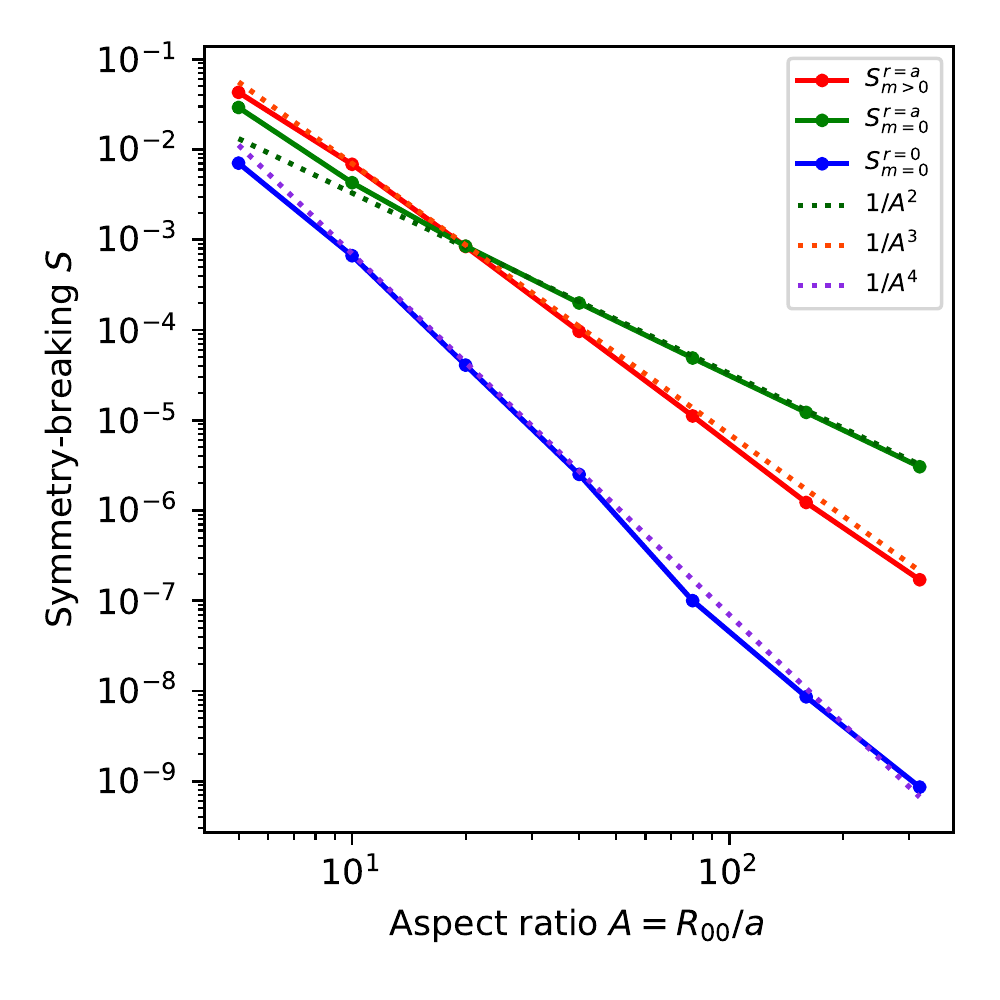}
  \caption{
 The measures of quasisymmetry-breaking (\ref{eq:S_def}), computed by running the VMEC and BOOZ\_XFORM codes inside the constructed boundary surfaces, scale as the expected power of aspect ratio. Data here are for the tokamak-stellarator hybrid example
 of section \ref{sec:hybrid}.
}
\label{fig:hybrid_scaling}
\end{figure}

To verify the construction for a case in which the plasma pressure and on-axis current are nonzero, we next consider a tokamak-stellarator hybrid configuration, in which both
nonaxisymmetric shaping and toroidal current contribute
to the rotational transform.
We again consider a two-field-period geometry, with axis shape
\begin{align}
R_0(\phi)  \;\mathrm{[m]}=& 1 + 0.09 \cos(2\phi)  , \\
z_0(\phi)  \;\mathrm{[m]}= & \hspace{0.08in}-0.09 \sin(2\phi)  .\nonumber
\end{align}
The parameters $\sigma(0)$ and $B_{2s}$ were again set to zero so the configuration is stellarator-symmetric.
The other input parameters were $\etabar=0.95$ m$^{-1}$, $I_2=0.9$ T$/$m, $p_2=-6\times 10^5$ Pa$/$m$^2$, and $B_{2c}=-0.7$ T$/$m$^2$.
For this value of $p_2$, the volume-averaged $\beta$ (plasma pressure $/$ magnetic pressure) for the configuration at $A=5$ is 2.9\%.
The resulting configuration has  
$\iota_0=0.960$. 
For comparison, a vacuum field inside the constructed $A=5$ boundary has an on-axis transform 
$\iota_0 = 0.214$.
This level of vacuum transform might be sufficient to provide stellarator-like stability.
The boundary shape for $A=5$ is shown in figure \ref{fig:tokamakStellaratorHybrid_example}. Figure \ref{fig:hybrid_spectrum} shows the Boozer spectrum of the finite-$\beta$ finite-current configuration inside this boundary. While the desired mode $B_{1,0}$ dominates, and it has a magnitude close to that predicted by the construction, the departures from quasisymmetry are larger than in the previous examples, associated with the smaller value of $A$ here.
Figure \ref{fig:hybrid_B20Convergence} shows that as $A$ is increased,
$[B_{m=0}(\varphi,r=a) - B(\varphi,r=0)]/a^2$ again
converges to the predicted function, $B_{20}(\varphi)$.
The scaling of the three symmetry-breaking measures with $A$
is plotted in figure \ref{fig:hybrid_scaling},
and again, they scale as the expected power of $A$.
Together, figures \ref{fig:hybrid_B20Convergence}-\ref{fig:hybrid_scaling} verify the $O((r/\mathcal{R})^2)$ construction behaves correctly when $I_2$ and $p_2$ terms are included.

\subsection{Quasi-helical symmetry}
\label{sec:QH}

\begin{figure}
  \centering
    \includegraphics[width=2.5in]{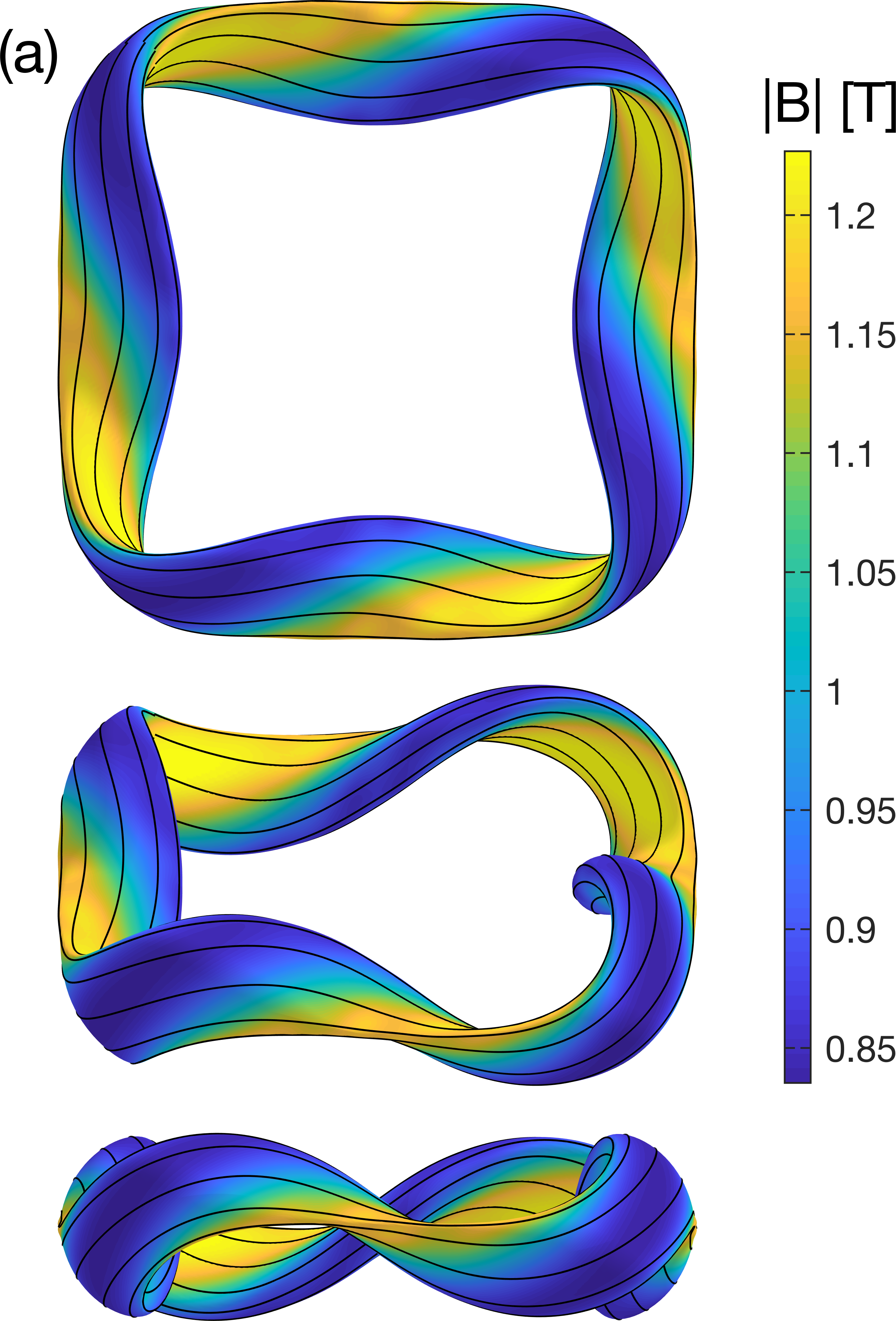}
  \raisebox{1.4in}{\includegraphics[width=2.5in]{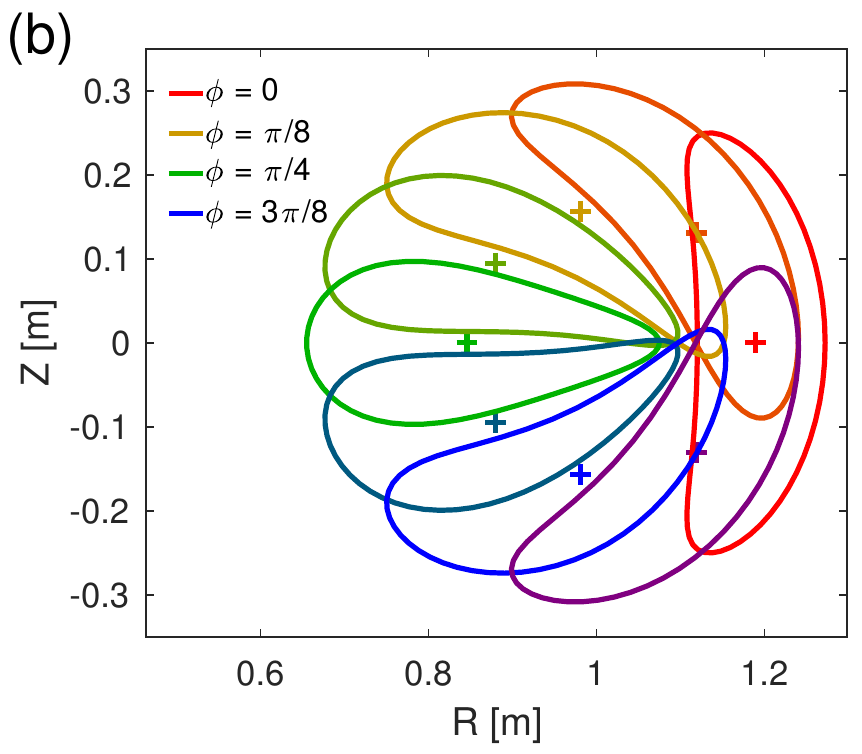}}
  \caption{
The quasi-helically symmetric example of section \ref{sec:QH}, 
for aspect ratio 
$A=8$, $A_{vmec}=7.14$.
The 3D surface shape in (a), shown from three angles, and the cross-sections in (b), are generated by the construction. In (a), magnetic field lines are shown as black lines, and color indicates the field strength computed by VMEC.}
\label{fig:QH_example}
\end{figure}

We next consider a quasi-helically symmetric configuration. The axis shape is taken to be
\begin{align}
R_0(\phi)  \;\mathrm{[m]}=& 1 + 0.1700 \cos(4\phi) + 0.01804 \cos(8\phi) + 0.001409 \cos(12\phi) + 0.00005877 \cos(16\phi), \nonumber\\
z_0(\phi)  \;\mathrm{[m]}= & \hspace{0.25in}0.1583 \sin(4\phi) + 0.01820 \sin(8\phi) + 0.001548 \sin(12\phi) + 0.00007772 \sin(16\phi),
\label{eq:QH_axis_shape}
\end{align}
with 
$\etabar=1.569$ m$^{-1}$ and $B_{2c}=0.1348$ T/m$^2$.
These values were obtained using the optimization procedure of section \ref{sec:optimization} to minimize $X_2$, $Y_2$, $X_3$, and $Y_3$.
For this axis shape, the normal vector rotates around the axis poloidally four times as the axis is traversed toroidally, so the construction yields quasi-helical symmetry rather than quasi-axisymmetry.
The parameters $\sigma(0)$ and $B_{2s}$ were set to zero so the configuration is stellarator-symmetric.
The other input parameters were $I_2=0$ and $p_2=0$.
The resulting configuration has 
$\iota_0=1.14$.
The constructed boundary shape for 
$A=8$ 
is shown in figure \ref{fig:QH_example}.

Compared to the case of quasi-axisymmetry, for quasi-helical symmetry it seems relatively hard to find sets of input parameters for which $X_2$, $Y_2$, $X_3$, and $Y_3$ are acceptably small.
If these quantities are not small, the boundary aspect ratio must be large, or else the symmetry-breaking errors tend to be large and the boundary surface may self-intersect.
This challenge for finding good quasi-helically symmetric configurations likely arises from the fact that they require significant helical excursion of the axis, implying larger $\tau$ and $\kappa$ compared to quasi-axisymmetric configurations, which act to drive larger $X_2$ and $Y_2$.
The configuration in this section manages to have small values of $\{X_2,Y_2,X_3,Y_3\}$ due to some delicate balances in the equations of appendix \ref{sec:2nd_order_field_strength}.
For instance, merely rounding the coefficients in the axis shape (\ref{eq:QH_axis_shape}) to 3 digits of precision rather than 4 causes significant increases in $X_3$ and $Y_3$ that result in visible changes to the boundary shape.

\begin{figure}
  \centering
    \includegraphics[width=3in]{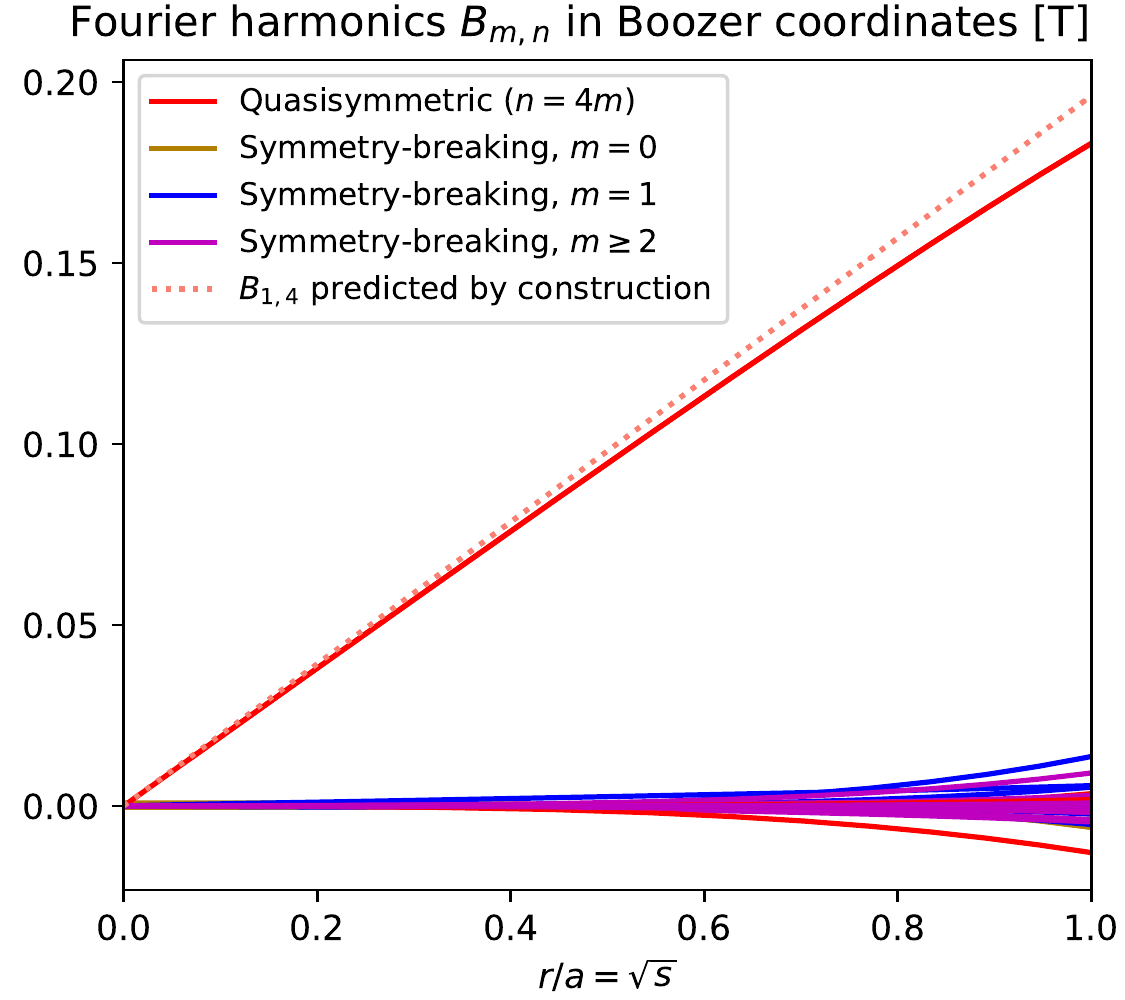}
  \caption{
The spectrum of $B$ for the quasi-helically symmetric example of section \ref{sec:QH}, computed by running the VMEC and BOOZ\_XFORM codes inside the constructed boundary surface for aspect ratio 
$A=8$, $A_{vmec}=7.14$.}
\label{fig:QH_spectrum}
\end{figure}

\begin{figure}
  \centering
    \includegraphics[width=3in]{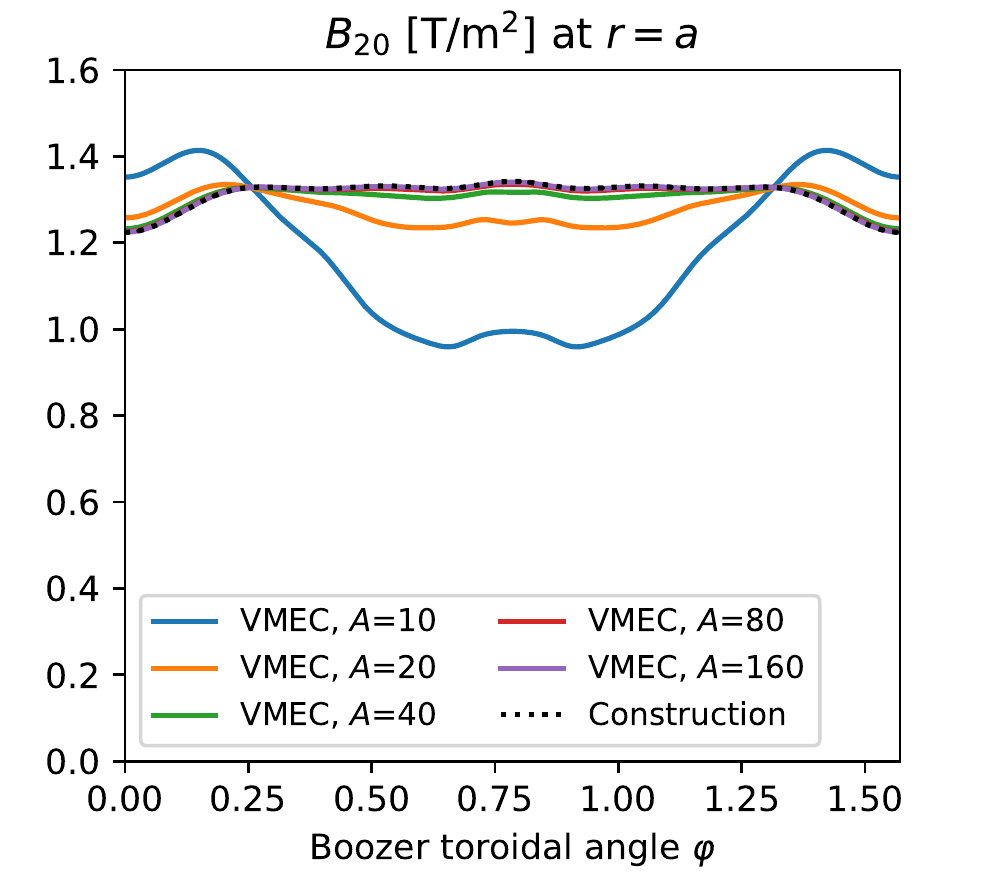}
  \caption{
As the aspect ratio $A$ increases, the $B_{20}(\varphi)$ component of the field strength of the numerical VMEC configurations converges to the function predicted by the Garren-Boozer construction. Data here are for the quasi-helically symmetric configuration of section \ref{sec:QH}.  }
\label{fig:QH_B20Convergence}
\end{figure}

\begin{figure}
  \centering
    \includegraphics[width=3in]{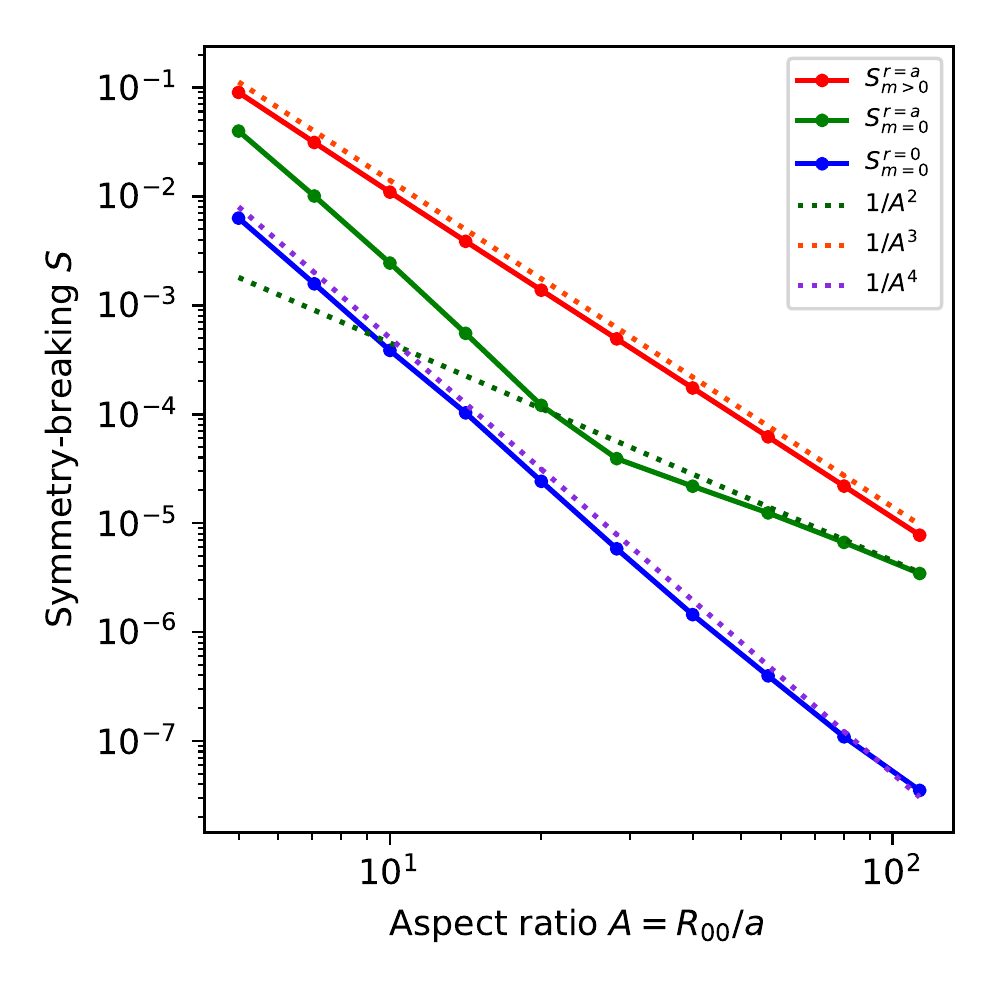}
  \caption{
 The measures of quasisymmetry-breaking (\ref{eq:S_def}), computed by running the VMEC and BOOZ\_XFORM codes inside the constructed boundary surfaces, scale as the expected power of aspect ratio. Data here are for the quasi-helically symmetric example of section \ref{sec:QH}.}
\label{fig:QH_scaling}
\end{figure}

As with the earlier configurations, VMEC and BOOZ\_XFORM calculations for this quasi-helically symmetric configuration confirm that the desired field strength is produced. One aspect of this verification is shown in figure \ref{fig:QH_spectrum}, which displays the Boozer spectrum inside the constructed $A=8$ boundary. This time the dominant mode is $B_{1,4}$, and the magnitude of this mode matches the prediction $r \etabar B_0$. Figure \ref{fig:QH_B20Convergence}
shows that as $A \to \infty$,
$[B_{m=0}(\varphi,r=a) - B(\varphi,r=0)]/a^2$ again
converges to the predicted function, $B_{20}(\varphi)$.
Figure \ref{fig:QH_scaling} shows that $S_{m>0}^{r=a}$, $S_{m=0}^{r=a}$, and $S_{m=0}^{r=0}$ scale approximately as expected ($1/A^3$, $1/A^4$ transitioning to $1/A^2$ at large $A$, and $1/A^4$), as for the previous configurations. 
\changed{
The total deviation from quasisymmetry $S_{tot}$ is also displayed in figure \ref{fig:GarrenBoozerScaling},
demonstrating again Garren and Boozer's predicted scaling. 
(The range of aspect ratio plotted differs from that for the quasi-axisymmetric configuration since at the highest $A$, it is difficult to obtain converged values from VMEC for the very small symmetry-breaking modes.)
}

\subsection{Testing all terms}
\label{sec:allTerms}

\begin{figure}
  \centering
  \includegraphics[width=2.5in]{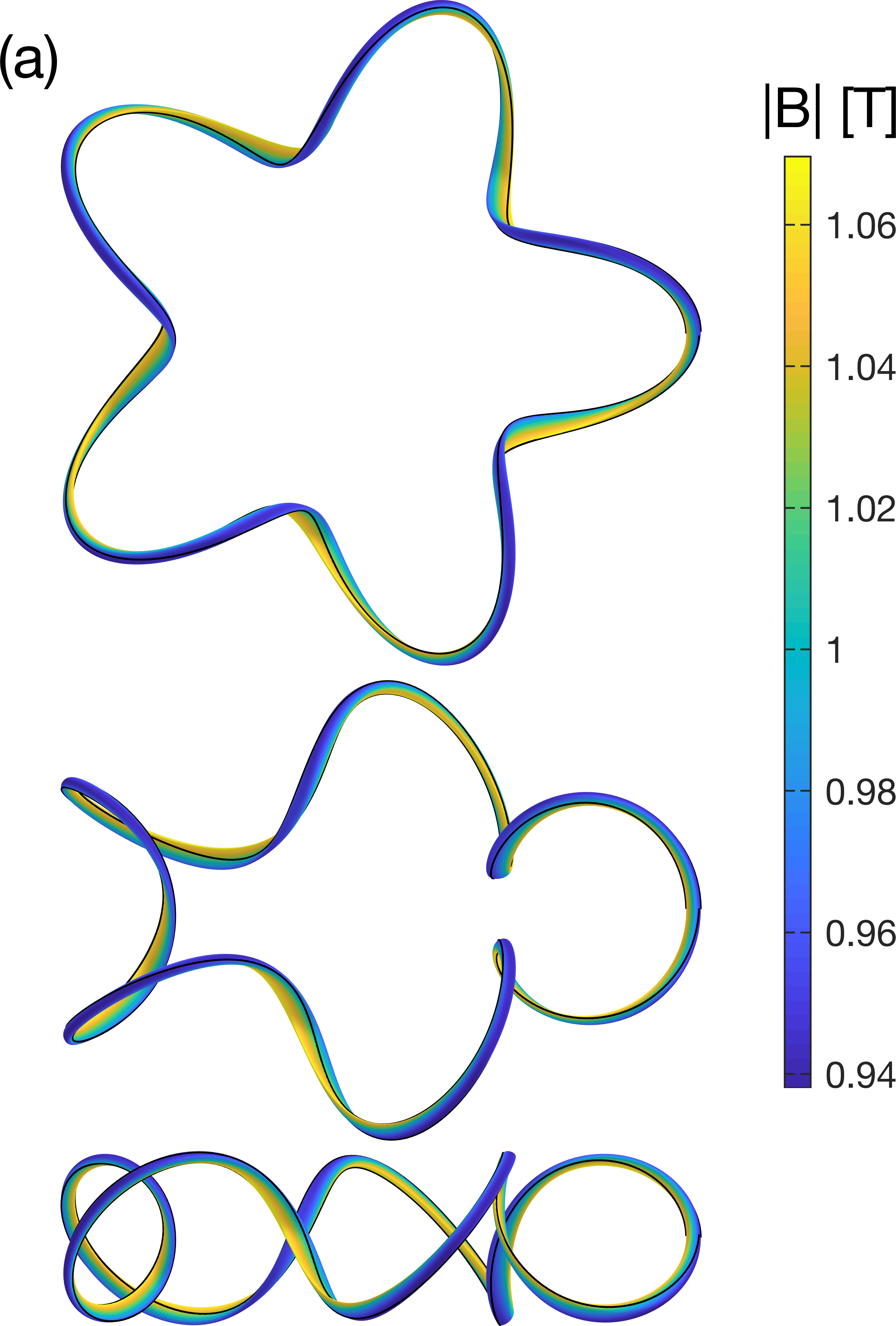}
  \raisebox{1.4in}{\includegraphics[width=2.5in]{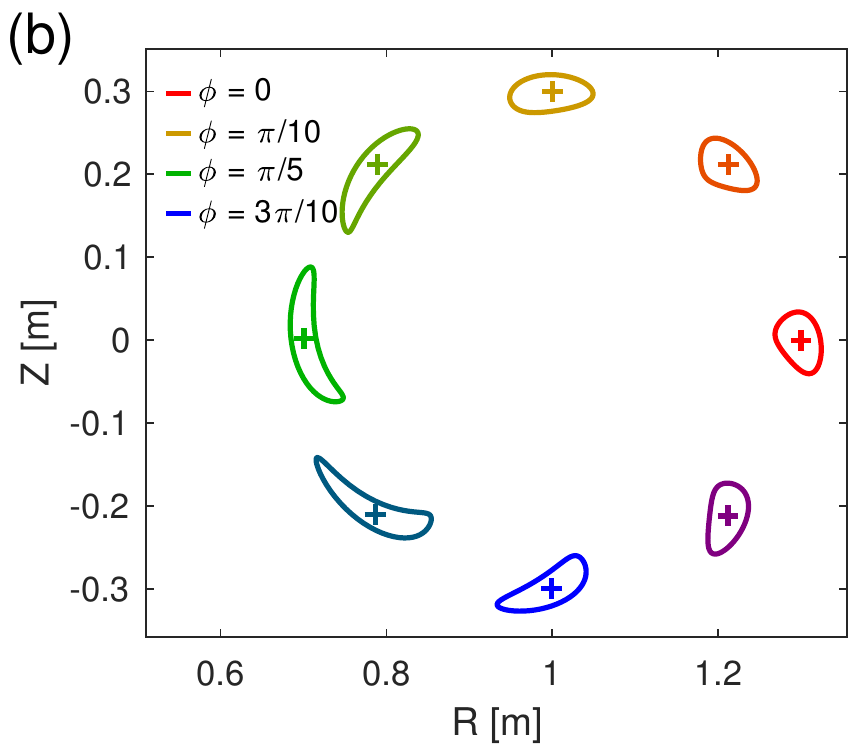}}
  \caption{
The non-stellarator-symmetric quasi-helically-symmetric example of section \ref{sec:allTerms}, 
for aspect ratio $A=40$, $A_{vmec}=28.5$.
The 3D surface shape in (a), shown from three angles, and the cross-sections in (b), are generated by the construction. In (a), magnetic field lines are shown as black lines, and color indicates the field strength computed by VMEC.}
\label{fig:allTerms_example}
\end{figure}

\begin{figure}
  \centering
  \includegraphics[width=3in]{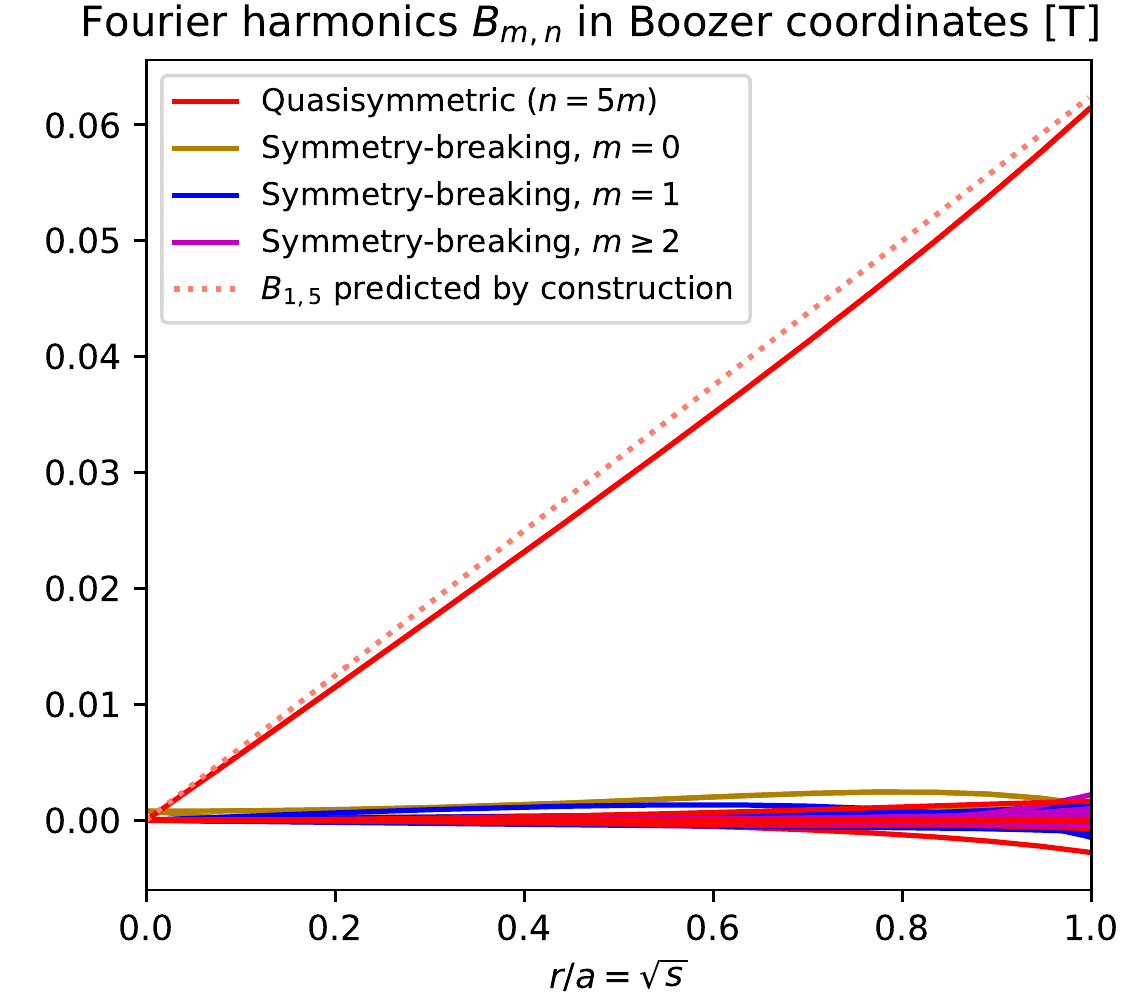}
  \caption{
The spectrum of $B$ (including both $\propto \cos(m\theta-n\varphi)$ and $\propto \sin(m\theta-n\varphi)$ modes) for the non-stellarator-symmetric quasi-helically-symmetric example of section \ref{sec:allTerms}, computed by running the VMEC and BOOZ\_XFORM codes inside the constructed boundary surface for aspect ratio $A=40$, $A_{vmec}=28.5$.}
\label{fig:allTerms_spectrum}
\end{figure}

\begin{figure}
  \centering
  \includegraphics[width=3in]{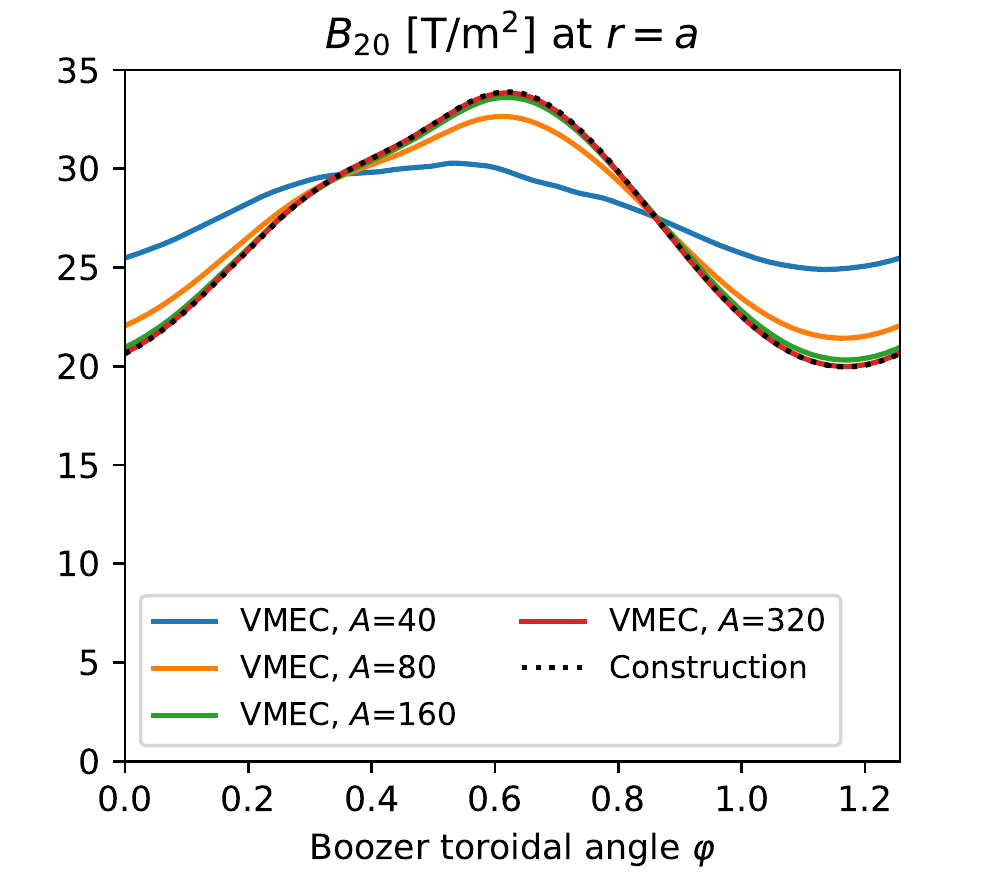}
  \caption{
As the aspect ratio $A$ increases, the $B_{20}(\varphi)$ component of the field strength of the numerical VMEC configurations converges to the function predicted by the Garren-Boozer construction. Data here are for the non-stellarator-symmetric quasi-helically-symmetric configuration of section \ref{sec:allTerms}.  }
\label{fig:allTerms_B20Convergence}
\end{figure}

\begin{figure}
  \centering
  \includegraphics[width=3in]{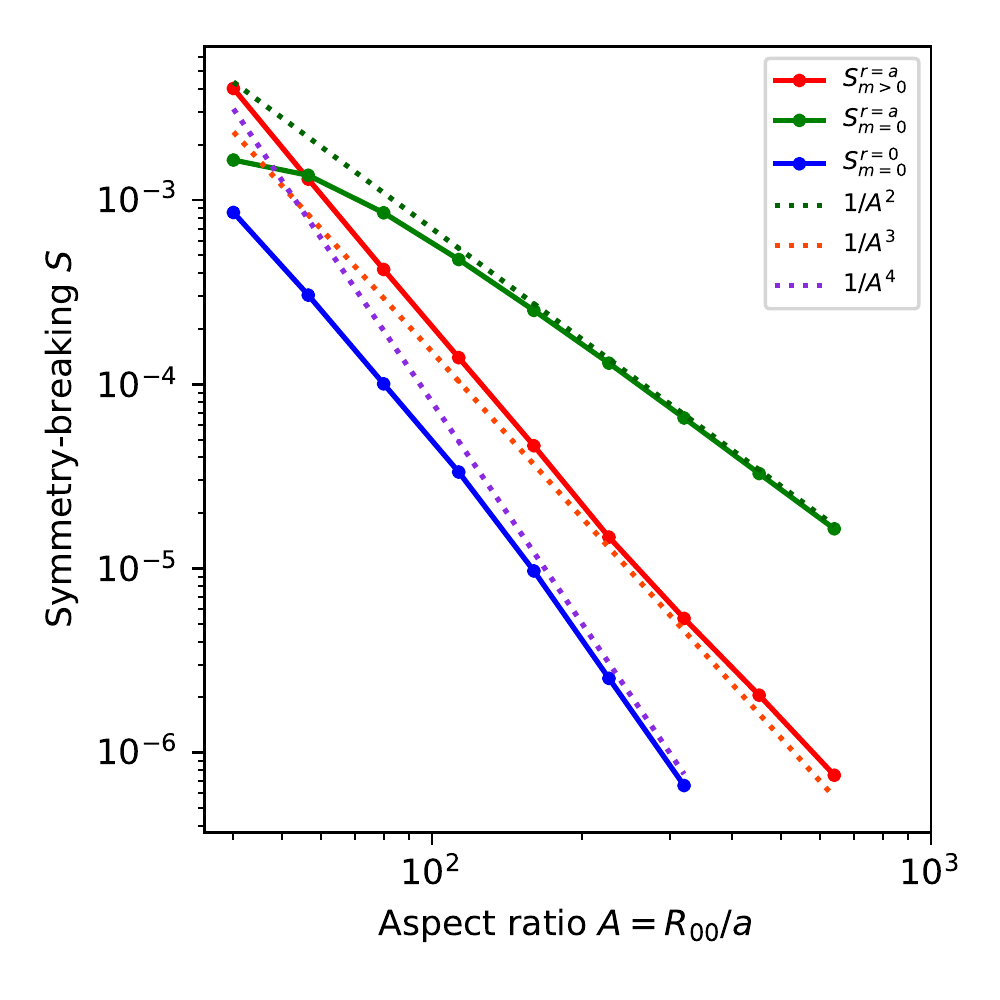}
  \caption{
 The measures of quasisymmetry-breaking (\ref{eq:S_def}), computed by running the VMEC and BOOZ\_XFORM codes inside the constructed boundary surfaces, scale as the expected power of aspect ratio. Data here are for the non-stellarator-symmetric quasi-helically-symmetric example of section \ref{sec:allTerms}.}
\label{fig:allTerms_scaling}
\end{figure}

For a final example, we present an example in which all the parameters of the near-axis model are nonzero. This example is limited to quite a large aspect ratio due to the large $X_2$ and $Y_2$ terms, and so is not interesting as an experimental design, but it is useful here as a challenging verification test. 
We choose the axis shape
\begin{align}
R_0(\phi)  \;\mathrm{[m]}=& 1 + 0.3 \cos(5\phi), \\
z_0(\phi)  \;\mathrm{[m]}= & \hspace{0.25in}0.3 \sin(5\phi),\nonumber
\end{align}
which yields quasi-helical symmetry with $N=5$.
The other input parameters are chosen to be $\etabar=2.5$ m$^{-1}$, $\sigma(0)=0.3$, $I_2=1.6$ T$/$m, $p_2=-5\times 10^6$ Pa$/$m$^2$, $B_{2c}=1$ T$/$m$^2$, and $B_{2s}=3$ T$/$m$^2$.
Note that stellarator symmetry is broken both by the nonzero value of $\sigma(0)$ and of $B_{2s}$. This resulting configuration has $\iota_0=0.829$. The constructed boundary
shape is shown in figure \ref{fig:allTerms_example} for $A=40$ ($A_{vmec}=28.5$), and it can be seen that the surface cross-sections are not stellarator-symmetric.
The amplitudes of the $\cos(m\theta-n\varphi)$ and $\sin(m\theta-n\varphi)$ modes of $B$ inside this boundary, as computed by VMEC
and BOOZ\_XFORM, are shown in figure \ref{fig:allTerms_spectrum}.
The $\cos(\theta-5\varphi)$ term dominates, as desired, and its amplitude agrees with the prediction of the near-axis equations.
Repeating the  VMEC
and BOOZ\_XFORM computations for this solution of the near-axis equations for a range of aspect ratios,
$[B_{m=0}(\varphi,r=a) - B(\varphi,r=0)]/a^2$ again
converges to the predicted function $B_{20}(\varphi)$,
as shown in figure \ref{fig:allTerms_B20Convergence}.
Figure \ref{fig:allTerms_scaling} shows that the symmetry-breaking modes scale as $1/A^3$ or better, as desired, except for the expected $1/A^2$ scaling
of $S_{m=0}^{r=a}$ associated with the toroidal variation of $B_{20}$.
(The rightmost blue points are missing since it did not seem possible to obtain values that were converged with respect to VMEC resolution parameters.)
Thus, finite-aspect-ratio VMEC calculations successfully match the near-axis solution even when all parameters of the latter are nonzero.

\section{Discussion and conclusions}
\label{sec:conclusions}

In this work, we have developed a new and fast method to generate quasisymmetric magnetic
fields with sophisticated shaping. In contrast to the traditional approach based on numerical optimization, the approach here uses a reduced set of equations relating the field strength in Boozer coordinates $B(r,\theta,\varphi)$
to the
three-dimensional shapes of the magnetic surfaces.
The shapes that are describable by the $O((r/\mathcal{R})^2)$ near-axis model here are sufficiently general that they can be quite reminiscent of stellarators that have been designed previously using numerical optimization. For instance, the configuration of section \ref{sec:QA} (figure \ref{fig:QA_example}) resembles CFQS \citep{CFQSShimizu,CFQSLiu} and the configuration of \citet{Henneberg}. Also the configuration
of section \ref{sec:QH} (figure \ref{fig:QH_example}) resembles the HSX experiment \citep{HSX}. 
Despite these similarities, the examples here were generated  independently
of any previously known configurations.
Since these shapes computed by our model are described analytically, they can be parameterized, evaluated rapidly, and differentiated.
As analytic expressions for the position vector in terms of both Boozer coordinates and cylindrical coordinates (appendix \ref{sec:transformation}) are available, one can evaluate virtually any quantity of interest, such as the geometric quantities appearing in the gyrokinetic model of turbulence.
Since the system of equations involves only one independent variable ($\varphi$), compared to three for general MHD equilibrium, the equations here are orders of magnitude faster to solve.

\begin{figure}
\centerline{
\includegraphics[width=5.4in]{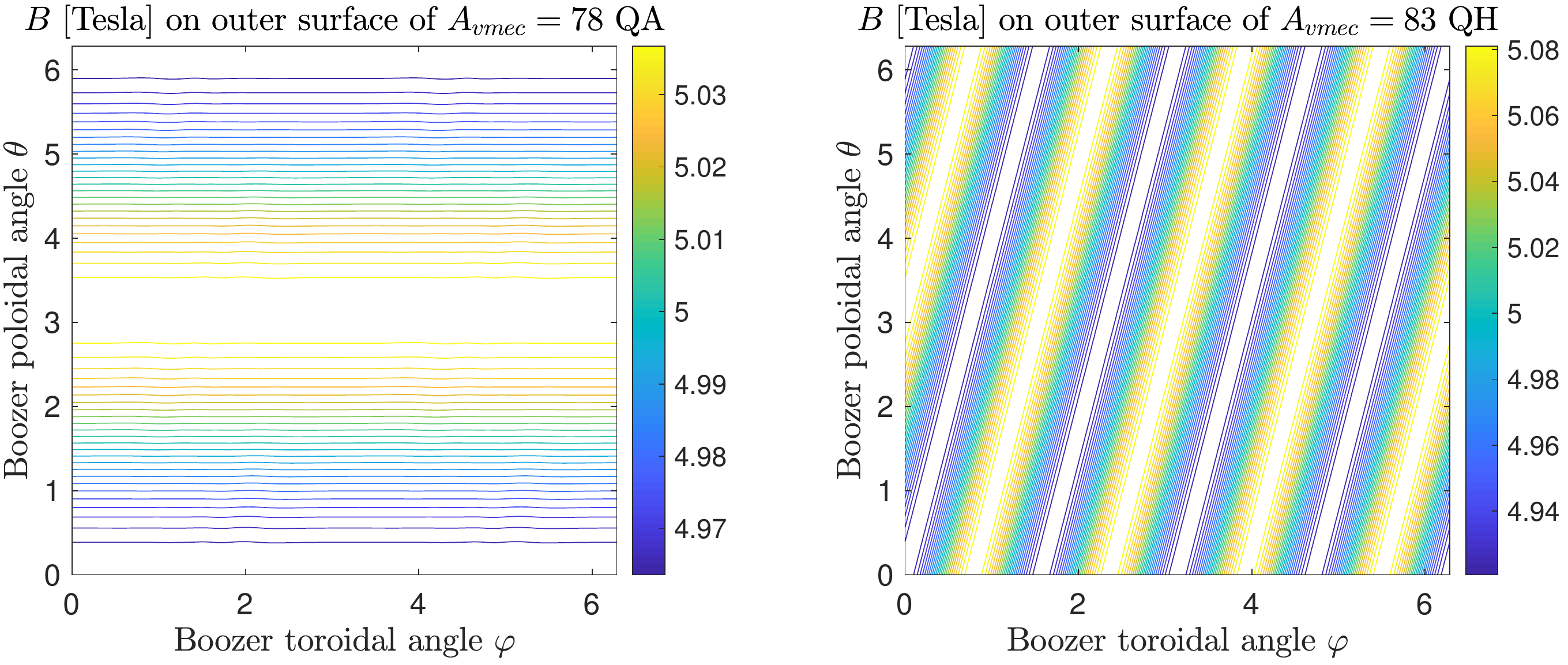}
}
\centerline{
\raisebox{0.25in}{\includegraphics[width=2.5in]{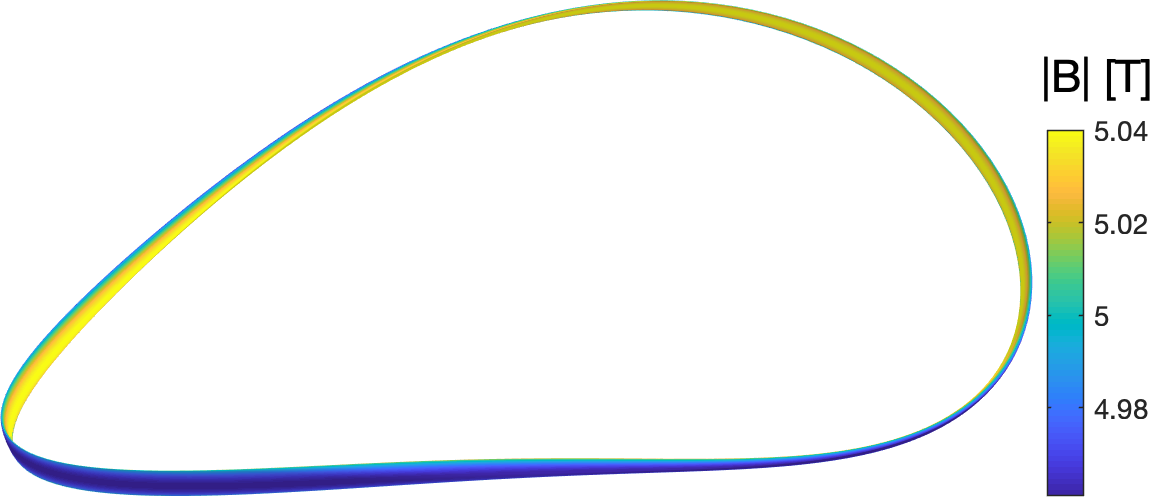}}
\hspace{0.2in}
\includegraphics[width=2.5in]{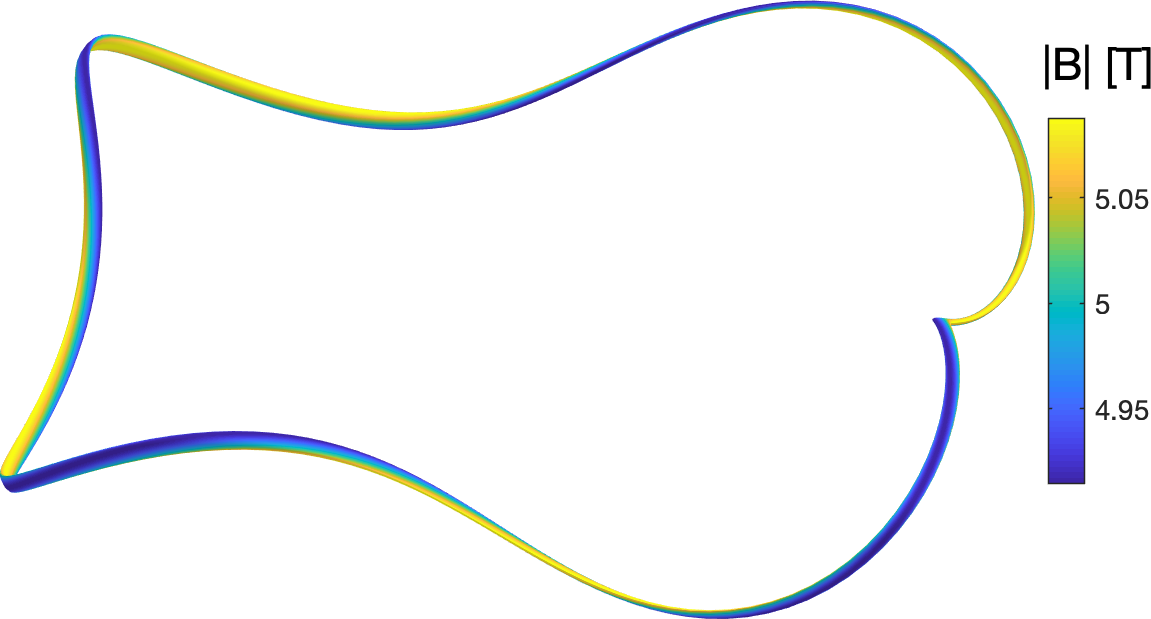}
}
\caption{Contours of $B(\theta,\zeta)$ at the boundaries of the configurations of sections \ref{sec:QA} and \ref{sec:QH}, scaled to a mean field of 5 Tesla, at the aspect ratio for which the largest symmetry-breaking Fourier modes have amplitude $0.5$ Gauss, the magnitude of Earth's magnetic field.
Departures from quasisymmetry are nearly imperceptible on the scale of the plots, demonstrating that quasisymmetry can be realized in strongly-nonaxisymmetric equilibria to very high accuracy, at least at high $A$.
\label{fig:A80}}
\end{figure}

Through the examples in section \ref{sec:results}, we have 
demonstrated that the approach here is a practical way to generate and parameterize both quasi-axisymmetric and quasi-helically symmetric configurations. For each of the examples, we showed that the departures from quasisymmetry computed by conventional codes scale with the aspect ratio as expected. 
In particular, we have demonstrated that
quasisymmetry can be achieved (without axisymmetry) to any desired precision, at sufficiently high aspect ratio.
Due to the high-order accuracy of the equations in our model, the quality of quasisymmetry can be extremely good. For example, the symmetry-breaking measures (\ref{eq:S_def}) are smaller than $4\times 10^{-7}$ for the rightmost `Config 1' point in figure \ref{fig:scaling}.
While at $A=320$ this configuration is not of great experimental interest, it does represent the 
most accurate realization of quasisymmetry in a 3D equilibrium ever reported.
\changed{While arbitrarily small departures from quasisymmetry can also be obtained with the $O((r/\mathcal{R})^1$) construction,
as demonstrated in figure 4 of \cite{PaperII}, higher aspect ratios would be required to obtain quasisymmetry to the same precision, due to the weaker scaling of symmetry-breaking with $1/A^2$ in that case. }
As an additional demonstration of the high accuracy to which quasisymmetry can be achieved by the $O((r/\mathcal{R})^2$) construction, figure \ref{fig:A80} shows the contours of $B$ on the boundary surfaces of the configurations of sections \ref{sec:QA} and \ref{sec:QH}
\changed{
with aspect ratios chosen by the following criterion: at a reactor-relevant mean field $B_{0,0}=5$ T, the largest quasisymmetry-breaking Fourier mode amplitudes are only 0.5 Gauss, the approximate magnitude of Earth's magnetic field.
This condition results in aspect ratios $A_{vmec} \sim$ 80.
There may be other configurations for which this condition can be met at lower aspect ratio; our goal here is merely to demonstrate that the condition can indeed be achieved.
}
The deviation from symmetry is nearly invisible in figure \ref{fig:A80}, and the $B$ contours are far more quasisymmetric than in other nominally quasisymmetric configurations reported previously, e.g. figures 5-7 of \citet{Beidler}.

Most of the solutions exhibited in this paper have a relatively high aspect ratio, which is not surprising since the
method is based on an expansion in aspect ratio. Several methods are likely to enable configurations of lower aspect ratio to be generated. First, by including more Fourier modes in the axis shape, $X_2$ and $Y_2$ could perhaps be further reduced, resulting in configurations for which the $O((r/\mathcal{R})^3)$ symmetry-breaking terms have a smaller leading constant. Second, the method of section 4.4 of \citet{PaperII} could be used to extrapolate outward from a high-$A$ configuration while preserving good quasisymmetry in the core. Third, a large value of $a$ could be used in the present approach, resulting in moderate deviations from quasisymmetry, which could then be reduced by conventional optimization. Lastly, a configuration with smaller $a$ and good quasisymmetry generated by the construction here could be used to initialize conventional optimization, in which the aspect ratio is included in the objective function for minimization.

The work here suggests many avenues for future study, some of which are enumerated here. (1) It was shown previously that practical quasisymmetric configurations obtained by optimization closely match the $O(r/\mathcal{R})$ near-axis construction \citep{fitToGarrenBoozer}, and the comparison should be repeated for the $O((r/\mathcal{R})^2)$ model. (2) An efficient procedure should be found to solve for model parameters such that $B_{20}$ is independent of $\varphi$. (3) While a precise understanding exists of the solution space for $O(r/\mathcal{R})$ quasisymmetry \citep{PaperII}, the same insight has yet to be developed for the $O((r/\mathcal{R})^2)$ model. It would be valuable to understand the space of solutions
to the $O((r/\mathcal{R})^2)$ model to be sure all the interesting regions of parameter space have been identified. 
(4) It was seen here that some toroidal variation of $B_{20}$ could be allowed without
$B_{20}$ becoming the dominant quasisymmetry-breaking mode, so the effect of allowing small toroidal variation of $B_{2c}$ or $B_{2s}$ should be examined.
(5) It should be investigated whether quasisymmetry could be optimized off-axis, by introducing small toroidal variation in $B_0$ that is canceled by $B_{20}$ at a certain radius.
(6) The space of configurations that are omnigenous to $O(r/\mathcal{R})$ was recently
examined \citep{PaperIII}, and the analysis could possibly be extended to $O((r/\mathcal{R})^2)$ omnigenity using results derived here.

Finally, extensions of the quasisymmetry model here to even higher order in $r$ could be pursued. One motivation for such an extension is that global magnetic shear first appears at $O((r/\mathcal{R})^3)$.
Although quasisymmetry cannot generally be achieved through $O((r/\mathcal{R})^3)$ \citep{GB2}, 
the size of the $O((r/\mathcal{R})^3)$ terms informs how rapidly quasisymmetry degrades with $r$,
so solutions could be sought in which these terms were minimized.



This work was supported by the
U.S. Department of Energy, Office of Science, Office of Fusion Energy Science,
under award numbers DE-FG02-93ER54197 and DE-FG02-86ER53223.
This work was also supported by a grant from the Simons Foundation (560651, ML).


\appendix
\section{Derivation of the equations at each order}
\label{sec:2nd_order_field_strength}

In this section we elaborate on section \ref{sec:expansion},
showing a streamlined method to derive the required equations at each order in $r/\mathcal{R}$.
It is possible to obtain the same final equations without the manipulations of section \ref{sec:Wricks_trick}, 
but at the cost of substantial additional algebra. In particular, to derive the equations for $\{X_j,Y_j,Z_j\}$ at a given order $j$ without these manipulations, it would be necessary to first derive equations involving $Z_{j+1}$,
and then form linear combinations to eliminate this higher-order quantity. The method of section \ref{sec:Wricks_trick}
enables the equations for $\{X_j,Y_j,Z_j\}$
to be obtained directly without introducing $Z_{j+1}$.

\subsection{Fundamental equations}
\label{sec:Wricks_trick}

To begin, note the product of (\ref{eq:straight_field_lines_h}) and (\ref{eq:Boozer_h}) gives the inverse Jacobian
\begin{align}
\nabla \psi \cdot \nabla\vartheta \times \nabla \varphi = \frac{1}{\sqrt{g}} = \frac{B^2}{G+\iota I}. 
\label{eq:Jacobian}
\end{align}
Then equating (\ref{eq:straight_field_lines_h}) and (\ref{eq:Boozer_h}), applying (\ref{eq:Frenet})-(\ref{eq:dual}), and 
using $\partial/\partial\psi = (r \bar{B})^{-1} \partial/\partial r$,
we obtain the following three scalar equations:
\begin{align}
\label{eq:YZ}
\{Y, \; Z \} = T_X, \\
\label{eq:ZX}
\{Z, \; X \} = T_Y, \\
\{X, \; Y \} = T_Z,
\label{eq:XY}
\end{align}
where $\{\ldots,\ldots\}$ denotes a Poisson bracket in the $(r,\vartheta)$ coordinates: 
\begin{align}
\{X, \; Y \} = \frac{\partial X}{\partial r} \frac{\partial Y}{\partial\vartheta}  - \frac{\partial X}{\partial\vartheta} \frac{\partial Y}{\partial r}.
\end{align}
The right-hand sides of (\ref{eq:YZ})-(\ref{eq:XY}) are
\begin{align}
T_X = \frac{1}{G+NI} \left[
 r \bar{B} \left(\Xi +   \iotaN \frac{\partial X}{\partial\vartheta} \right)
-I \left(\Upsilon \frac{\partial Z}{\partial r} - \Lambda \frac{\partial Y}{\partial r} \right) 
-\beta r \bar{B} \left( \Lambda \frac{\partial Y}{\partial \vartheta} - \Upsilon \frac{\partial Z}{\partial \vartheta}\right)
\right],
\label{eq:normal}
\end{align}
\begin{align}
T_Y = \frac{1}{G+NI} \left[
 r \bar{B} \left( \Upsilon +  \iotaN \frac{\partial Y}{\partial \vartheta} \right)
-I \left( \Lambda \frac{\partial X}{\partial r} - \Xi \frac{\partial Z}{\partial r}\right)
-\beta r \bar{B} \left( \Xi \frac{\partial Z}{\partial \vartheta} - \Lambda \frac{\partial X}{\partial\vartheta}\right)
\right],
\label{eq:binormal}
\end{align}
and
\begin{align}
T_Z =\frac{1}{G+NI} \left[
  r \bar{B} \left( \Lambda +   \iotaN \frac{\partial Z}{\partial \vartheta} \right)
-I\left( \Xi \frac{\partial Y}{\partial r} - \Upsilon \frac{\partial X}{\partial r}\right)
-\beta r \bar{B} \left( \Upsilon \frac{\partial X}{\partial\vartheta} - \Xi \frac{\partial Y}{\partial \vartheta}\right)
\right],
\label{eq:tangent}
\end{align}
where
\begin{align}
\label{eq:Lambda}
\Lambda
&= \frac{\partial Z}{\partial \varphi} +  (1 -  X \kappa) \ell', \\
\Xi
&= \frac{\partial X}{\partial \varphi} +   \left( -Y \tau + Z \kappa\right) \ell', \\
\Upsilon
&= \frac{\partial Y}{\partial \varphi} +   X \tau \ell',
\label{eq:Upsilon}
\end{align}
and $\ell' =\sqrt{(d\vect{r}_0 / d\varphi)^2}$.

As alluded to above, it turns out to be inconvenient to solve (\ref{eq:YZ}) and (\ref{eq:ZX}) as written,
since these equations involve $Z$ to one higher order than $X$ or $Y$. The reason
is that $X_1$ and $Y_1$ are nonzero while $Z_1$ turns out to vanish (shown in the next subsection),
so the left-hand sides at $O((r/\mathcal{R})^j r)$ include terms $\{rY_1,r^{j+1}Z_{j+1}\}$ and $\{r^{j+1}Z_{j+1},rX_1\}$, while the highest orders
of $X$ and $Y$ appear through $\{r^j Y_j, r^2 Z_2\}$ and $\{r^2 Z_2, r^j X_j\}$.
Therefore it turns out to be convenient to form two combinations of (\ref{eq:YZ}) and (\ref{eq:ZX}),
one in which $Z$ is given explicitly in terms of lower-order quantities, and the other in which
the higher-order $Z$ terms are eliminated to give a constraint on the lower-order quantities.
To form the first desired combination, we start by writing
(\ref{eq:YZ})-(\ref{eq:ZX}) as
\begin{align}
\begin{pmatrix} 
-\partial Y/\partial \vartheta && \partial Y/\partial r \\
\partial X/\partial\vartheta && -\partial X/\partial r
\end{pmatrix}
\begin{pmatrix} \partial Z/\partial r \\ \partial Z/\partial \vartheta \end{pmatrix}
=
\begin{pmatrix} T_X \\ T_Y \end{pmatrix}.
\end{align}
This linear system can be solved
to give
\begin{align}
\begin{pmatrix} \partial Z/\partial r \\ \partial Z/\partial \vartheta \end{pmatrix}
=
-\left[\frac{\partial X}{\partial r} \frac{\partial Y}{\partial \vartheta} - \frac{\partial X}{\partial\vartheta}\frac{\partial Y}{\partial r}\right]^{-1}
\begin{pmatrix} 
\partial X/\partial r && \partial Y/\partial r \\
\partial X/\partial\vartheta && \partial Y/\partial \vartheta
\end{pmatrix}
\begin{pmatrix} T_X \\ T_Y \end{pmatrix},
\label{eq:Z_system}
\end{align}
where the determinant
can be recognized from (\ref{eq:XY}) as $T_Z$.
The top row of 
(\ref{eq:Z_system}) 
then gives an equation that will tell us $Z$ at each order in terms of lower-order quantities:
\begin{align}
\frac{\partial Z}{\partial r} = -\frac{1}{T_Z}\left(\frac{\partial X}{\partial r} T_X + \frac{\partial Y}{\partial r} T_Y\right) .
\label{eq:Zr}
\end{align}
The equality of mixed partial derivatives $\partial^2 Z/\partial r \partial\vartheta = \partial^2 Z / \partial\vartheta \partial r$ can be used with (\ref{eq:Z_system})
to obtain
\begin{align}
\{X, \; T_X/T_Z \} + \{Y, \; T_Y/T_Z \} = 0.
\label{eq:mixedPartialsOld}
\end{align}
This latter equation is the second desired combination of (\ref{eq:YZ}) and (\ref{eq:ZX}), giving a constraint on $X$ and $Y$ at each order without introducing $Z$ at the next order.

We can also derive a different combination of (\ref{eq:YZ}) and (\ref{eq:ZX})
with the same property, as an equivalent alternative to (\ref{eq:mixedPartialsOld}), which more closely corresponds to the equations of
Garren and Boozer at $O((r/\mathcal{R})^2)$. This second approach begins with the observation that the problematic terms that introduce $Z$ at
higher order than $X$ and $Y$ are $\{r Y_1, r^{j+1} Z_{j+1}\}$ and $\{r^{j+1} Z_{j+1}, r X_1\}$. To separate out these terms, we introduce $X_{>1} = X - rX_1$ and $Y_{>1} = Y - r Y_1$, so (\ref{eq:YZ})-(\ref{eq:ZX}) give
\begin{align}
\label{eq:TXexpanded}
T_X - \{ rY_1,Z\}-\{Y_{>1},Z\}=0,
\hspace{0.5in}
T_Y - \{Z, r X_1\} - \{Z, X_{>1}\}=0. 
\end{align}
We look for a combination of these equations in which the problematic terms $\{ rY_1,Z\}$ and $\{Z, r X_1\}$ are
annihilated. To this end, it can be verified that
\begin{align}
\{ r X_1, \{ rY_1,Z\}/r\} + \{ r Y_1, \{Z, r X_1\}/r\} = 0.
\end{align}
Forming the analogous combination of (\ref{eq:TXexpanded}) then gives the desired relation,
in which $Z$ appears at no higher order than $X$ or $Y$:
\begin{align}
\left\{ r X_1, \frac{T_X - \{Y_{>1},Z\}}{r} \right\} + \left\{ r Y_1, \frac{T_Y - \{Z,X_{>1}\}}{r}\right\}=0.
\label{eq:mixedPartials}
\end{align}

Finally, we obtain an expression for the magnetic field strength by squaring (\ref{eq:straight_field_lines_h}), and using (\ref{eq:Jacobian}):
\begin{equation}
\frac{(G+\iota I)^2}{B^2}
= \left( \Lambda + \iotaN \frac{\partial Z}{\partial\vartheta}\right)^2
+ \left(\Xi + \iotaN \frac{\partial X}{\partial\vartheta} \right)^2
+ \left( \Upsilon +\iotaN \frac{\partial Y}{\partial\vartheta} \right) ^2.
\label{eq:modB}
\end{equation}
Equations (\ref{eq:XY}), (\ref{eq:Zr}),  (\ref{eq:mixedPartials}), and (\ref{eq:modB})
are the four equations we will solve at each order for the corresponding unknowns $X$, $Y$,  $Z$, and $B$.


\subsection{Equations through $O((r/\mathcal{R})^2)$}
\label{sec:equations_through_r2}

We now evaluate the first few orders of the $r/\mathcal{R}$ expansion, without assuming quasisymmetry.
At $O((r/\mathcal{R})^0)$, (\ref{eq:modB}) gives
\begin{align}
G_0 = s_G B_0 \, \ell',
\label{eq:G0}
\end{align}
where $s_G=\pm 1=\mathrm{sign}(G_0)$, and (\ref{eq:Zr}) gives $Z_1=0$. Eq (\ref{eq:XY}) and (\ref{eq:mixedPartials}) have no terms of this order.
At $O((r/\mathcal{R})^1)$, (\ref{eq:XY}) gives
\begin{align}
\label{eq:flux_area}
X_{1c} Y_{1s} - X_{1s} Y_{1c} = \frac{s_G \bar{B}}{B_0},
\end{align}
the $\sin\vartheta$ and $\cos\vartheta$ modes of (\ref{eq:modB}) give
\begin{align}
B_{1s} = \kappa X_{1s} B_0,
\hspace{0.5in}
B_{1c} = \kappa X_{1c} B_0,
\label{eq:B1}
\end{align}
and (\ref{eq:mixedPartials}) gives
\begin{align}
\label{eq:sigma}
 \iota_{N0} V_1 =  X_{1c} X'_{1s} - X_{1s}   X'_{1c}  
+  Y_{1c} Y'_{1s} - Y_{1s} Y'_{1c}
+2 \left( \frac{I_2}{\bar{B}} - \tau \right) \frac{ G_0\bar{B}}{ B_0^2},
\end{align}
where primes denote $d/d\varphi$ and
\begin{align}
V_1 = X_{1s}^2 + X_{1c}^2 + Y_{1s}^2 + Y_{1c}^2.
\label{eq:V1}
\end{align}
It is convenient to introduce $\sigma(\varphi) = (B_{1s} Y_{1s} + B_{1c} Y_{1c})/(s_G \bar{B} \kappa)$, in which case (\ref{eq:flux_area})-(\ref{eq:B1}) imply
\begin{align}
Y_{1s} = \frac{(B_{1c}+ B_{1s}\sigma)s_G \bar{B} \kappa}{B_{1s}^2 + B_{1c}^2},
\hspace{0.5in}
Y_{1c} = \frac{(-B_{1s}+ B_{1c}\sigma)s_G \bar{B} \kappa}{B_{1s}^2 + B_{1c}^2},
\end{align}
and (\ref{eq:sigma})-(\ref{eq:V1}) can be written
\begin{align}
\label{eq:sigma_general}
&
\sigma'
+ \left[ \frac{(B_{1s}^2+B_{1c}^2)^2}{B_0^2 \bar{B}^2 \kappa^4} + 1 + \sigma^2 \right]
\left[\iota_{N0} + \frac{B_{1s}  B'_{1c} - B_{1c} B'_{1s}}{B_{1s}^2 + B_{1c}^2} \right] 
\\
& \hspace{1.6in}
-2\left( \frac{I_2}{\bar{B}}-\tau\right) \frac{G_0 \left( B_{1s}^2 + B_{1c}^2\right)}{\bar{B} B_0^2 \kappa^2}=0.
\nonumber
\end{align}

Next, the $\vartheta$-independent, $\sin 2\vartheta$, and $\cos 2\vartheta$ modes of (\ref{eq:Zr}) give
\begin{align}
\label{eq:Z20}
Z_{20} &= \frac{\beta_0 \bar{B} \ell' }{2 G_0} 
-\frac{V'_1}{8 \ell'} ,
\\
\label{eq:Z2s}
Z_{2s} &= 
-\frac{1}{8 \ell'}  \left( V'_2 - 2 \iota_{N0} V_3\right),
\\
\label{eq:Z2c}
Z_{2c} &= 
-\frac{1}{8 \ell'}  \left( V'_3 + 2 \iota_{N0} V_2\right),
\end{align}
where
\begin{align}
V_2 &= 2 \left[ X_{1s} X_{1c} + Y_{1s} Y_{1c} \right], \\
V_3 &= X_{1c}^2 - X_{1s}^2 + Y_{1c}^2 - Y_{1s}^2.
\end{align}

At $O((r/\mathcal{R})^2)$, the $\sin\vartheta$ and $\cos\vartheta$ terms of (\ref{eq:XY}) are
\begin{align}
-\frac{\Gsign \bar{B}}{2 B_0}X_{1s} \kappa
&=
-X_{1s} Y_{2s} - X_{1c} Y_{2c} + X_{1c} Y_{20}
+X_{2s} Y_{1s} + X_{2c} Y_{1c} - X_{20} Y_{1c},
\label{eq:2nd_order_tangent_sin}
\\
-\frac{\Gsign \bar{B}}{2 B_0}X_{1c} \kappa
&=
-X_{1s} Y_{2c} + X_{1c} Y_{2s} - X_{1s} Y_{20}
+X_{2c} Y_{1s} - X_{2s} Y_{1c} + X_{20} Y_{1s}.
\label{eq:2nd_order_tangent_cos}
\end{align}
The $\vartheta$-independent, $\sin 2\vartheta$, and $\cos 2\vartheta$ modes of (\ref{eq:modB}) at $O((r/\mathcal{R})^2)$ give
\begin{align}
\label{eq:X20} 
X_{20} =&
\frac{1}{\kappa \ell'}
\left\{ Z'_{20}-\frac{1}{\ell'}\left[
-\frac{G_0^2 B_{20}}{B_0^3} + \frac{3 G_0^2(B_{1c}^2+B_{1s}^2)}{4 B_0^4}
 \right.\right. \\
&\hspace{0.7in}\left.\left.
+\frac{G_0(G_2+\iota_0 I_2)}{B_0^2}
-\frac{X_{1c}^2+X_{1s}^2}{4} \left( \kappa \ell' \right)^2 - \frac{q_c^2+q_s^2+r_c^2+r_s^2}{4}
\right]
\right\},
\nonumber
\\
\label{eq:X2s} 
X_{2s} =&
\frac{1}{\kappa \ell'}
\left\{ Z'_{2s} - 2 \iota_{N0} Z_{2c} -\frac{1}{\ell'}\left[
-\frac{G_0^2 B_{2s}}{B_0^3} + \frac{3 G_0^2 B_{1c} B_{1s}}{2 B_0^4}
 \right.\right. \\
&\hspace{2.3in}\left.\left.
-\frac{X_{1c} X_{1s}}{2} \left( \kappa \ell' \right)^2 - \frac{q_c q_s+r_c r_s}{2}
\right]
\right\},\nonumber \\
\label{eq:X2c} 
X_{2c} =&
\frac{1}{\kappa \ell'}
\left\{ Z'_{2c} + 2 \iota_{N0} Z_{2s} -\frac{1}{\ell'}\left[
-\frac{G_0^2 B_{2c}}{B_0^3} + \frac{3 G_0^2(B_{1c}^2-B_{1s}^2)}{4 B_0^4}
 \right.\right. \\
&\hspace{1.8in}\left.\left.
-\frac{X_{1c}^2-X_{1s}^2}{4} \left( \kappa \ell' \right)^2 - \frac{q_c^2-q_s^2+r_c^2-r_s^2}{4}
\right]
\right\},\nonumber
\end{align}
where
\begin{align}
q_s &= X'_{1s} - \iota_{N0} X_{1c} - Y_{1s} \tau \ell' , \\
q_c &= X'_{1c} + \iota_{N0} X_{1s} - Y_{1c} \tau \ell' , \\
r_s &= Y'_{1s} - \iota_{N0} Y_{1c} + X_{1s} \tau \ell' , \\
r_c &= Y'_{1c} + \iota_{N0} Y_{1s} + X_{1c} \tau \ell' .
\label{eq:rc}
\end{align}
The $\cos\vartheta$ and $\sin\vartheta$ terms of (\ref{eq:mixedPartials}) at $O((r/\mathcal{R})^2)$ are
\begin{align}
\label{eq:3rd_order_1}
-X_{1s}f_{X0}
+X_{1c}f_{Xs}
-X_{1s}f_{Xc}
-Y_{1s}f_{Y0}
+Y_{1c}f_{Ys}
-Y_{1s}f_{Yc}
=0
\end{align}
and
\begin{align}
\label{eq:3rd_order_2}
-X_{1c}f_{X0}
+X_{1s}f_{Xs}
+X_{1c}f_{Xc}
-Y_{1c}f_{Y0}
+Y_{1s}f_{Ys}
+Y_{1c}f_{Yc}
=0,
\end{align}
where
 \begin{align}
\label{eq:normal30}
f_{X0}=& X'_{20} - \tau \ell'  Y_{20} + \kappa  \ell'  Z_{20}
-\frac{4G_0}{\bar{B}}\left(Y_{2c} Z_{2s} - Y_{2s} Z_{2c} \right)
\\
&-\frac{I_2}{\bar{B}}\left(
\frac{\kappa}{2} [X_{1s} Y_{1s}+X_{1c} Y_{1c}] - 2 Y_{20} \right) \ell' 
- \frac{\beta_0 \kappa}{2}\ell'  \left( 
X_{1s}Y_{1c} - X_{1c} Y_{1s}  \right)
\nonumber \\
&-\frac{1}{2}\ell'  (\beta_{1c} Y_{1s} - \beta_{1s} Y_{1c}),
\nonumber
\end{align}
\begin{align}
\label{eq:normal3s}
f_{Xs}=& X'_{2s} - 2 \iota_{N0} X_{2c}- \tau \ell'  Y_{2s} + \kappa  \ell'  Z_{2s}
-\frac{4G_0}{\bar{B}}\left(
-Y_{20} Z_{2c} + Y_{2c} Z_{20}\right) 
\\
&-\frac{I_2}{\bar{B}}\left(
\frac{\kappa}{2} [X_{1s} Y_{1c}+X_{1c} Y_{1s}] - 2 Y_{2s} \right) 
\ell' 
-\beta_0\ell'   \left( -2 Y_{2c} + \frac{ \kappa}{2}[
X_{1c}Y_{1c} - X_{1s} Y_{1s}]  \right)
\nonumber \\
&-\frac{1}{2}\ell'  (\beta_{1s} Y_{1s} - \beta_{1c} Y_{1c}),
\nonumber
\end{align}
\begin{align}
\label{eq:normal3c}
f_{Xc} =& X'_{2c} + 2 \iota_{N0} X_{2s}- \tau \ell'  Y_{2c} + \kappa  \ell'  Z_{2c}
-\frac{4G_0}{\bar{B}}\left(
Y_{20} Z_{2s} - Y_{2s} Z_{20}\right)
\\
&-\frac{I_2}{\bar{B}}\left(
\frac{\kappa}{2} [X_{1c} Y_{1c}-X_{1s} Y_{1s}] - 2 Y_{2c} \right) 
\ell' 
- \beta_0\ell'   \left( 2 Y_{2s} - \frac{ \kappa}{2}[
X_{1c}Y_{1s} + X_{1s} Y_{1c}]  \right)
\nonumber \\
&-\frac{1}{2}\ell'  (\beta_{1c} Y_{1s} + \beta_{1s} Y_{1c}),
\nonumber
\end{align}
\begin{align}
\label{eq:binormal30}
f_{Y0}=& Y'_{20} + \tau \ell'  X_{20} 
-\frac{4G_0}{\bar{B}}\left(
X_{2s} Z_{2c} - X_{2c} Z_{2s} \right)
\\
&-\frac{I_2}{\bar{B}}\left(-
\frac{\kappa}{2} [X_{1s}^2+X_{1c}^2] + 2 X_{20} \right) 
\ell' 
-\frac{1}{2}\ell'  (\beta_{1s} X_{1c} - \beta_{1c} X_{1s}),
\nonumber
\end{align}
\begin{align}
\label{eq:binormal3s}
f_{Ys}=& Y'_{2s} - 2 \iota_{N0} Y_{2c}
+ \tau \ell'  X_{2s}
-\frac{4G_0}{\bar{B}}\left(
X_{20} Z_{2c} - X_{2c} Z_{20}\right)
\\
&-\frac{I_2}{\bar{B}}\left(
-\kappa X_{1s} X_{1c}+ 2 X_{2s} \right) 
\ell' 
-\beta_0\ell'   \left( 2 X_{2c} + \frac{ \kappa}{2}[
X_{1s}^2 - X_{1c}^2 ]  \right)
\nonumber \\
&-\frac{1}{2}\ell'  (\beta_{1c} X_{1c} - \beta_{1s} X_{1s}),
\nonumber
\end{align}
\begin{align}
\label{eq:binormal3c}
f_{Yc}=& Y'_{2c} + 2 \iota_{N0} Y_{2s}
+ \tau \ell'  X_{2c}
-\frac{4G_0}{\bar{B}}\left(
X_{2s} Z_{20} - X_{20} Z_{2s}\right)
\\
&-\frac{I_2}{\bar{B}}\left(
\frac{\kappa}{2}[ X_{1s}^2-X_{1c}^2]+ 2 X_{2c} \right) 
\ell' 
-\beta_0\ell'   \left( -2 X_{2s} + \kappa
X_{1s} X_{1c} \right)
\nonumber \\
&+\frac{1}{2}\ell'  (\beta_{1c} X_{1s} + \beta_{1s} X_{1c}).
\nonumber
 \end{align}
We will not need the $O((r/\mathcal{R})^2)$ terms of (\ref{eq:Zr}), which give $Z_3$.

Finally, for the analysis in section \ref{sec:finite_r}, we need the independent-of-$\vartheta$ mode
of (\ref{eq:XY}) at $O((r/\mathcal{R})^3)$, which gives
\begin{align}
\label{eq:XY3}
X_{3s1} Y_{1c} - X_{3c1}Y_{1s} + X_{1s} Y_{3c1} - X_{1c} Y_{3s1} = Q ,
\end{align}
with $Q$ given by (\ref{eq:Q}).

The averaged equilibrium condition (\ref{eq:MHD_avg}) gives
\begin{align}
G_2 + \iota_0 I_2 = -\frac{\mu_0 p_2 G_0}{2\pi} \int_0^{2\pi} \frac{d\varphi}{B_0^2}.
\label{eq:MHD_avg_leading}
\end{align}
The remaining equilibrium condition (\ref{eq:beta}) gives
\begin{align}
\beta'_0 = \frac{2 \mu_0 p_2 G_0}{\bar{B}} \left[ \frac{1}{B_0^2} - \frac{1}{2\pi} \int_0^{2\pi}\frac{d\varphi}{B_0^2}\right]
\label{eq:beta0}
\end{align}
at $O((r/\mathcal{R})^0)$, and 
\begin{align}
\label{eq:beta1}
\frac{\partial\beta_1}{\partial\varphi} +\iota_{N0} \frac{\partial \beta_1}{\partial\vartheta}= -\frac{4 \mu_0 p_2 G_0 B_1}{\bar{B} B_0^3}
\end{align} 
at $O((r/\mathcal{R})^1)$. 


\subsection{Reduction for quasisymmetry}
\label{sec:symmetry_reduction}

The equations of the previous section simplify slightly in the case of quasi-axisymmetric or quasi-helical symmetry. (We will not consider quasi-poloidal symmetry since it cannot exist at $O((r/\mathcal{R})^1)$.)
Since $B_0(\varphi)$ must be constant, it is convenient to take $\bar{B}=\psisign B_0$ where $\psisign = \mathrm{sign}(\psi)$.
That is, we take the reference magnetic field used to define effective minor radius equal in magnitude to the on-axis field.
Also, the averaged equilibrium condition (\ref{eq:MHD_avg_leading}) simplifies to
\begin{align}
G_2 =- \iota_0 I_2  - \mu_0 p_2 G_0 / B_0^2,
\label{eq:MHD_avg_QS}
\end{align}
and (\ref{eq:beta0}) gives $\beta'_0=0$. Without loss of generality we can then take $\beta_0=0$, since a shift to the origin of $\varphi$ shifts $\beta$ by a flux function.

The origin of the $\vartheta$ coordinate can be chosen such that $B_{1s}=0$, hence $X_{1s}=0$ from (\ref{eq:B1}).
Introducing the constant $\etabar=B_{1c}/B_0$, 
eq (\ref{eq:position_vector_r1}) follows, and (\ref{eq:sigma_general}) reduces to (\ref{eq:sigma_quasisymmetry}).
Examining the $\sin\vartheta$ and $\cos\vartheta$ components of (\ref{eq:beta1}), we find
$\beta_{1c}=\beta_{1c}^{(c)}\cos (\iota_{N0}\varphi) + \beta_{1c}^{(s)}\sin (\iota_{N0}\varphi)$ for some constants $ \beta_{1c}^{(c)}$ and $\beta_{1c}^{(s)}$. To avoid large magnetic islands near the axis, we assume $\iota_{N0}$ is not an integer, in which case the only periodic solution for $\beta_{1c}$ is  $\beta_{1c}=0$. Then (\ref{eq:beta1}) gives
\begin{align}
\beta_{1s}= -\frac{4 \psisign \mu_0 p_2 G_0 \bar{\eta}}{\iota_{N0} B_0^3}.
\label{eq:beta1s}
\end{align} 

Then, noting $\beta_0 = \beta_{1c}=X_{1s}=0$, (\ref{eq:G0})-(\ref{eq:beta1s}) are equivalent to the equations in the appendix of \cite{GB2}, up to the following differences. The sign of $\tau$ is everywhere flipped due to the opposite sign convention. Terms $\propto I_2$ are omitted in (A29)-(A34) of \cite{GB2}. Several expressions differ by factors of $\sqrt{2}$ or 2  since our expansion parameter $r/\mathcal{R}$ differs from the one in \cite{GB2} by $\sqrt{2}$. A $+$ sign is missing in (A10) of \cite{GB2}. A factor of 2 is missing in each of the terms $\propto (\iota_0-N)$ in (A30)-(A31) and (A33)-(A34) of \cite{GB2}, and the left hand side of (A34) should read $f_{y,2c}(\varphi)$.

Finally, $Q$ in (\ref{eq:Q}) and (\ref{eq:XY3}) simplifies to
\begin{align}
\label{eq:Q_QS}
Q(\varphi) =& -\frac{s_\psi B_0}{2G_0^2} \ell'  \left(\iota_{N0} I_2 + \frac{\mu_0 p_2 G_0}{B_0^2}\right)
  +2(X_{2c} Y_{2s} - X_{2s} Y_{2c}) \\
& + \frac{s_\psi B_0 }{2G_0} \left( \ell'  X_{20} \kappa - Z'_{20}\right) 
 +\frac{I_2}{4G_0} \left(  -\ell'   \tau V_1 + Y_{1c} X'_{1c} - X_{1c}Y'_{1c} \right).
 \nonumber
\end{align}


\section{Effect of a finite value of the expansion parameter}
\label{sec:finite_r_appendix}

\subsection{Preliminaries}

In this section, a detailed derivation is given of (\ref{eq:constructedB})-(\ref{eq:BHat}).
Quasisymmetry is not assumed, so the analysis here applies equally well if the Garren-Boozer equations are used to construct a geometry possessing omnigenity
or some other desired pattern of field strength.

We first complete the formulation of the problem.
The profile functions $I(r)$ and $p(r)$ are
assumed to be identical in the tilde and non-tilde configurations, since these profiles are typically inputs to an MHD equilibrium calculation, so in a finite-minor-radius calculation they can be matched exactly to the ideal (non-tilde) profiles. However, we should allow the profiles $G(r)$ and $\iota(r)$ to differ in the tilde configurations, writing
\begin{align}
\tilde{G}(a,r) = \sum_{j=0}^{\infty} r^{2j} \tilde{G}_{2j}(a),
\hspace{0.5in}
\tilde{\iota}(a,r) = \sum_{j=0}^{\infty} r^{2j} \tilde{\iota}_{2j}(a),
\end{align}
where
\begin{align}
\tilde{G}_j(a) = \sum_{k=0}^{\infty} a^{k} \tilde{G}_{j}^{(k)},
&\hspace{0.5in}
\tilde{\iota}_j(a) = \sum_{k=0}^{\infty} a^{k} \tilde{\iota}_{j}^{(a)}.
\end{align}
Finally, we are free to add a constant to the angles $(\tilde\vartheta,\tilde\varphi)$, and it is convenient to eliminate this degeneracy by requiring that the angle differences vanish on average:
\begin{align}
\label{eq:angle_average_0}
\int_0^{2\pi}d\vartheta \int_0^{2\pi}d\varphi\; t(a,\vartheta,\varphi) = 0,
\hspace{0.5in}
\int_0^{2\pi}d\vartheta \int_0^{2\pi}d\varphi\; p(a,\vartheta,\varphi) = 0.
\end{align}


\subsection{$O(r/\mathcal{R})$ construction}

We begin with the $O((r/\mathcal{R})^0)$ terms of (\ref{eq:positionVector_tilde}),
\begin{align}
\vect{r}_0(\varphi) = \tilde{\vect{r}}_0^{(0)}\left(\varphi + f^{(0)}(\vartheta,\varphi)\right),
\label{eq:x0}
\end{align}
which implies $f^{(0)}(\vartheta,\varphi) = f^{(0)}(\varphi)$.
Applying $d/d\varphi$ to (\ref{eq:x0}) and applying it to the $O((r/\mathcal{R})^0)$ Frenet relations of the tilde configuration, one finds $\tilde{\vect{t}}^{(0)}$, $\tilde{\vect{n}}^{(0)}$, $\tilde{\vect{b}}^{(0)}$, $\kappa^{(0)}$, and $\tau^{(0)}$ match the corresponding non-tilde quantities, e.g. $\vect{n}(\varphi) = \tilde{\vect{n}}^{(0)}(\bar\varphi)$ where $\bar\varphi = \varphi + f^{(0)}(\varphi)$.

Proceeding to the $O(r/\mathcal{R})$ terms of  (\ref{eq:positionVector_tilde}),
\begin{align}
\label{eq:a1}
&\left[X_{1s}(\varphi)\sin\vartheta + X_{1c}(\varphi)\cos\vartheta\right]\vect{n}(\varphi)
+
\left[Y_{1s}(\varphi)\sin\vartheta + Y_{1c}(\varphi)\cos\vartheta\right]\vect{b}(\varphi) \\
&=
\tilde{\vect{r}}_0^{(1)}\left(\bar\varphi\right)
+f^{(1)}(\vartheta,\varphi) \left[ 1+ f^{(0)\prime}(\varphi)\right]^{-1}\vect{t}(\varphi)
\left[\vect{r}'_0(\varphi) \cdot \vect{r}'_0(\varphi)\right]^{1/2} \nonumber \\
&+\left[ \tilde{X}_{1s}^{(0)}\left(\bar\varphi\right) \sin\left( \vartheta + t^{(0)}(\vartheta,\varphi)\right)
+\tilde{X}_{1c}^{(0)}\left(\bar\varphi\right) \cos\left( \vartheta + t^{(0)}(\vartheta,\varphi)\right) \right] \vect{n}(\varphi)
 \nonumber \\
&+\left[ \tilde{Y}_{1s}^{(0)}\left(\bar\varphi\right) \sin\left( \vartheta + t^{(0)}(\vartheta,\varphi)\right)
+\tilde{Y}_{1c}^{(0)}\left(\bar\varphi\right) \cos\left( \vartheta + t^{(0)}(\vartheta,\varphi)\right) \right] \vect{b}(\varphi).
\nonumber
\end{align}
It can be shown from either the $\vect{n}$ or $\vect{b}$ component that $t^{(0)}(\vartheta,\varphi) = t^{(0)}(\varphi)$. This can be done by applying $\partial/\partial\vartheta$ to the $\vect{n}$ component, squaring the result, adding the square of the $\vect{n}$ component, and eliminating $t^{(0)}$ where it is not differentiated. Evaluating the result at $\vartheta = \mathrm{atan}(X_{1s}/X_{1c})$ and adding or subtracting the result at $\vartheta = \pi + \mathrm{atan}(X_{1s}/X_{1c})$, one finds $\partial t^{(0)}(\vartheta,\varphi) / \partial\vartheta = 0$.

Next, the $\vect{t}$ component of (\ref{eq:a1}) implies $f^{(1)}(\vartheta,\varphi) = f^{(1)}(\varphi)$. The average of (\ref{eq:a1}) over $\vartheta$ then gives
\begin{align}
\label{eq:r01}
\tilde{\vect{r}}_0^{(1)}\left(\bar\varphi\right)
+f^{(1)}(\varphi) \left[ 1+ f^{(0)\prime}(\varphi)\right]^{-1}\vect{t}(\varphi)
\left[\vect{r}'_0(\varphi) \cdot \vect{r}'_0(\varphi)\right]^{1/2} 
=0.
\end{align}
The $\vect{n}$ component of (\ref{eq:a1}) gives
\begin{align}
\label{eq:X_rotation}
\begin{pmatrix} \tilde{X}_{1s}^{(0)}(\bar\varphi) \\ \tilde{X}_{1c}^{(0)}(\bar\varphi)
\end{pmatrix}
=\begin{pmatrix} \cos\left( t^{(0)}(\varphi)\right) & \sin\left( t^{(0)}(\varphi)\right) \\
-\sin\left( t^{(0)}(\varphi)\right) & \cos\left( t^{(0)}(\varphi)\right)
\end{pmatrix}
\begin{pmatrix} X_{1s}(\varphi) \\ X_{1c}(\varphi)
\end{pmatrix},
\end{align}
and the $\vect{b}$ component of (\ref{eq:a1}) gives the same result but with $X \to Y$.
Plugging these results into the $O((r/\mathcal{R})^0)$ terms in the tilde version of (\ref{eq:flux_area}),
\begin{align}
\tilde{X}_{1c}^{(0)}(\bar\varphi) \tilde{Y}_{1s}^{(0)}(\bar\varphi) 
-\tilde{X}_{1s}^{(0)}(\bar\varphi) \tilde{Y}_{1c}^{(0)}(\bar\varphi) 
=\frac{s_G \bar{B}}{ \tilde{B}_0^{(0)}(\bar\varphi)},
\end{align}
and comparing to the non-tilde version of (\ref{eq:flux_area}),
we conclude $\tilde{B}_0^{(0)}(\bar\varphi) = B_0(\varphi)$.
Applying this result and the derivative of (\ref{eq:x0}) in the $O((r/\mathcal{R})^0)$ tilde version of (\ref{eq:G0}),
\begin{align}
\tilde{G}_0^{(0)} = s_G \tilde{B}_0^{(0)}(\bar\varphi)
\left[ \tilde{\vect{r}}_0^{(0)\prime}(\bar\varphi) \cdot \tilde{\vect{r}}_0^{(0)\prime}(\bar\varphi) \right]^{1/2},
\end{align}
we obtain $\tilde{G}_0^{(0)} = G_0 / \left[ 1+ f^{(0)\prime}(\varphi)\right]$,
which implies $f^{(0)\prime}(\varphi)=0$. From (\ref{eq:angle_average_0}), then $f^{(0)}=0$, so $\tilde{G}_0^{(0)}=G_0$. Since $\bar\varphi=\varphi$, we can simplify notation in the remainder of this Appendix: functions of a single argument can be assumed to have argument $\varphi$ .

Next, $t^{(0)}$ can be constrained using the $O((r/\mathcal{R})^0)$ terms in the tilde version of (\ref{eq:sigma}):
\begin{align}
& (\tilde\iota_0^{(0)} - N) \left[
\tilde{X}_{1s}^{(0)2}
+\tilde{X}_{1c}^{(0)2}
+\tilde{Y}_{1s}^{(0)2}
+\tilde{Y}_{1c}^{(0)2}
\right]
\\
&=  \tilde{X}^{(0)}_{1c}   \tilde{X}^{(0)\prime}_{1s}
 - \tilde{X}^{(0)}_{1s} \tilde{X}^{(0)\prime}_{1c}  
+  \tilde{Y}^{(0)}_{1c} \tilde{Y}^{(0)\prime}_{1s}
- \tilde{Y}^{(0)}_{1s}  \tilde{Y}^{(0)\prime}_{1c}
+2 \left( \frac{I_2}{\bar{B}} - \tilde{\tau}^{(0)} \right) \frac{ G_0\bar{B}}{ B_0^2}.\nonumber
\nonumber
\end{align}
Substituting (\ref{eq:X_rotation})  (and its $X \to Y$ equivalent) and subtracting the non-tilde version of (\ref{eq:sigma}), one finds $\tilde\iota_0^{(0)} - \iota_0 = t^{(0)\prime}(\varphi)$,
which implies $\tilde\iota_0^{(0)} = \iota_0$ and $t^{(0)\prime}(\varphi)=0$.
From (\ref{eq:angle_average_0}), then $t^{(0)}=0$. Then
(\ref{eq:X_rotation}) gives
\begin{align}
\tilde{X}_{1s}^{(0)} = X_{1s},
\hspace{0.2in}
\tilde{X}_{1c}^{(0)} = X_{1c},
\hspace{0.2in}
\tilde{Y}_{1s}^{(0)} = Y_{1s},
\hspace{0.2in}
\tilde{Y}_{1c}^{(0)} = Y_{1c}.
\end{align}
The $O((r/\mathcal{R})^0)$ terms in the tilde version of (\ref{eq:B1}) then give $\tilde{B}_1^{(0)}(\varphi) = B_1(\varphi)$.

We proceed to the $O((r/\mathcal{R})^2)$ terms in (\ref{eq:positionVector_tilde}):
\begin{align}
\label{eq:a2}
&X_2(\vartheta,\varphi) \vect{n} + Y_2(\vartheta,\varphi)\vect{b}+ Z_2(\vartheta,\varphi)\vect{t} \\
&= \tilde{\vect{r}}_0^{(2)}
+f^{(1)} \tilde{\vect{r}}_0^{(1)\prime}
+f^{(2)}(\vartheta,\varphi) \vect{r}'_0
+\frac{1}{2} f^{(1)2} \vect{r}''_0 \nonumber\\
&+\tilde{X}_2^{(0)}(\vartheta,\varphi)\vect{n}
+\tilde{X}_1^{(1)}(\vartheta,\varphi)\vect{n}
+X_1(\vartheta,\varphi)\tilde{\vect{n}}^{(1)}
+f^{(1)}X_1(\vartheta,\varphi)\vect{n}'
\nonumber \\
&+t^{(1)}(\vartheta,\varphi) \vect{n} \partial_1 X_1(\vartheta,\varphi)
+f^{(1)} \vect{n} \partial_2 X_1(\vartheta,\varphi)
\nonumber \\
&+\tilde{Y}_2^{(0)}(\vartheta,\varphi)\vect{b}
+\tilde{Y}_1^{(1)}(\vartheta,\varphi)\vect{b}
+Y_1(\vartheta,\varphi)\tilde{\vect{b}}^{(1)}
+f^{(1)}Y_1(\vartheta,\varphi)\vect{b}'
\nonumber \\
&+t^{(1)}(\vartheta,\varphi) \vect{b} \partial_1 Y_1(\vartheta,\varphi)
+f^{(1)} \vect{b} \partial_2 Y_1(\vartheta,\varphi)
+\tilde{Z}_2^{(0)}(\vartheta,\varphi)\vect{t}
\nonumber 
,
\end{align}
where $\partial_1$ and $\partial_2$ indicate partial derivatives with respect to the first or second argument.
(If the construction is done only through $O((r/\mathcal{R})^1)$, the left-hand side is zero.)
Applying $\pi^{-1} \int_0^{2\pi}d\vartheta (\sin\vartheta)(\ldots)$ and $\pi^{-1} \int_0^{2\pi}d\vartheta (\cos\vartheta)(\ldots)$ to the $\vect{n}$ component,
\begin{align}
\label{eq:X1s1}
\tilde{X}_{1s}^{(1)}
=& - X_{1s} t_{sc}^{(1)} + X_{1c} t_{ss}^{(1)} - f^{(1)} X'_{1s} 
-Y_{1s}\vect{n}\cdot \tilde{\vect{b}}^{(1)} - f^{(1)} Y_{1s} \vect{n} \cdot \vect{b}'
, \\
\tilde{X}_{1c}^{(1)}
=& - X_{1s} t_{cc}^{(1)} + X_{1c} t_{sc}^{(1)} - f^{(1)} X'_{1c} 
-Y_{1c}\vect{n}\cdot \tilde{\vect{b}}^{(1)} - f^{(1)} Y_{1c} \vect{n} \cdot \vect{b}',
\end{align}
where
\begin{align}
\label{eq:tss}
t_{ss}^{(1)}(\varphi) =& \pi^{-1} \int_0^{2\pi}d\vartheta \, t^{(1)}(\vartheta,\varphi) \sin^2\vartheta, \\
t_{sc}^{(1)}(\varphi) =& \pi^{-1} \int_0^{2\pi}d\vartheta \, t^{(1)}(\vartheta,\varphi) \sin\vartheta \cos\vartheta, \\
t_{cc}^{(1)}(\varphi) =& \pi^{-1} \int_0^{2\pi}d\vartheta \, t^{(1)}(\vartheta,\varphi) \cos^2\vartheta.
\label{eq:tcc}
\end{align}
We have used $\vect{n} \cdot \tilde{\vect{n}}^{(1)}=0$ since this is the $O(r/\mathcal{R})$ term in $| \tilde{\vect{n}}|=1$.
Similarly, from the $\vect{b}$ component of (\ref{eq:a2}),
\begin{align}
\tilde{Y}_{1s}^{(1)}
=& - Y_{1s} t_{sc}^{(1)} + Y_{1c} t_{ss}^{(1)} - f^{(1)} Y'_{1s} 
-X_{1s}\vect{b}\cdot \tilde{\vect{n}}^{(1)} - f^{(1)} X_{1s} \vect{b} \cdot \vect{n}'
, \\
\label{eq:Y1c1}
\tilde{Y}_{1c}^{(1)}
=& - Y_{1s} t_{cc}^{(1)} + Y_{1c} t_{sc}^{(1)} - f^{(1)} Y'_{1c} 
-X_{1c}\vect{b}\cdot \tilde{\vect{n}}^{(1)} - f^{(1)} X_{1c} \vect{b} \cdot \vect{n}'
.
\end{align}
Equations (\ref{eq:X1s1})-(\ref{eq:Y1c1}) are substituted into the $O(r/\mathcal{R})$ terms in the tilde version of (\ref{eq:flux_area}),
\begin{align}
\label{eq:flux_area_1}
X_{1c} \tilde{Y}_{1s}^{(1)} + \tilde{X}_{1c}^{(1)} Y_{1s}
-X_{1s} \tilde{Y}_{1c}^{(1)} - \tilde{X}_{1s}^{(1)} Y_{1c}
=-\frac{s_G \bar{B} \tilde{B}_0^{(1)}}{B_0^2}, 
\end{align}
which after many cancellations gives
\begin{align}
s_G \bar{B} \tilde{B}_0^{(1)} / B_0^2
=f^{(1)}  \left[ X_{1c} Y_{1s} - X_{1s} Y_{1c} \right]'.
\end{align}
From (\ref{eq:flux_area}), 
\begin{align}
\label{eq:B01}
\tilde{B}_0^{(1)} = -f^{(1)} B'_0.
\end{align}
To determine $f^{(1)}$, this result and (\ref{eq:r01}) are substituted into the $O((r/\mathcal{R})^1)$ terms of the tilde version of (\ref{eq:G0}):
\begin{align}
\label{eq:G01}
\tilde{G}_0^{(1)}
=s_G \tilde{B}_0^{(1)} \ell'  + \frac{s_G B_0}{\ell' }
\vect{r}'_0 \cdot \tilde{\vect{r}}_0^{(1)\prime}.
\end{align}
As a result we find
$\tilde{G}_0^{(1)} =  - f^{(1)\prime}(\varphi) G_0 $. Since $f^{(1)}(\varphi)$ is single-valued, $\tilde{G}_0^{(1)}=0$ and $f^{(1)\prime}(\varphi)=0$. Then by (\ref{eq:angle_average_0}), $f^{(1)}=0$. Then (\ref{eq:B01}) gives $\tilde{B}_0^{(1)}(\varphi)=0$.

At this point, we have proved the first set of assertions following (\ref{eq:constructedB}): when a finite $r=a$
 is plugged into a solution of the $O((r/\mathcal{R})^1)$ Garren-Boozer equations, the real finite-minor-radius MHD equilibrium inside
 the constructed boundary has the desired magnetic field as a function of Boozer coordinates through $O(r/\mathcal{R})$. 
 

\subsection{$O((r/\mathcal{R})^2)$ construction}

We now proceed to evaluate $\tilde{B}(a,r,\tilde\vartheta,\tilde\varphi)$ through $O((r/\mathcal{R})^2)$.
First, $f^{(1)}=0$ and (\ref{eq:r01}) imply $\tilde{\vect{r}}_0^{(1)}=0$. The $O(r/\mathcal{R})$ terms in the Frenet relations for the tilde configuration
then imply $\tilde{\vect{t}}^{(1)} = \tilde{\vect{n}}^{(1)} = \tilde{\vect{b}}^{(1)} =  \kappa^{(1)}=\tau^{(1)}=0$.

At this point, we assume the construction is done through at least $O((r/\mathcal{R})^2)$.
Since $Z_2$ is given by a unique function of $X_1$ and $Y_1$ by (\ref{eq:Z20})-(\ref{eq:Z2c}), then $\tilde{Z}_2^{(0)}(\vartheta,\varphi) = Z_2(\vartheta,\varphi)$.
Then the $\vect{t}$ component of (\ref{eq:a2}) implies $f^{(2)}(\vartheta,\varphi) = f^{(2)}(\varphi)$. 

The  terms remaining in the $\vect{n}$ component of (\ref{eq:a2}) are
\begin{align}
\label{eq:a2_normal}
X_2(\vartheta,\varphi) = &\vect{n}\cdot \tilde{\vect{r}}_0^{(2)}
+\tilde{X}_2^{(0)}(\vartheta,\varphi) 
+\tilde{X}_1^{(1)}(\vartheta,\varphi) 
+ t^{(1)}(\vartheta,\varphi) \left[X_{1s}\cos\vartheta - X_{1c} \sin\vartheta\right].
\end{align}
We are free to define 
\begin{align}
\label{eq:tbar}
\bar{t}(\vartheta,\varphi)
=& t^{(1)}(\vartheta,\varphi)
-\frac{2}{X_{1s}^2+X_{1c}^2} \left[ \left(X_{2s}- \tilde{X}_{2s}^{(0)}\right)X_{1s}
+\left( X_{2c} - \tilde{X}_{2c}^{(0)}\right) X_{1c}\right] \sin\vartheta \nonumber\\
&-\frac{2}{X_{1s}^2+X_{1c}^2} \left[ \left(X_{2c}- \tilde{X}_{2c}^{(0)}\right)X_{1s}
-\left( X_{2s} - \tilde{X}_{2s}^{(0)}\right) X_{1c}\right] \cos\vartheta.
\end{align}
Then (\ref{eq:a2_normal}) can be written
\begin{align}
\label{eq:C}
0=C(\varphi) + \tilde{X}_{1s}^{(1)}\sin\vartheta + \tilde{X}_{1c}^{(1)}\cos\vartheta
+\bar{t}(\vartheta,\varphi)\left[X_{1s}\cos\vartheta - X_{1c} \sin\vartheta\right]
\end{align}
for a $\vartheta$-independent function $C(\varphi)$.
Evaluating (\ref{eq:C}) at $\vartheta = \mathrm{atan}(X_{1s}/X_{1c})$,
and adding or subtracting (\ref{eq:C}) at $\vartheta = \pi + \mathrm{atan}(X_{1s}/X_{1c})$, we find $C=0$ and $\tilde{X}_{1s}^{(1)} X_{1s} + \tilde{Y}_{1c}^{(1)} Y_{1c}=0$.
Then (\ref{eq:C}) implies $\tilde{X}_{1s}^{(1)} - \bar{t}(\vartheta,\varphi) X_{1c}=0$, so $\bar{t}(\vartheta,\varphi)=\bar{t}(\varphi)$.

Repeating the analysis from (\ref{eq:a2_normal}) with the $\vect{b}$ component
of (\ref{eq:a2}), we find (\ref{eq:tbar}) with $X \to Y$. Comparing the $\sin\vartheta$ and $\cos\vartheta$ modes of this result with those of (\ref{eq:tbar}), then
\begin{align}
\frac{\xi_{s} X_{1s}+\xi_c X_{1c} }{X_{1s}^2 + X_{1c}^2}
=&
\frac{\gamma_{s} Y_{1s}+\gamma_c Y_{1c} }{Y_{1s}^2 + Y_{1c}^2},\\
\frac{\xi_{c} X_{1s}-\xi_s X_{1c} }{X_{1s}^2 + X_{1c}^2}
=&
\frac{\gamma_{c} Y_{1s}-\gamma_s Y_{1c} }{Y_{1s}^2 + Y_{1c}^2},
\end{align}
where $\xi_s = X_{2s} - \tilde{X}_{2s}^{(0)}$, 
$\xi_c = X_{2c} - \tilde{X}_{2c}^{(0)}$, 
$\gamma_s = Y_{2s} - \tilde{Y}_{2s}^{(0)}$, and
$\gamma_c = Y_{2c} - \tilde{Y}_{2c}^{(0)}$.
We obtain four other linear homogeneous equations relating $\{\xi_s, \xi_c, \gamma_s,\gamma_c\}$ by taking the non-tilde versions of (\ref{eq:3rd_order_1})-(\ref{eq:3rd_order_2}) and (\ref{eq:2nd_order_tangent_sin})-(\ref{eq:2nd_order_tangent_cos}), and subtracting the $O((r/\mathcal{R})^0)$ tilde versions.
These four equations also involve $\xi_0=X_{20} - \tilde{X}_{20}^{(0)}$ and $\gamma_0=Y_{20} - \tilde{Y}_{20}^{(0)}$. We thus have six linear homogeneous equations relating the six unknowns $\{\xi_0,\xi_s, \xi_c, \gamma_0,\gamma_s,\gamma_c\}$. A valid solution is the one in which all six quantities vanish. The six equations are generally linearly independent, and so this is the unique solution.
Hence, $\tilde{X}_2^{(0)}(\vartheta,\varphi) = X_2(\vartheta,\varphi)$ and $\tilde{Y}_2^{(0)}(\vartheta,\varphi) = Y_2(\vartheta,\varphi)$. Comparing the non-tilde and the $O((r/\mathcal{R})^0)$ tilde versions of (\ref{eq:X20})-(\ref{eq:X2c}), we conclude $\tilde{B}_{2}^{(0)}(\vartheta,\varphi) = B_2(\vartheta,\varphi)$. This completes the evaluation of one more term in (\ref{eq:constructedB}).

Knowing now that $t^{(1)} (\vartheta,\varphi)= t^{(1)}(\varphi)$, $f^{(1)}=0$,
$\tilde{\vect{n}}^{(1)}=0$, and $\tilde{\vect{b}}^{(1)}=0$, we return to (\ref{eq:X1s1})-(\ref{eq:Y1c1}), which become
\begin{align}
\tilde{X}_{1s}^{(1)} = t^{(1)} X_{1c},&
\hspace{0.5in}
\tilde{X}_{1c}^{(1)} = - t^{(1)} X_{1s}, \\
\tilde{Y}_{1s}^{(1)} = t^{(1)} Y_{1c},&
\hspace{0.5in}
\tilde{Y}_{1c}^{(1)} = - t^{(1)} Y_{1s}. \nonumber
\end{align}
These expressions are substituted into the $O(r/\mathcal{R})$ terms in the tilde version of (\ref{eq:sigma}):
\begin{align}
&\tilde{\iota}_0^{(1)} \left[
X_{1s}^2+X_{1c}^2 + Y_{1s}^2 + Y_{1c}^2 \right]
+2 \iota_0 \left[
X_{1s} \tilde{X}_{1s}^{(1)} + X_{1c} \tilde{X}_{1c}^{(1)}
+Y_{1s} \tilde{Y}_{1s}^{(1)} + Y_{1c} \tilde{Y}_{1c}^{(1)} \right] 
\\
&= X_{1c} \tilde{X}_{1s}^{(1)\prime} - X_{1s} \tilde{X}_{1c}^{(1)\prime}
+Y_{1c} \tilde{Y}_{1s}^{(1)\prime} - Y_{1s} \tilde{Y}_{1c}^{(1)\prime}
\nonumber \\
&+\tilde{X}_{1c}^{(1)} X'_{1s} - \tilde{X}_{1s}^{(1)} X'_{1c}
+\tilde{Y}_{1c}^{(1)} Y'_{1s} - \tilde{Y}_{1s}^{(1)} Y'_{1c}.
\nonumber
\end{align}
The result is $\tilde{\iota}_0^{(1)} = t^{(1)\prime}(\varphi)$, implying $\tilde{\iota}_0^{(1)} =0$ and 
$t^{(1)}(\varphi)=$constant. From (\ref{eq:angle_average_0}), then, $t^{(1)}=0$,
so $\tilde{X}_{1s}^{(1)} = \tilde{X}_{1c}^{(1)} = \tilde{Y}_{1s}^{(1)} = \tilde{Y}_{1c}^{(1)} = 0$.
The $O(r/\mathcal{R})$ terms in (\ref{eq:B1}) can now be evaluated to give one more term we need in (\ref{eq:constructedB}):
$\tilde{B}_1^{(1)}(\vartheta,\varphi) = 0$.
The only remaining terms in (\ref{eq:a2}) give 
\begin{align}
\label{eq:r02}
\tilde{\vect{r}}_0^{(2)} = -f^{(2)} \vect{r}'_0.
\end{align}

We proceed to the $O((r/\mathcal{R})^3)$ terms in (\ref{eq:positionVector_tilde}):
\begin{align}
&X_3(\vartheta,\varphi)\vect{n} + Y_3(\vartheta,\varphi)\vect{b} + Z_3(\vartheta,\varphi)\vect{t}
=\tilde{\vect{r}}_0^{(3)} + f^{(3)}(\vartheta,\varphi) \vect{r}'_0 
+ \vect{t} \tilde{Z}_3^{(0)}(\vartheta,\varphi)
+ \vect{t} \tilde{Z}_2^{(1)}(\vartheta,\varphi) \nonumber \\
&+\vect{n} \tilde{X}_3^{(0)}(\vartheta,\varphi) + \tilde{\vect{n}}^{(2)}X_1(\vartheta,\varphi)
+\vect{n}\tilde{X}_1^{(2)}(\vartheta,\varphi)+\vect{n}\tilde{X}_2^{(1)}(\vartheta,\varphi)
\nonumber \\
&+\vect{b} \tilde{Y}_3^{(0)}(\vartheta,\varphi) + \tilde{\vect{b}}^{(2)}Y_1(\vartheta,\varphi)
+\vect{b}\tilde{Y}_1^{(2)}(\vartheta,\varphi)+\vect{b}\tilde{Y}_2^{(1)}(\vartheta,\varphi)
\nonumber \\
&+f^{(2)} Y_1(\vartheta,\varphi) \vect{b}' + f^{(2)}X_1(\vartheta,\varphi) \vect{n}' 
+\vect{n} f^{(2)} \partial_2 X_1(\vartheta,\varphi) + \vect{b} f^{(2)} \partial_2 Y_1 (\vartheta,\varphi)
\nonumber \\
&+ t^{(2)}(\vartheta,\varphi) \vect{n} \partial_1 X_1(\vartheta,\varphi) + t^{(2)}(\vartheta,\varphi) \vect{b} \partial_1 Y_1 (\vartheta,\varphi).
\label{eq:a3}
\end{align}
We take the $\vect{n}$ component, noting $\vect{n}\cdot\vect{n}'=0$ and $\vect{n}\cdot \tilde{\vect{n}}^{(2)}=0$,
since the latter is the $O((r/\mathcal{R})^2)$ term in $|\tilde{\vect{n}}|^2=1$. The $\sin\vartheta$ and $\cos\vartheta$ modes of the result are
\begin{align}
\label{eq:X1s2}
\tilde{X}_{1s}^{(2)} = X_{3s1} - \tilde{X}_{3s1}^{(0)} - \vect{n}\cdot\tilde{\vect{b}}^{(2)}Y_{1s}
-f^{(2)}Y_{1s} \vect{n}\cdot\vect{b}' - f^{(2)} X'_{1s} - X_{1s} t_{sc}^{(2)} + X_{1c} t_{ss}^{(2)},
\\
\tilde{X}_{1c}^{(2)} = X_{3c1} - \tilde{X}_{3c1}^{(0)} - \vect{n}\cdot\tilde{\vect{b}}^{(2)}Y_{1c}
-f^{(2)}Y_{1c} \vect{n}\cdot\vect{b}' - f^{(2)} X'_{1c} - X_{1s} t_{cc}^{(2)}+ X_{1c} t_{sc}^{(2)}, \nonumber
\end{align}
where $t_{ss}^{(2)}(\varphi)$, $t_{sc}^{(2)}(\varphi)$, and $t_{cc}^{(2)}(\varphi)$ are defined exactly as in (\ref{eq:tss})-(\ref{eq:tcc})
but with $t^{(1)} \to t^{(2)}$.
Similarly, the $\sin\vartheta$ and $\cos\vartheta$ modes of the $\vect{b}$ component of (\ref{eq:a3}) are
\begin{align}
\label{eq:Y1s2}
\tilde{Y}_{1s}^{(2)} = Y_{3s1} - \tilde{Y}_{3s1}^{(0)} - \vect{b}\cdot\tilde{\vect{n}}^{(2)}X_{1s}
-f^{(2)}X_{1s} \vect{b}\cdot\vect{n}' - f^{(2)} Y'_{1s} - Y_{1s} t_{sc}^{(2)} + Y_{1c} t_{ss}^{(2)},
\\
\tilde{Y}_{1c}^{(2)} = Y_{3c1} - \tilde{Y}_{3c1}^{(0)} - \vect{b}\cdot\tilde{\vect{n}}^{(2)}X_{1c}
-f^{(2)}X_{1c} \vect{b}\cdot\vect{n}' - f^{(2)} Y'_{1c} - Y_{1s} t_{cc}^{(2)} + Y_{1c} t_{sc}^{(2)}. \nonumber
\end{align}
Note that even if the expansion for the construction is truncated such that $X_3=Y_3=0$, generally $\tilde{X}_3$ and $\tilde{Y}_3$
will be nonzero since the tilde expansion always includes all orders in $r/\mathcal{R}$.
Equations (\ref{eq:X1s2})-(\ref{eq:Y1s2}) are substituted into the $O((r/\mathcal{R})^2)$ terms of (\ref{eq:flux_area}) ((\ref{eq:flux_area_1}) with $^{(1)} \to \,^{(2)}$),
using the fact that (\ref{eq:XY3}) is satisfied in the tilde configuration.
The result is (\ref{eq:B02})-(\ref{eq:Q}).
Then, we consider the $O((r/\mathcal{R})^2)$ terms of the tilde version of (\ref{eq:G0}), which give (\ref{eq:G01}) with $^{(1)} \to \,^{(2)}$.
Applying (\ref{eq:r02}), we find $\tilde{G}_0^{(2)} = G_0\hat{B} - f^{(2)\prime} G_0$.
Averaging over $\varphi$ gives $\tilde{G}_0^{(2)}$, which in turn gives (\ref{eq:p2}).


\section{Conversion to cylindrical coordinates}
\label{sec:transformation}


In this Appendix, we derive a method by which a magnetic surface described by (\ref{eq:positionVector}) can be
converted to a representation in cylindrical coordinates, as needed to specify input to some equilibrium codes such as VMEC.
The method here is based on relating $\varphi$ to the standard toroidal angle $\phi$ to the requisite accuracy in the $r/\mathcal{R}$ expansion.
Compared to direct evaluation of (\ref{eq:positionVector}), which requires solution of a nonlinear root-finding problem at each point on the surface to find $\varphi(\phi,\vartheta)$, the method here requires only application of linear operations to a solution $\{X_1,Y_1,X_2,Y_2,Z_2,X_3,Y_3\}$.

In cylindrical coordinates $(R,\phi,z)$ we can write the position vector $\vect{r}$ as
\begin{align}
\vect{r} = R(r,\vartheta,\phi) \vect{e}_R(\phi) + z(r,\vartheta,\phi) \vect{e}_z.
\label{eq:cylindrical_representation}
\end{align}
We are free to continue to use the helical Boozer angle $\vartheta$ to parameterize the surfaces.
Then $R(r,\vartheta,\phi)$ and $z(r,\vartheta,\phi)$ are expanded in the same way as $B$ and $\beta$,
except with $\varphi \to \phi$:
\begin{align}
R(r,\vartheta,\phi) = R_0(\phi) + r R_1(\vartheta,\phi) + r^2 R_2(\vartheta,\phi)+r^3 R_3(\vartheta,\phi)
+\ldots
\label{eq:RExpansion1}
\end{align}
where
\begin{align}
\label{eq:RExpansion2}
R_1(\vartheta,\phi) =& R_{1s}(\phi)\sin\vartheta + R_{1c}(\phi)\cos\vartheta, \\
R_2(\vartheta,\phi) =& R_{20}(\phi) + R_{2s}(\phi)\sin 2\vartheta + R_{2c}(\phi)\cos 2\vartheta, 
\nonumber\\
R_3(\vartheta,\phi) =& R_{3s3}(\phi)\sin 3\vartheta + R_{3s1}(\phi)\sin \vartheta + R_{3c3}(\phi)\cos 3\vartheta + R_{3c1}(\phi)\cos\vartheta. 
\nonumber
\end{align}
The same representation with $R\to z$ is also used. 
We also define $\nu(r,\vartheta,\phi)$ to be the difference between the Boozer and cylindrical toroidal angle: $\varphi = \phi + \nu$, and $\nu$ is expanded in the same way as $R$ in (\ref{eq:RExpansion1})-(\ref{eq:RExpansion2}). We define $\varphi_0(\phi) = \phi + \nu_0(\phi)$. To relate the Frenet-Serret and cylindrical representations of the position vector, each quantity in the former that depends on $\varphi$ is Taylor-expanded about $\varphi_0$. For instance, the position vector along the magnetic axis is written
\begin{align}
\vect{r}_0(\varphi) = \vect{r}_0(\varphi_0)
+ (\varphi - \varphi_0) \vect{r}'_0
+ \frac{(\varphi-\varphi_0)^2}{2} \vect{r}''_0
+ \frac{(\varphi-\varphi_0)^3}{6} \vect{r}'''_0+\ldots
\end{align}
where the derivatives $d^n \vect{r}_0/d\varphi^n$ denoted with primes are all evaluated at $\varphi_0$.
These derivatives of $\vect{r}_0$ are then written in terms of the Frenet-Serret vectors, e.g.
\begin{align}
\vect{r}''_0 =  \left( \ell'  \vect{t} \right)'
= \left( \ell'\right)^2 \kappa \vect{n} + \ell''\vect{t},
\end{align}
where $\ell'' = 0$ for quasisymmetry.

We next equate the representation (\ref{eq:cylindrical_representation}) of the position vector in
cylindrical coordinates to the representation (\ref{eq:positionVector}) of the position vector in the Frenet-Serret frame,
identifying terms at each order in $r/\mathcal{R}$. At $O((r/\mathcal{R})^0)$,
\begin{align}
R_0(\phi)\vect{e}_R(\phi) + z_0(\phi) \vect{e}_z = \vect{r}_0(\varphi_0) .
\end{align}
At $O((r/\mathcal{R})^1)$,
\begin{align}
R_1(\vartheta,\phi)\vect{e}_R(\phi)+z_1(\vartheta,\phi)\vect{e}_z
=\ell'  \vect{t}(\varphi_0) \nu_1(\vartheta,\phi) + X_1(\vartheta,\varphi_0) \vect{n}(\varphi_0)
+ Y_1(\vartheta,\varphi_0) \vect{b}(\varphi_0).
\label{eq:cylindricalConversionOrder1}
\end{align}
The $\vect{n}(\varphi_0)$ and $\vect{b}(\varphi_0)$ components of this equation give the linear system
\begin{align}
\begin{pmatrix} n_R & n_z \\ b_R & b_z \end{pmatrix}
\begin{pmatrix} R_1 \\ z_1 \end{pmatrix}
=
\begin{pmatrix} X_1 \\ Y_1 \end{pmatrix},
\label{eq:order1system}
\end{align}
where $n_R = \vect{n}(\varphi_0)\cdot\vect{e}_R(\phi)$, $b_z = \vect{b}(\varphi_0)\cdot\vect{e}_z$, etc.,
and quantities are understood to depend on $\phi$ or $\varphi_0$. As explained preceding (4.5) of \cite{PaperI}, the matrix in (\ref{eq:order1system}) has determinant $-R_0/(d\ell/d\phi)$ where
$d\ell/d\phi = [(dR_0/d\phi)^2+R_0^2+(dz_0/d\phi)^2]^{1/2}$,
so (\ref{eq:order1system})
can be inverted to give
\begin{align}
\begin{pmatrix} R_1 \\ z_1 \end{pmatrix}
=
\frac{d\ell/d\phi}{R_0}
\begin{pmatrix} -b_z & n_z \\ b_R & -n_R \end{pmatrix}
\begin{pmatrix} X_1 \\ Y_1 \end{pmatrix}.
\label{eq:order1systemInverted}
\end{align}
Furthermore, the $\vect{t}(\varphi_0)$ component of (\ref{eq:cylindricalConversionOrder1}) gives
\begin{align}
\label{eq:nu1}
\nu_1 = \left(\frac{d \ell}{d\phi} \ell' \right)^{-1}
\left( R_1 \frac{dR_0}{d\phi} +  z_1 \frac{dz_0}{d\phi} \right).
\end{align}
Both (\ref{eq:order1systemInverted}) and (\ref{eq:nu1})
have $\sin\vartheta$ and $\cos\vartheta$ components that are satisfied independently. Thus,
(\ref{eq:order1systemInverted})-(\ref{eq:nu1}) give $R_{1s}$, $R_{1c}$, $z_{1s}$,  $z_{1c}$, $\nu_{1s}$, and $\nu_{1c}$ in terms
of $X_{1s}$, $X_{1c}$, $Y_{1s}$, and $Y_{1c}$.
 
Next, the $O((r/\mathcal{R})^2)$ terms arising when (\ref{eq:cylindrical_representation}) and (\ref{eq:positionVector}) are equated give
\begin{align}
\label{eq:cylindricalConversionOrder2}
R_2 \vect{e}_R + z_2 \vect{e}_z = &\nu_2 \ell' \vect{t} + \frac{\nu_1^2}{2} \left(\ell' \right)^2 \kappa \vect{n}
+X_2 \vect{n}+\frac{\partial X_1}{\partial \varphi} \nu_1 \vect{n}+X_1 \vect{n}'\nu_1 \\
&+Y_2 \vect{b} + \frac{\partial Y_1}{\partial\varphi} \nu_1 \vect{b} + Y_1 \vect{b}' \nu_1
+ Z_2 \vect{t}.
\nonumber
\end{align}
Here, we have set $\ell''=0$ due to quasisymmetry.
The $\vect{n}$ and $\vect{b}$ components of (\ref{eq:cylindricalConversionOrder2})
give
\begin{align}
\begin{pmatrix} n_R & n_z \\ b_R & b_z \end{pmatrix}
\begin{pmatrix} R_2 \\ z_2 \end{pmatrix}
=
\begin{pmatrix} S_{2n}\\ S_{2b}\end{pmatrix},
\label{eq:order2system}
\end{align}
where
\begin{align}
S_{2n} = &X_2 + \frac{\nu_1^2 \kappa}{2} \left(\ell' \right)^2 + \nu_1 \frac{\partial X_1}{\partial\varphi}
-\nu_1 \tau  \ell'  Y_1, \\
S_{2b} = & Y_2 + \nu_1 \frac{\partial Y_1}{\partial\varphi} + \nu_1 \tau \ell'  X_1.
\nonumber
\end{align}
The matrix in (\ref{eq:order2system}) is the same one that arose in (\ref{eq:order1system}),
so it may be inverted as before to give
\begin{align}
\begin{pmatrix} R_2 \\ z_2 \end{pmatrix}
=
\frac{d\ell/d\phi}{R_0}
\begin{pmatrix} -b_z & n_z \\ b_R & -n_R \end{pmatrix}
\begin{pmatrix} S_{2n} \\ S_{2b} \end{pmatrix}.
\label{eq:order2systemInverted}
\end{align}
We will also need the $\vect{t}$ component of (\ref{eq:cylindricalConversionOrder2}):
\begin{align}
\label{eq:nu2}
\nu_2 = \left( \frac{d\ell}{d\phi} \ell' \right)^{-1} \left(R_2 \frac{dR_0}{d\phi} + z_2 \frac{dz_0}{d\phi}\right) - \frac{Z_2}{\ell' }
+\kappa \nu_1 X_1.
\end{align}
Both (\ref{eq:order2systemInverted}) and (\ref{eq:nu2}) have components with $\vartheta$ dependence
$\propto \sin 2\vartheta$, $\propto \cos 2\vartheta$, and $\propto 1$. To evaluate these components,
we can note that for any quantities $P_1 = P_{1s}\sin\vartheta + P_{1c}\cos\vartheta$ and $Q_1 = Q_{1s}\sin\vartheta + Q_{1c}\cos\vartheta$,
\begin{align}
P_1 Q_1 = \frac{P_{1c} Q_{1c} + P_{1s}Q_{1s}}{2}
+ \frac{P_{1c} Q_{1s} + P_{1s} Q_{1c}}{2}\sin 2\vartheta
+ \frac{P_{1c} Q_{1c} - P_{1s} Q_{1s}}{2}\cos 2\vartheta.
\end{align}
Thus, (\ref{eq:order2systemInverted}) and (\ref{eq:nu2}) give the Fourier modes of $R_2$, $z_2$, and $\nu_2$ in terms of the Fourier modes of $X_2$, $Y_2$, and $Z_2$.

Finally, we identify the $O((r/\mathcal{R})^3)$ terms arising when (\ref{eq:cylindrical_representation}) and (\ref{eq:positionVector}) are equated. The $\vect{n}(\varphi_0)$ and $\vect{b}(\varphi_0)$ components have the same form as (\ref{eq:order2system}) but with
$R_2 \to R_3$, $z_2 \to z_3$, $S_{2n} \to S_{3n}$, and $S_{2b} \to S_{3b}$
where
\begin{align}
S_{3n} = &\nu_1 \nu_2 \left(\ell' \right)^2 \kappa
+\frac{\nu_1^3}{6}\left(\ell' \right)^2 \kappa'
+\nu_2 \frac{\partial X_1}{\partial \varphi}
+\frac{\nu_1^2}{2} \frac{\partial^2 X_1}{\partial\varphi^2} + \nu_1 \frac{\partial X_2}{\partial \varphi} + X_3
\\ &
-\frac{\nu_1^2}{2} \left(\ell' \right)^2 (\kappa^2+\tau^2) X_1
-\nu_1^2 \tau \ell'  \frac{\partial Y_1}{\partial\varphi}
-\nu_1 \tau \ell'  Y_2 - \nu_2 \tau \ell'  Y_1
\nonumber \\ &
-\frac{\nu_1^2}{2}\ell'  \tau'Y_1 + \nu_1 \kappa \ell'  Z_2
\nonumber,
\end{align}
\begin{align}
S_{3b} = &
\frac{\nu_1^3}{6} \kappa\tau \left(\ell' \right)^3
+\nu_1^2 \tau \ell' \frac{\partial X_1}{\partial\varphi}
+\nu_1\tau \ell'  X_2
+\nu_2 \tau \ell'  X_1
\\ &
+\frac{\nu_1^2}{2} \tau' \ell'  X_1
+\nu_2 \frac{\partial Y_1}{\partial\varphi}
+\frac{\nu_1^2}{2} \frac{\partial^2 Y_1}{\partial \varphi^2}
+\nu_1 \frac{\partial Y_2}{\partial\varphi}
+Y_3 - \frac{\nu_1^2}{2} \tau^2 \left(\ell' \right)^2 Y_1.
\nonumber
\end{align}
The system 
can be solved as in the previous orders to give
\begin{align}
\begin{pmatrix} R_3 \\ z_3 \end{pmatrix}
=
\frac{d\ell/d\phi}{R_0}
\begin{pmatrix} -b_z & n_z \\ b_R & -n_R \end{pmatrix}
\begin{pmatrix} S_{3n} \\ S_{3b} \end{pmatrix}.
\label{eq:order3systemInverted}
\end{align}
In (\ref{eq:order3systemInverted}) the Fourier components
$\propto \sin\vartheta$, $\propto \sin 3\vartheta$, $\propto \cos\vartheta$, and $\propto \cos 3\vartheta$
are each satisfied independently.



\bibliographystyle{jpp}
\bibliography{ConstructingQuasisymmetricStellaratorsTo2ndOrder}

\begin{thebibliography}{30}
\expandafter\ifx\csname natexlab\endcsname\relax\def\natexlab#1{#1}\fi
\def\au#1{#1} \def\ed#1{#1} \def\yr#1{#1}\def\at#1{#1}\def\jt#1{\textit{#1}}
  \def\bt#1{#1}\def\bvol#1{\textbf{#1}} \def\vol#1{#1} \def\pg#1{#1}
  \def\publ#1{#1}\def\arxiv#1{#1}\def\org#1{#1}\def\st#1{\textit{#1}}

\bibitem[Anderson {\em et~al.\/}(1995)Anderson, Almagri, Anderson, Matthews,
  Talmadge \& Shohet]{HSX}
{\sc \au{Anderson, F S~B}, \au{Almagri, A~F}, \au{Anderson, D~T}, \au{Matthews,
  P~G}, \au{Talmadge, J~N} \& \au{Shohet, J~L}} \yr{1995}  \at{The helically
  symmetric experiment ({HSX}): Goals, design, and status}.  \jt{Fusion Tech.}
  \bvol{27},  \pg{273}.

\bibitem[Beidler {\em et~al.\/}(2011)Beidler, Allmaier, Isaev, Kasilov,
  Kernbichler, Leitold, Maassberg, Mikkelsen, Murakami, Schmidt, Spong,
  Tribaldos \& Wakasa]{Beidler}
{\sc \au{Beidler, C.D.}, \au{Allmaier, K.}, \au{Isaev, M.~Yu.}, \au{Kasilov,
  S.V.}, \au{Kernbichler, W.}, \au{Leitold, G.O.}, \au{Maassberg, H.},
  \au{Mikkelsen, D.R.}, \au{Murakami, S.}, \au{Schmidt, M.}, \au{Spong, D.A.},
  \au{Tribaldos, V.} \& \au{Wakasa, A.}} \yr{2011}  \at{Benchmarking of the
  mono-energetic transport coefficients-results from the international
  collaboration on neoclassical transport in stellarators {(ICNTS)}}.
  \jt{Nucl. Fusion}  \bvol{51},  \pg{076001}.

\bibitem[Boozer(1983)]{Boozer83}
{\sc \au{Boozer, A~H}} \yr{1983}  \at{Transport and isomorphic equilibria}.
  \jt{Phys. Fluids}  \bvol{26},  \pg{496}.

\bibitem[Boozer(1995)]{Boozer1995}
{\sc \au{Boozer, A~H}} \yr{1995}  \at{Quasi-helical symmetry in stellarators}.
  \jt{Plasma Phys. Controlled Fusion}  \bvol{37},  \pg{A103}.

\bibitem[Drevlak {\em et~al.\/}(2019)Drevlak, Beidler, Geiger, Helander \&
  Turkin]{ROSE}
{\sc \au{Drevlak, M.}, \au{Beidler, C.~D.}, \au{Geiger, J.}, \au{Helander, P.}
  \& \au{Turkin, Y.}} \yr{2019}  \at{{Optimisation of stellarator equilibria
  with ROSE}}.  \jt{Nucl. Fusion}  \bvol{59},  \pg{016010}.

\bibitem[Drevlak {\em et~al.\/}(2013)Drevlak, Brochard, Helander, Kisslinger,
  Mikhailov, N\"{u}hrenberg, N\"{u}hrenberg \& Turkin]{ESTELL}
{\sc \au{Drevlak, M}, \au{Brochard, F}, \au{Helander, P}, \au{Kisslinger, J},
  \au{Mikhailov, M}, \au{N\"{u}hrenberg, C}, \au{N\"{u}hrenberg, J} \&
  \au{Turkin, Y}} \yr{2013}  \at{{ESTELL}: A quasi-toroidally symmetric
  stellarator}.  \jt{Contrib. Plasma Phys.}  \bvol{53},  \pg{459}.

\bibitem[Garabedian(1996)]{Garabedian}
{\sc \au{Garabedian, P~R}} \yr{1996}  \at{Stellarators with the magnetic
  symmetry of a tokamak}.  \jt{Phys. Plasmas}  \bvol{3},  \pg{2483}.

\bibitem[Garren \& Boozer(1991{\natexlab{{\em a\/}}})]{GB2}
{\sc \au{Garren, D~A} \& \au{Boozer, A~H}} \yr{1991{\natexlab{{\em a\/}}}}
  \at{Existence of quasihelically symmetric stellarators}.  \jt{Phys. Fluids B}
   \bvol{3},  \pg{2822}.

\bibitem[Garren \& Boozer(1991{\natexlab{{\em b\/}}})]{GB1}
{\sc \au{Garren, D~A} \& \au{Boozer, A~H}} \yr{1991{\natexlab{{\em b\/}}}}
  \at{Magnetic field strength of toroidal plasma equilibria}.  \jt{Phys. Fluids
  B}  \bvol{3},  \pg{2805}.

\bibitem[Helander(2014)]{HelanderReview}
{\sc \au{Helander, P}} \yr{2014}  \at{Theory of plasma confinement in
  non-axisymmetric magnetic fields}.  \jt{Rep. Prog. Phys.}  \bvol{77},
  \pg{087001}.

\bibitem[Henneberg {\em et~al.\/}(2018)Henneberg, Drevlak, N{\"{u}}hrenberg,
  Beidler, Turkin, Loizu \& Helander]{Henneberg}
{\sc \au{Henneberg, S.~A.}, \au{Drevlak, M.}, \au{N{\"{u}}hrenberg, C.},
  \au{Beidler, C.~D.}, \au{Turkin, Y.}, \au{Loizu, J.} \& \au{Helander, P.}}
  \yr{2018}  \at{Properties of a new quasi-axisymmetric configuration}.
  \jt{arXiv:1810.04914} .

\bibitem[Hirshman \& Whitson(1983)]{VMEC1983}
{\sc \au{Hirshman, S~P} \& \au{Whitson, J~C}} \yr{1983}  \at{Steepest-descent
  moment method for three-dimensional magnetohydrodynamic equilibria}.
  \jt{Phys. Fluids}  \bvol{26},  \pg{3553}.

\bibitem[Ku \& Boozer(2011)]{KuBoozerQHS}
{\sc \au{Ku, L~P} \& \au{Boozer, A~H}} \yr{2011}  \at{New classes of
  quasi-helically symmetric stellarators}.  \jt{Nucl. Fusion}  \bvol{51},
  \pg{013004}.

\bibitem[Lagarias {\em et~al.\/}(1998)Lagarias, Reeds, Wright \&
  Wright]{Lagarias}
{\sc \au{Lagarias, J~C}, \au{Reeds, J.~A.}, \au{Wright, M.~H.} \& \au{Wright,
  P.~E.}} \yr{1998}  \at{{Convergence Properties of the Nelder-Mead Simplex
  Method in Low Dimensions}}.  \jt{SIAM J. Opt.}  \bvol{9},  \pg{112}.

\bibitem[Landreman(2019)]{fitToGarrenBoozer}
{\sc \au{Landreman, M}} \yr{2019}  \at{{Optimized quasisymmetric stellarators
  are consistent with the Garren-Boozer construction}}.  \jt{Plasma Phys.
  Controlled Fusion}  \bvol{61},  \pg{075001}.

\bibitem[Landreman \& Sengupta(2018)]{PaperI}
{\sc \au{Landreman, M} \& \au{Sengupta, W}} \yr{2018}  \at{{Direct construction
  of optimized stellarator shapes. I. Theory in cylindrical coordinates}}.
  \jt{J. Plasma Phys.}  \bvol{84},  \pg{905840616}.

\bibitem[Landreman {\em et~al.\/}(2019)Landreman, Sengupta \& Plunk]{PaperII}
{\sc \au{Landreman, M}, \au{Sengupta, W} \& \au{Plunk, G~G}} \yr{2019}
  \at{{Direct construction of optimized stellarator shapes. II. Numerical
  quasisymmetric solutions}}.  \jt{J. Plasma Phys.}  \bvol{85},
  \pg{905850103}.

\bibitem[Liu {\em et~al.\/}(2018)Liu, Shimizu, Isobe, Okamura, Nishimura,
  Suzuki, Xu, Zhang, Liu, Huang, Wang, Liu, Tang, Yin, Wan \& {the CFQS
  team}]{CFQSLiu}
{\sc \au{Liu, H}, \au{Shimizu, A}, \au{Isobe, M}, \au{Okamura, S},
  \au{Nishimura, S}, \au{Suzuki, C}, \au{Xu, Y}, \au{Zhang, X}, \au{Liu, B},
  \au{Huang, J}, \au{Wang, X}, \au{Liu, H}, \au{Tang, C}, \au{Yin, D}, \au{Wan,
  Y} \& \au{{the CFQS team}}} \yr{2018}  \at{Magnetic configuration and modular
  coil design for the chinese first quasi-axisymmetric stellarator}.
  \jt{Plasma Fusion Res.}  \bvol{13},  \pg{3405067}.

\bibitem[Lortz \& N\"{u}hrenberg(1976)]{LortzNuhrenberg}
{\sc \au{Lortz, D} \& \au{N\"{u}hrenberg, J}} \yr{1976}  \at{{Equilibrium and
  Stability of a Three-Dimensional Toroidal {MHD} Configuration Near its
  Magnetic Axis}}.  \jt{Z. Naturforsch.}  \bvol{31a},  \pg{1277}.

\bibitem[Mercier(1964)]{Mercier}
{\sc \au{Mercier, C}} \yr{1964}  \at{Equilibrium and stability of a toroidal
  magnetohydrodynamic system in the neighbourhood of a magnetic axis}.
  \jt{Nucl. Fusion}  \bvol{4},  \pg{213}.

\bibitem[N\"{u}hrenberg \& Zille(1988)]{NuhrenbergZille}
{\sc \au{N\"{u}hrenberg, J} \& \au{Zille, R}} \yr{1988}  \at{Quasi-helically
  symmetric toroidal stellarators}.  \jt{Phys. Lett. A}  \bvol{129},  \pg{113}.

\bibitem[Pedersen {\em et~al.\/}(2012)Pedersen, Danielson, Hugenschmidt, Marx,
  Sarasola, Schauer, Schweikhard, Surko \& Winkler]{Pedersen}
{\sc \au{Pedersen, T~Sunn}, \au{Danielson, J~R}, \au{Hugenschmidt, C},
  \au{Marx, G}, \au{Sarasola, X}, \au{Schauer, F}, \au{Schweikhard, L},
  \au{Surko, C~M} \& \au{Winkler, E}} \yr{2012}  \at{Plans for the creation and
  studies of electron–positron plasmas in a stellarator}.  \jt{New J. Phys.}
  \bvol{14},  \pg{035010}.

\bibitem[Plunk \& Helander(2018)]{PlunkHelander}
{\sc \au{Plunk, G~G} \& \au{Helander, P}} \yr{2018}  \at{Quasi-axisymmetric
  magnetic fields: weakly non-axisymmetric case in a vacuum}.  \jt{J. Plasma
  Phys.}  \bvol{84},  \pg{905840205}.

\bibitem[Plunk {\em et~al.\/}(2019)Plunk, Landreman \& Helander]{PaperIII}
{\sc \au{Plunk, G~G}, \au{Landreman, M} \& \au{Helander, P}} \yr{2019}
  \at{{Direct construction of optimized stellarator shapes. III. Omnigenity
  near the magnetic axis}}.  \jt{arXiv:1909.08919, accepted for publication in
  J Plasma Phys.} .

\bibitem[Reiman {\em et~al.\/}(1999)Reiman, Fu, Hirshman, Ku, Monticello,
  Mynick, Redi, Spong, Zarnstorff, Blackwell, Boozer, Brooks, Cooper, Drevlak,
  Goldston, Harris, Isaev, Kessel, Lin, Lyon, Merkel, Mikhailov, Miner,
  Nakajima, Neilson, N\"{u}hrenberg, Okamoto, Pomphrey, Reiersen, Sanchez,
  Schmidt, Subbotin, Valanju, Watanabe \& White]{StelloptReiman}
{\sc \au{Reiman, A}, \au{Fu, G}, \au{Hirshman, S}, \au{Ku, L}, \au{Monticello,
  D}, \au{Mynick, H}, \au{Redi, M}, \au{Spong, D}, \au{Zarnstorff, M},
  \au{Blackwell, B}, \au{Boozer, A}, \au{Brooks, A}, \au{Cooper, W~A},
  \au{Drevlak, M}, \au{Goldston, R}, \au{Harris, J}, \au{Isaev, M}, \au{Kessel,
  C}, \au{Lin, Z}, \au{Lyon, J~F}, \au{Merkel, P}, \au{Mikhailov, M},
  \au{Miner, W}, \au{Nakajima, N}, \au{Neilson, G}, \au{N\"{u}hrenberg, C},
  \au{Okamoto, M}, \au{Pomphrey, N}, \au{Reiersen, W}, \au{Sanchez, R},
  \au{Schmidt, J}, \au{Subbotin, A}, \au{Valanju, P}, \au{Watanabe, K~Y} \&
  \au{White, R}} \yr{1999}  \at{{Physics design of a high-$\beta$
  quasi-axisymmetric stellarator}}.  \jt{Plasma Phys. Controlled Fusion}
  \bvol{41},  \pg{B273}.

\bibitem[Sanchez {\em et~al.\/}(2000)Sanchez, Hirshman, Ware, Berry \&
  Spong]{Sanchez}
{\sc \au{Sanchez, R}, \au{Hirshman, S~P}, \au{Ware, A~S}, \au{Berry, L~A} \&
  \au{Spong, D~A}} \yr{2000}  \at{Ballooning stability optimization of
  low-aspect-ratio stellarators}.  \jt{Plasma Phys. Controlled Fusion}
  \bvol{42},  \pg{641}.

\bibitem[Shimizu {\em et~al.\/}(2018)Shimizu, Liu, Isobe, Okamura, Nishimura,
  Suzuki, Xu, Zhang, Liu, Huang, Wang, Liu, Tang \& {the CFQS
  team}]{CFQSShimizu}
{\sc \au{Shimizu, A}, \au{Liu, H}, \au{Isobe, M}, \au{Okamura, S},
  \au{Nishimura, S}, \au{Suzuki, C}, \au{Xu, Y}, \au{Zhang, X}, \au{Liu, B},
  \au{Huang, J}, \au{Wang, X}, \au{Liu, H}, \au{Tang, C} \& \au{{the CFQS
  team}}} \yr{2018}  \at{Configuration property of the chinese first
  quasi-axisymmetric stellarator}.  \jt{Plasma Fusion Res.}  \bvol{13},
  \pg{3403123}.

\bibitem[Solov'ev \& Shafranov(1970)]{SolovevShafranov}
{\sc \au{Solov'ev, L~S} \& \au{Shafranov, V~D}} \yr{1970} {\em {Reviews of
  Plasma Physics 5}\/}.  \publ{New York - London: Consultants Bureau}.

\bibitem[Spong {\em et~al.\/}(1998)Spong, Hirshman, Whitson, Batchelor,
  Carreras, Lynch \& Rome]{StelloptSpong}
{\sc \au{Spong, D.~A.}, \au{Hirshman, S.~P.}, \au{Whitson, J.~C.},
  \au{Batchelor, D.~B.}, \au{Carreras, B.~A.}, \au{Lynch, V.~E.} \& \au{Rome,
  J.~A.}} \yr{1998}  \at{{$J^*$ optimization of small aspect ratio
  stellarator/tokamak hybrid devices}}.  \jt{Phys. Plasmas}  \bvol{5},
  \pg{1752}.

\bibitem[Zarnstorff {\em et~al.\/}(2001)Zarnstorff, Berry, Brooks, Fredrickson,
  Fu, Hirshman, Hudson, Ku, Lazarus, Mikkelsen, Monticello, Neilson, Pomphrey,
  Reiman, Spong, Strickler, Boozer, Cooper, Goldston, Hatcher, Isaev, Kessel,
  Lewandowski, Lyon, Merkel, Mynick, Nelson, Nuehrenberg, Redi, Reiersen,
  Rutherford, Sanchez, Schmidt \& White]{NCSX}
{\sc \au{Zarnstorff, M~C}, \au{Berry, L~A}, \au{Brooks, A}, \au{Fredrickson,
  E}, \au{Fu, {G-Y}}, \au{Hirshman, S}, \au{Hudson, S}, \au{Ku, {L-P}},
  \au{Lazarus, E}, \au{Mikkelsen, D}, \au{Monticello, D}, \au{Neilson, G~H},
  \au{Pomphrey, N}, \au{Reiman, A}, \au{Spong, D}, \au{Strickler, D},
  \au{Boozer, A}, \au{Cooper, W~A}, \au{Goldston, R}, \au{Hatcher, R},
  \au{Isaev, M}, \au{Kessel, C}, \au{Lewandowski, J}, \au{Lyon, J~F},
  \au{Merkel, P}, \au{Mynick, H}, \au{Nelson, B~E}, \au{Nuehrenberg, C},
  \au{Redi, M}, \au{Reiersen, W}, \au{Rutherford, P}, \au{Sanchez, R},
  \au{Schmidt, J} \& \au{White, R~B}} \yr{2001}  \at{Physics of the compact
  advanced stellarator {NCSX}}.  \jt{Plasma Phys. Controlled Fusion}
  \bvol{43},  \pg{A237}.

\end{thebibliography}

\end{document}